# The intuitionistic fragment of computability logic at the propositional level

Giorgi Japaridze*

Department of Computing Sciences, Villanova University
800 Lancaster Avenue, Villanova, PA 19085, USA
Email: giorgi.japaridze@villanova.edu
URL: http://www.csc.villanova.edu/~japaridz/

**Abstract**

This paper presents a soundness and completeness proof for propositional intuitionistic calculus with respect to the semantics of computability logic. The latter interprets formulas as interactive computational problems, formalized as games between a machine and its environment. Intuitionistic implication is understood as algorithmic reduction in the weakest possible — and hence most natural — sense, disjunction and conjunction as deterministic-choice combinations of problems (disjunction = machine's choice, conjunction = environment's choice), and "absurd" as a computational problem of universal strength.

# Contents

*This material is based upon work supported by the National Science Foundation under Grant No. 0208816.

# 1 Introduction

*Computability logic* (CL), introduced in [6], is a framework and program for redeveloping logic as a formal theory of computability, as opposed to the formal theory of truth which it has more traditionally been. Unlike many nonclassical visions, the most notable examples being intuitionistic and linear logics, CL is not an axiomatically conceived approach. Rather, it was born as a pure semantics, leaving the task of finding corresponding axiomatizations (syntax) as a challenge for future efforts. While [6] did not prove any syntactic results, it did take the risk of stating conjectures regarding the soundness and completeness (*adequacy*) of certain deductive systems with respect to the semantics of CL. Among the systems conjectured to adequately axiomatize a fragment of otherwise much more expressive CL was Heyting's intuitionistic calculus **INT** in its full predicate-level language. The first step toward a positive verification of that conjecture was made in [12], where its soundness part was proven. The next step forward came in [13], where the implicative fragment of **INT** was proven to be complete. Generalizing and further advancing the methods used in [13], the present paper tells the success story of a third step, proving the completeness of the full propositional fragment **Int** of **INT** with respect to the semantics of CL. To reduce the degree of reliance on prior (and not yet officially published) work, this paper also includes a soundness proof for **Int**, the result already known from [12] for the more expressive **INT**. In a reasonably optimistic scenario, the last step toward a full verification of the above conjecture will be extending the present propositional-level completeness proof to the predicate level, probably building upon the ideas and techniques developed in this article.

The sheer size of the present paper suggests that the CL-adequacy proof for intuitionistic logic is far from trivial. This may give rise to a naive question in the following style. There are already many semantics — Kripke models, algebraic or topological interpretations, etc. — with respect to which intuitionistic logic has been proven to be adequate. So why bother studying yet another semantics, with apparently by an order of magnitude more complex proofs of the corresponding adequacy theorems? Furthermore, the CL semantics is in fact a game semantics. Then why not just be happy with, say, Lorenzen's [17] game semantics, which has been around since long ago?

CL has the following words for the doubtful. If simplicity is most important, why not indeed put aside this (and any other) paper and focus on ... the empty logic with the semantics that ... falsifies everything. Or if, for some reason, you want to deal with intuitionistic and only intuitionistic logic, here is an idea: consider the semantics where truth, by definition, means provability in Heyting's calculus. You got a one-line-long definition of the whole semantics together with a proof of adequacy. Or if, for some reason, you want the semantics to be by all means a game semantics, how about this one: Every formula $F$ is a game between Proponent and Opponent. The first move is by Proponent, consisting in presenting a tree $T$ of formulas with $F$ at its root and claiming that $T$ is a proof of $F$. The next move is by Opponent, consisting in choosing a node $G$ of $T$ and claiming that $T$ has an error at $G$. Opponent wins if and only if $G$ (indeed) is neither an axiom nor does it follow from its children by any of the inference rules of Heyting's calculus. Sure adequacy is guaranteed: Proponent has a winning strategy for $F$ if and only if $F$ is provable. Or maybe you want something (pretending to be) more "serious", such as, say, games involving traversing a potential proof tree $T$ for $F$ starting from its root, where, at each step, depending on what rule is claimed (by Proponent) to justify the node, either Proponent or Opponent chooses a child? No problem, give me ten minutes and half a page, I will produce precise definitions and an adequacy proof. Or maybe ...

OK, that's enough. Hopefully the point has been made. It is always possible to come up with some sort of a "semantics" with respect to which the system under question is sound and complete. And there are no limits to how simple — or the opposite — both the semantics and the corresponding adequacy proofs can be. But the point is: what is the point? The term *semantics* has often been abused in the logical literature, resulting in a depreciation of its originally noble meaning. What has often been sold as the real thing has in fact been not too far from the parodies of semantics seen in the previous paragraph. This sort of unfair game has typically been motivated by the desire to justify, at any cost, one or another popular syntactic construction.

Wrong. According to the credo of CL, it is syntax that should be a faithful servant to a meaningful and motivated semantics rather than vice versa. This is so because CL sees logic as a tool for "successfully navigating the real world" ([10]). And the true purpose of a semantics — of one with a capital 'S' — should be providing a bridge between that real, outside world and the man-made world of otherwise meaningless symbolic expressions and proofs of formal logic. Without achieving this goal the semantics, whether technically a game semantics or not, would be nothing but just a game. Adjusting a semantics to the needs of some beautiful or otherwise appealing syntax, which has been done so often by so many logicians, would be fine if it was pos-sible to simultaneously make adjustments in that real, outside world. Unfortunately, however, the real world — at least at the level of abstraction suitable for logic — remains largely beyond the control of a man.

The semantics of classical logic is the model example of a semantics with a capital 'S'. Among the semantics for intuitionistic logic that are on the "capital-S" side one could name the formulas-as-types approach (Curry-Howard isomorphism) and Artemov's [1] semantics of proofs, even though, in their general spirits and philosophies, these two are very different from the approach of CL. The algebraic or Kripke semantics fall under another category, to which [10] refers as "*lowercase 's' semantics*". These are often very useful in analyzing and understanding given syntactic constructions, but in most cases cannot be treated as independent justifications for those constructions: an attempt to see more than syntax-serving technical tools in these sorts of semantics, as pointed out in [10], may yield a vicious circle. The same obviously applies to Lorenzen's game semantics for intuitionistic logic. To many questions in the style "but why this and not another way?", it essentially has no other answer but the circular "because this is necessary to achieve the adequacy of Heyting's calculus". Another serious disadvantage of Lorenzen's semantics before those of classical or computability logics is that, while it talks about validity, a concept of truth is inherently absent in it, signifying that applied theories (say, arithmetic) could not be meaningfully based on that semantics, even if pure logic could.

Blass's [2, 3] game semantics, which happens to be the closest precursor of the semantics of computability logic, deserves a special comment. A detailed discussion of what is common and different between the two is given in Sections 13 and 27 of [6], and we will not elaborate on this topic here. It should be noted that, unlike Lorenzen's game semantics, there are no artificially tuned elements in Blass's semantics, to which it owes a great degree of naturalness and appeal. And, as early as in 1972, the propositional fragment of intuitionistic logic was found to be sound with respect to it. Unfortunately, no further technical progress has been made in that promising direction, and the questions on predicate-level soundness or whatever-level completeness have been remaining unanswered since then.

Computability logic is not modest about claiming that it offers a capital-'S' semantics in the proper sense. Trying to convince the reader that there are good reasons for such a claim is outside the scope of the present paper. This job has been done in [6, 9, 10]. Below we only give an atrociously brief (by the standards of CL) overview of the intuitions pertaining to the relevant fragment of CL.

That relevant fragment — let us call it the *intuitionistic fragment* — is obtained by restricting the otherwise more expressive (and, in fact, open-ended) formalism of CL to the logical vocabulary $\{\circ\!-, \sqcup, \sqcap, \bigsqcup, \bigsqcap, \$\}$, with these six items intended as the readings of the intuitionistic implication, disjunction, conjunction, existential quantifier, universal quantifier and absurd, respectively. Negation is not formally present as a primitive symbol here, but is rather understood as an abbreviation defined by

$$\textit{The intuitionistic negation of } F \;=\; F \circ\!- \$.$$

The semantics of computability logic is indeed a game semantics. And games in it are indeed not "just games". They are seen as formal equivalents of our broadest intuition of computational problems, specifically, computational problems in their most general — interactive — sense. Computability, or algorithmic solvability, of such problems is understood as existence of an interactive Turing machine that wins the game no matter how the other player, which is the environment or the user, behaves. In this vision of logic, computability replaces the classical concept of truth, operations on games=problems replace Boolean connectives and quantifiers, and the understanding of logical formulas as (representing) computational problems comes in the place of their classical understanding as true/false statements, with valid formulas now describing universally correct principles of computability ("always computable" problems), and the logic in whole providing a systematic answer to the question on what and how can be algorithmically solved.

There is a host of the most basic and natural operations on interactive computational problems, and those represented by the operators of intuitionistic logic are among them. So, computability logic and its intuitionistic fragment in particular are logics of problems in a direct and proper sense. This reminds us of the famous thesis set forth by Kolmogorov [16] in 1932, and a number of unsuccessful (in that no adequacy was achieved) attempts to materialize it, such as Kleene's realizability [14], Gödel's Dialectica interpretation [5] or Medvedev's finite-problem semantics [18]. According to Kolmogorov's thesis, stated at the philosophical level rather than in clear technical terms, *intuitionistic logic is a logic of problems.* CL materializes this abstract philosophy by turning it into a precise mathematics, only for a more expressive formal language than that of intuitionistic logic, with the latter being a modest fragment of the former.

In what follows we briefly survey the "intuitionistic" operations of computability logic. Among the other operations officially introduced so far within the framework of CL are *negation* $\neg$, *parallel connectives* $\vee$,$\wedge$, *parallel quantifiers* $\bigvee$, $\bigwedge$, *strict reduction* $\rightarrow$, *blind quantifiers* $\exists$, $\forall$, *parallel recurrences* $\curlyvee$, $\curlywedge$, *branching recurrences* ⫯, ⫰. Detailed intuitive explanations and formal definitions for all these operations, along with the "intuitionistic" ones, can be found in [10].

The operators $\sqcap, \sqcup, \bigsqcap, \bigsqcup$ are called *choice operators* in CL, because they represent (deterministic) choice steps in the course of interaction between a machine and its environment. The choice disjunction $A_1 \sqcup A_2$ of games $A_1$ and $A_2$ is the game where the first move is by the machine. The move should be '1' or '2', amounting to selecting the left or the right disjunct. After such a move $i \in \{1, 2\}$ is made, the play continues and the winner is determined according to the rules of the chosen component $A_i$. And if the machine simply fails to make any move/choice, then it is considered to have lost. Choice conjunction $A_1 \sqcap A_2$ is defined in exactly the same way, only with the roles of the machine and the environment interchanged. That is, the initial move $i \in \{1, 2\}$ in $A_1 \sqcap A_2$ is the environment's privilege and obligation. As for choice quantifiers, with $\{1, 2, 3, \ldots\}$ being the universe of discourse, they can be understood as infinite versions of $\sqcup$ and $\sqcap$, defined by

$$\bigsqcup x A(x) \;=\; A(1) \sqcup A(2) \sqcup A(3) \sqcup \ldots$$

and

$$\bigsqcap x A(x) \;=\; A(1) \sqcap A(2) \sqcap A(3) \sqcap \ldots$$

For instance, where $f(x)$ is a function, $\sqcap x \sqcup y\big(y = f(x)\big)$ is the game in which the first move is by the environment, consisting in specifying a particular value $m$ for $x$. Such a move, which intuitively can be seen as asking the machine the question "*what is the value of* $f(m)$*?*" brings the game down to the position $\sqcup y\big(y = f(m)\big)$. The next step is by the machine, who should specify a value $n$ for $y$, further bringing the game down to $n = f(m)$. The latter is understood as a moveless position won by the machine if true and lost if false.[1] The machine's move $n$ can thus be seen as claiming that $n$ is the value of $f(m)$. Now it must be clear that $\sqcap x \sqcup y\big(y = f(x)\big)$ represents the problem of computing $f$, with the machine having an algorithmic winning strategy for this game iff $f$ is a computable function. Similarly, where $S$ is a set, $\sqcap x(x \in S \sqcup x \notin S)$ represents the problem of deciding $S$: here, again, the first move is by the environment, consisting in choosing a value $m$ for $x$ (asking "*is* $m$ *an element of* $S$*?*"); and the next step is by the machine who, in order to win, should choose the true disjunct of $m \in S \sqcup m \notin S$, i.e. correctly tell whether $m$ is an element of $S$ or not.

Next, $\circ\!-$ is an operation of reducing one problem to another in the weakest possible — and hence most natural — sense, for which reason [6] calls it *weak reduction*,[2] as opposed to strong (strict, linear) reduction denoted in CL by the different symbol $\rightarrow$. As a reduction operation, $\circ\!-$ induces — and in certain contexts can be identified with — a reducibility relation. Specifically, a problem $B$ is *weakly reducible* to a problem $A$ iff the problem $A \circ\!- B$ has an algorithmic solution. Intuitively, $A \circ\!- B$ is a problem solving which means solving $B$ while having $A$ as an external computational resource, with an "external computational resource" meaning a to-be-solved-by-the-environment computational problem. More precisely, in the antecedent the roles of the problem-solving machine and its environment are interchanged. Acting in the role of an environment there, the machine may observe how the antecedent is being solved, and utilize this information in its own solving the consequent.

An example might help. Let $T$ be a finitely axiomatized applied theory based on classical first-order logic, such as, say, Robinson's arithmetic, and let $x$ range over the formulas of the language of $T$. Next, let $Pr_T(x)$ be the predicate "$x$ *is a theorem of* $T$", $Pr(x)$ the predicate "$x$ *is provable in classical predicate calculus*", and $\neg Pr_T(x)$ and $\neg Pr(x)$ the ordinary negations of these predicates. As a decision problem, the $T$-theoremhood problem would be expressed by $\sqcap x\big(Pr_T(x) \sqcup \neg Pr_T(x)\big)$. This is generally undecidable, yet algorithmically reducible to the problem $\sqcap x\big(Pr(x) \sqcup \neg Pr(x)\big)$ of provability in predicate calculus. The problem of reducing the former to the latter can then be expressed by

$$\sqcap x\big(Pr(x) \sqcup \neg Pr(x)\big) \quad \circ\!- \quad \sqcap x\big(Pr_T(x) \sqcup \neg Pr_T(x)\big). \tag{1}$$

The duty of a machine solving the above compound problem is to solve $\sqcap x\big(Pr_T(x) \sqcup \neg Pr_T(x)\big)$, i.e. answer any given question of the type "*is* $F$ *a theorem of* $T$*?*" asked by the environment. Yet, the machine is expected to do so only on the condition that the environment does not fail to correctly solve the similar problem $\sqcap x\big(Pr(x) \sqcup \neg Pr(x)\big)$ in the antecedent, where the roles are switched and it is the machine who can ask questions like "*is H provable in predicate calculus?*". So, here is an algorithmic strategy for the machine. Wait until the environment asks "*is F a theorem of T?*" for some particular $F$. Then ask the counterquestion "*is $Ax \rightarrow F$ provable in predicate calculus?*", where $Ax$ is the conjunction of all nonlogical axioms of $T$. The environment will have to provide a correct yes/no answer, or else it loses. Whatever the answer of the environment in the antecedent is, repeat the same answer in the consequent, and rest your case. A success of this strategy is guaranteed by the deduction theorem for classical logic.

[1] This should not suggest that the atoms of intuitionistic logic are required to be interpreted as moveless games/positions. Rather, they represent arbitrary interactive computational problems.

[2] The symbol used for $\circ\!-$ in [6] was $\Rightarrow$.

Observe that the above explanation of the meaning of $A \circ\!\!- B$ is exactly what the Turing reduction of $B$ to $A$ is all about. The latter — more precisely, the *Turing reducibility* of $B$ to $A$ — is defined as existence of a Turing machine that solves $B$ when having an oracle correctly answering any questions regarding $A$. We can see such an oracle as (a part of) the environment solving $A$ for the machine, thus providing an external computational resource. The only difference is that the resource provided by an oracle is always a simple question-answering type of a task such as the above $\sqcap x\bigl(Pr(x) \sqcup \neg Pr(x)\bigr)$, while the antecedent (as well as the consequent, of course) of a $\circ\!\!-$-implication can have an arbitrarily complex interaction interface. For instance, (1) is already a problem with a non-standard interface, and CL, unlike traditional approaches, allows us to meaningfully talk about reducing (1) itself to yet another problem, or reducing another problem to it. As a reducibility relation, $\circ\!\!-$ is thus a conservative generalization of the kind old textbook concept of Turing reducibility from the traditional, input/output sorts of problems to problems of arbitrary degrees of interactivity.

A relevant question here is whether the antecedent, as a resource, is allowed to be reused. The answer is *yes*. Turing reduction does not impose any limits on how many times the oracle can be queried. Similarly, if our strategy for (1) had a need to repeatedly ask "*is H provable in predicate calculus?*" (for various $H$s), it would have been able to do so. This essentially means that multiple copies rather than a single copy of $A$ are played in the antecedent of $A \circ\!\!- B$.

Computability logic, of course, does have a stronger reduction operation which forbids repeated usage of the antecedent. It is the already mentioned $\rightarrow$. Not surprisingly, the logical behavior of $\rightarrow$ is dramatically different from that of $\circ\!\!-$: informally speaking, the former is (plainly) resource-sensitive while the latter is not. The following two intuitionistic principles are examples of valid formulas that stop being valid if $\circ\!\!-$ is replaced by $\rightarrow$:

$$P \sqcap (P \circ\!\!- Q) \circ\!\!- Q;$$

$$\bigl(P \circ\!\!- (Q \circ\!\!- R)\bigr) \circ\!\!- \bigl((P \circ\!\!- Q) \circ\!\!- (P \circ\!\!- R)\bigr).$$

As for \$, CL understands it as a computational problem of universal strength, i.e. a problem to which any other problem of interest is algorithmically reducible. Turing completeness, NP-completeness, etc. are good intuitive examples of "being of universal strength". Subsection 2.3 explains this concept in greater detail.

The present introductory section has been written for a general audience, but the same cannot be said about the rest of the paper, which is largely meant only for those who are already sufficiently familiar with computability logic, or are willing to do some parallel reading. While still very young, due to the dynamic development ([6]-[13]) since its birth in 2003, CL has reached the point where it is no longer feasible or expedient to define (let alone explain and motivate) all over again each of the already introduced concepts relevant to a given technical paper. This article does rely, both technically and motivationally, on some prior material. Specifically, the proofs presented here should be read after or in parallel with the relevant parts of [10], which serves as a source of all special notation, terminology and concepts for the present paper (call it an appendix if you wish). While long, [10] is easy to read as it is written in a semitutorial style, without assuming any preliminary knowledge of the subject. Every unfamiliar term or notation used but not defined in the present paper can and should be looked up in [10], which has a convenient index of all terms and symbols. Familiarity with or parallel reading of [10] is a necessary and sufficient condition for understanding the technical parts of this paper.

# 2 Intuitionistic logic

## 2.1 Syntax

The language of propositional intuitionistic logic **Int** has one **logical atom** \$ and infinitely many **nonlogical atoms**, for which we use the metavariables $P$, $Q$, $R$, $X$, $Y$, $Z$, $T$, $W$, possibly with indices. Its formulas, that will be referred to as **Int-formulas**, are built from atoms and the binary operators $\circ\!-$, $\sqcup$, $\sqcap$ in the standard way. We will be using the letters $E, F, G, H, K$[3] (possibly with indices) as metavariables for formulas, and underlined letters such as $\underline{G}$ as metavariables for finite sequences of formulas. The meaning of an expression such as $\underline{G}, F, \underline{H}$ should be clear: this is the result of appending $F$ to $\underline{G}$, and then appending $\underline{H}$ to the resulting sequence.

An **Int-sequent** is a pair $\underline{G} \Rightarrow E$, where $\underline{G}$, called the **antecedent**, is any finite sequence of **Int**-formulas, and $E$, called the **succedent**, is a (one single) **Int**-formula.

Below is a Gentzen-style axiomatization of **Int**. A formula $K$ is considered provable in it iff the empty-antecedent sequent $\Rightarrow K$ is provable. We will be writing **Int** $\vdash$ for provability in **Int**.

The **axioms** of **Int** are all **Int**-sequents of the form

$$K \Rightarrow K \quad \text{or} \quad \$ \Rightarrow K,$$

and the **rules of inference**, with $i$ ranging over $\{1, 2\}$, are given by the following schemata:

[3] And sometimes some other capital letters as well, such as $A, B, C, D, I$.

**Exchange**

$$\frac{\underline{G}, E, F, \underline{H} \Rightarrow K}{\underline{G}, F, E, \underline{H} \Rightarrow K}$$

**Weakening**

$$\frac{\underline{G} \Rightarrow K}{\underline{G}, E \Rightarrow K}$$

**Contraction**

$$\frac{\underline{G}, E, E \Rightarrow K}{\underline{G}, E \Rightarrow K}$$

**Left** $\circ\!-$

$$\frac{\underline{G}, E \Rightarrow K_1 \qquad \underline{H} \Rightarrow K_2}{\underline{G}, \underline{H}, K_2 \circ\!- E \Rightarrow K_1}$$

**Right** $\circ\!-$

$$\frac{\underline{G}, E \Rightarrow K}{\underline{G} \Rightarrow E \circ\!- K}$$

**Left** $\sqcup$

$$\frac{\underline{G}, E_1 \Rightarrow K \qquad \underline{G}, E_2 \Rightarrow K}{\underline{G}, E_1 \sqcup E_2 \Rightarrow K}$$

**Right** $\sqcup$

$$\frac{\underline{G} \Rightarrow K_i}{\underline{G} \Rightarrow K_1 \sqcup K_2}$$

**Left** $\sqcap$

$$\frac{\underline{G}, E_i \Rightarrow K}{\underline{G}, E_1 \sqcap E_2 \Rightarrow K}$$

**Right** $\sqcap$

$$\frac{\underline{G} \Rightarrow K_1 \qquad \underline{G} \Rightarrow K_2}{\underline{G} \Rightarrow K_1 \sqcap K_2}$$

## 2.2 Kripke semantics

A **Kripke model** [15] — or simply a **model** — is a triple $\mathcal{K} = (\mathcal{W}, \mathcal{R}, \models)$, where:

- $\mathcal{W}$ (here) is the set $\{1, \ldots, n\}$ of the first $n \geq 1$ positive integers, called the **worlds** of $\mathcal{K}$. 1 is said to be the **root** of $\mathcal{K}$, and $n$ is said to be the **size** of $\mathcal{K}$.
- $\mathcal{R}$, called the **accessibility relation**, is a transitive, reflexive and antisymmetric relation between worlds, such that, for all $p, q_1, q_2 \in \mathcal{W}$, the following conditions are satisfied:
  - $1\mathcal{R}p$;
  - whenever $q_1\mathcal{R}p$ and $q_2\mathcal{R}p$, we have $q_1\mathcal{R}q_2$ or $q_2\mathcal{R}q_1$.

  In other words, $(\mathcal{W}, \mathcal{R})$ forms a tree with 1 at its root. When $p\mathcal{R}q$, we say that $q$ is **accessible** (in $\mathcal{K}$) from $p$.

- $\models$, called the **forcing relation** of $\mathcal{K}$, is a relation between worlds and **Int**-formulas, for all $p \in \mathcal{W}$ and all formulas $E, F$ satisfying the following conditions:
  - if $p \models E$ and $p\mathcal{R}q$, then $q \models E$;
  - $p \not\models \$$;
  - $p \models E \sqcup F$ iff $p \models E$ or $p \models F$ (or both);
  - $p \models E \sqcap F$ iff $p \models E$ and $p \models F$;
  - $p \models E \circ\!\!- F$ iff, for every world $q$ accessible from $p$, if $q \models E$, then $q \models F$.

  Where $\underline{G}$ is a sequence of formulas, we write $p \models \underline{G}$ to mean that $p \models G$ for every formula $G$ of $\underline{G}$. The relation $\models$ further extends to sequents by stipulating that $p \models \underline{G} \Rightarrow E$ iff, for every world $q$ accessible from $p$, if $q \models \underline{G}$, then $q \models E$. We read $p \models \varsigma$ (whatever $\varsigma$ is) as "$p$ **forces** $\varsigma$ (in $\mathcal{K}$)".

Let $\mathcal{K} = (\mathcal{W}, \mathcal{R}, \models)$ be a model. Where $\mathcal{F}$ is an **Int**-formula, or a sequence of **Int**-formulas, or an **Int**-sequent, we write $\mathcal{K} \Vdash\!\!= \mathcal{F}$ to mean that $1 \models \mathcal{F}$. Note that this immediately implies that $p \models \mathcal{F}$ for all $p \in \mathcal{W}$. And we say that an **Int**-formula $E$ is $\mathcal{K}$**-equivalent** to an **Int**-formula $F$ — symbolically $\mathcal{K} \Vdash\!\!= E \circ\!\!-\!\!\circ F$ — iff $\mathcal{K} \Vdash\!\!= E \circ\!\!- F$ and $\mathcal{K} \Vdash\!\!= F \circ\!\!- E$. Obviously $\mathcal{K} \Vdash\!\!= E \circ\!\!-\!\!\circ F$ means that, for each world $p$ of $\mathcal{K}$, $p \models E$ iff $p \models F$, so that "$\mathcal{K}$ does not see any difference between $E$ and $F$". Next, we will write simply $\Vdash\!\!= \mathcal{F}$ or $\Vdash\!\!= E \circ\!\!-\!\!\circ F$ to mean that $\mathcal{K} \Vdash\!\!= \mathcal{F}$ or $\mathcal{K} \Vdash\!\!= E \circ\!\!-\!\!\circ F$ for every Kripke model $\mathcal{K}$. The symbols $\not\models$ and $\not\Vdash\!\!=$, as expected, will be used for the negations of $\models$ and $\Vdash\!\!=$, respectively.

The following lemma is obvious:

**Lemma 2.1** *Assume $\mathcal{K}$ is a Kripke model, and $E_1, E_2, G_1, G_2$ are* **Int***-formulas such that:*

- *$E_2$ is the result of replacing in $E_1$ some occurrence(s) of $G_1$ by $G_2$;*
- $\mathcal{K} \Vdash\!\!= G_1 \circ\!\!-\!\!\circ G_2$.

*Then $\mathcal{K} \Vdash\!\!= E_1 \circ\!\!-\!\!\circ E_2$.*

The following well known ([15]) fact will play an important role in our proof of the completeness of **Int** with respect to the semantics of computability logic. We will be referring to it as *the soundness and completeness of* **Int** *with respect to Kripke semantics*:

**Fact 2.2** *An* **Int***-sequent or* **Int***-formula $\mathcal{S}$ is provable in* **Int** *if (***completeness***) and only if (***soundness***) $\Vdash\!\!= \mathcal{S}$.*

## 2.3 Computability-logic semantics

If the logical vocabulary of **Int** did not include $\$$, this subsection would be redundant, as the reader is assumed to be already familiar, from Sections 7 and 8 of [10], with the CL semantics for a much more expressive language. However, $\$$ was not officially present in that language, so we need to clarify how the semantics of CL extends to the $\$$-containing formalism of **Int**. As a logical constant, $\$$ was introduced in the very first paper [6] on computability logic, but since it has been promised that [10] would be the only external source on which the present paper relies, here we reproduce all necessary definitions and some discussions from [6] pertaining to how $\$$ is treated in CL.

To understand the intuitions underlying our interpretation of $\$$, let us, for a moment, imagine classical logic with its $\bot$ written as $\$$, and $\neg E$ understood as an abbreviation of $E \to \$$. How could one characterize $\$$ in philosophical terms? The standard answer here is that $\$$ is an always false formula. Yet, there would be nothing in an axiomatization of the above version of classical logic directly implying that $\$$ cannot be true. Indeed, the (essentially) only postulated or derived scheme characterizing $\$$ would be $\$ \to E$ for all formulas $E$. If, however, every nonlogical atom is interpreted as a true proposition, then $\$ \to E$ remains true even when $\$$ is true. Thus, a more general characterization of $\$$ would be that it represents a "strongest" — rather than necessarily false — proposition, i.e. a proposition that implies every other proposition expressible in the language. Still, interpretations where all atoms and hence all formulas are true would be rather pathological in the context of classical logic, and hardly of any interest. Indeed, when everything we can say is true, there is no need for a logic at all. So, the standard interpretation of $\$$ as $\bot$ ("always false formula") in classical logic is a natural choice essentially yielding no loss of generality even under the cautious philosophical vision '$\$$ = *strongest proposition*'. For, as long as there is at least one false statement expressible in the language — and in all reasonable applications there indeed would be such statements — being a strongest proposition automatically means being false, i.e. being $\bot$.

Things are quite different in computability logic though. It is more general than classical logic, and one should expect that $\$$, as a "strongest formula", would also naturally call for a more general or liberal interpretation than classical logic does. In computability logic, as we know, formulas stand for computational problems rather than just true/false statements, and $\models A \to B$ means that $B$ is (algorithmically) reducible to $A$ rather than just that $B$ is "implied" by $A$. So, $\$$ here becomes a "strongest problem", i.e. a problem to which every other problem of interest is reducible. Let us call such a problem a *universal problem*. Sure $\bot$ is one — the simplest — sort of a universal problem: everything is trivially reducible to it. But $\bot$ is a very special sort of a universal problem, and by far not the only meaningful interpretation for $\$$. What makes $\bot$ special is that it is an elementary problem, while all other atoms of **Int** generally represent problems of arbitrary degrees of interactivity — static games of arbitrary depths, that is. Insisting on interpreting $\$$ as an elementary problem (or anything equivalent to such) while all other atoms and hence all other formulas represent not-necessarily-elementary problems might create an unnatural hybrid, possibly giving rise to some logical anomalies and disharmony.

Another feature that makes $\bot$ less than general enough as the only legitimate logical representative of universal problems is that $\bot$ is what [6] (Section 18) calls *negative*, while two other equally meaningful possibilities for a universal problem are to be *positive* or *indetermined*. Informally speaking, a positive problem is one that has a solution yet not necessarily an algorithmic one; a negative problem is one whose negation is positive; and an indetermined problem is one that is neither positive nor negative.[4] Let us set aside indeterminism and focus on positiveness, here seeing it as a computability-logic counterpart of classical truth. As noted earlier, in the context of classical logic, dealing with an interpretation of the language that makes all formulas true/positive would be less than interesting. The same, however, is far from being the case for computability-logic-based applied languages. Considering languages — more precisely, interpretations — that only allow us to express positive problems does make perfect sense, and in no way does it generally make logic worthless or trivial. For example, note that the standard computability or complexity theories are exclusively concerned with positive problems. In fact, positiveness is inherent to the traditional, narrow understanding of the term "problem": in that understanding, a problem usually means a game of the form $\sqcap x\bigl(A(x) \sqcup \neg A(x)\bigr)$ or $\sqcap x \sqcup y\bigl(y = f(x)\bigr)$, where $A$ is a predicate and $f$ is a function. Such problems are automatically positive: they do have solutions, even though not always algorithmic or efficient ones. If, say, CL was used to formalize a fragment of computability theory that is focused on recursively enumerable problems (in our terms, games of the form $\sqcap x\bigl(A(x) \sqcup \neg A(x)\bigr)$ where $A$ is a recursively enumerable predicate), with the nonlogical general atoms thus ranging over such problems, then one of the interesting — more so than $\bot$ — interpretations of $\$$ would be understanding it as the acceptance problem for the universal Turing machine (UTM). Intuitively, the reason why this positive problem is universal is that, if I know what inputs UTM accepts and what inputs it does not accept, then, for any recursively enumerable predicate $A(x)$ and any particular value $m$ for $x$, I can effectively tell whether $A(m)$ is true or false. Going a little bit beyond the present scope of computability logic and imagining that $\rightarrow$ means polynomial-time reduction (in a yet-to-be defined precise sense), if one is exclusively concerned with problems from the class NP, then a natural candidate for a universal problem $\$$ would be the 3-satisfiability or any other NP-complete problem — or, for safety, let us say some polynomially-bounded version of the branching or parallel recurrence of such a problem. Then $\models E \rightarrow \$$ would mean that $E$ (too) is NP-complete, rather than that $E$ is a "false" or "unsolvable even by God" problem, i.e. that $E$ is a problem which the traditional complexity theory would not really consider meaningful and legitimate.

To summarize, the approach of computability logic, for adopting which it has good reasons, is to interpret $\$$ as a universal problem, i.e. a problem of universal strength, i.e. a problem to which all other problems are reducible (in the sense of Section 6 of [10]). This approach insists on making no additional assumptions regarding $\$$ as a universal problem, such as its being elementary, or negative, or non-positive, etc.

Of course, the above is only a philosophical characterization, and what, exactly, a universal problem means is yet to be formally defined. First of all, the range of "all" in

[4] Yes, indetermined problems do exists, although, according to Theorem 18.7 of [6], not among perifinite-depth ones.

"all other problems" should be specified. Otherwise, $\bot$ would always be among "other problems", which would make $\bot$ (modulo equivalence) the only possible candidate for a universal problem. That is, the concept of universal problem is relative — relative to a set of problems, called the *problem universe*, that are considered the only "legitimate", or "of interest" problems within a given treatment. We reasonably require any problem universe to be countable, and come in the form of a fixed list of its elements rather than just as the set of those elements. For simplicity we also require that such a list be infinite. There is no loss of generality here, because any finite list can be painlessly extended to an "essentially the same" yet formally infinite list by replicating its last element infinitely many times. Here comes the formal definition:

**Definition 2.3** A **problem universe** is an infinite sequence $U = \langle A_1, A_2, A_3, \ldots \rangle$ of static games. A **universal problem** of such a $U$ is any static game $B$ satisfying the following conditions:

1. For each $i \geq 1$, $A_i$ is reducible (in the sense of Section 6 of [10]) to $B$ and,
2. furthermore, there is an effective procedure that, for each $i \geq 1$, constructs a reduction (in the sense of Section 6 of [10]) of $A_i$ to $B$.

As noted, the intuition behind the concept of universal problem $B$ is that this is a problem of universal strength: having it as a computational resource means being able to solve any of the problems of the universe, as every problem of the universe is algorithmically reducible to $B$. However, just existence of those algorithmic reductions would have little value if there was no way to find them. The 'furthermore' clause of the above definition guarantees that reduction strategies not only exist, but can be effectively found. Of course, when the problem universe is "essentially finite", that clause is redundant.

There are many potential problem universes, and the next question to clarify is which one should be considered *the* problem universe $U$ in any given case, i.e. in any particular application of **Int** or CL in general. In precise terms, a particular application is nothing but the particular, "actual" interpretation $^*$ chosen for the language — the function that sends every formula $F$ to the computational problem $F^*$ represented by $F$. An answer comes very naturally. The nonlogical atoms of **Int** represent the computational problems of interest in a given application, so the interpretation $P^*$ of each such atom $P$ should be in $U$. And, since there is no shortage of such atoms, we may assume that vice versa is also always true: every element of $U$ is $P^*$ for some nonlogical atom $P$. The order in which the elements of $U$ are arranged is just a typically irrelevant technicality, and among the natural arrangements is the one determined by the lexicographic order of the $P$s. So, $U = \langle P_1^*, P_2^*, P_3^*, \ldots \rangle$, where $P_1, P_2, P_3, \ldots$ is the lexicographic list of all nonlogical atoms, presents itself as a natural choice for the problem universe for any given application — interpretation $^*$, that is — of the language of **Int**. An alternative approach would be to stipulate that $U = \langle F_1^*, F_2^*, F_3^*, \ldots \rangle$, where $F_1, F_2, F_3, \ldots$ is the lexicographic list of all **Int**-formulas rather than just nonlogical atoms. But this technical alternative is not worth considering. It follows from the later-proven Lemma 4.1 that every universal problem of $\langle P_1^*, P_2^*, P_3^*, \ldots \rangle$ is automatically also a universal problem of $\langle F_1^*, F_2^*, F_3^*, \ldots\rangle$ and vice versa. This means that the two approaches would be the same in every reasonable sense.

The agreed-on problem universe $U = \langle P_1^*, P_2^*, P_3^*, \ldots\rangle$ would, however, have many universal problems, and the next question to ask is which one should count as *the* universal problem $\$^*$ of $U$. Our answer is:

$$\$ \textit{ can be interpreted as any — up to equivalence — universal problem of } U. \qquad (2)$$

Let us clarify this. Why "*any*" is obvious: depending on needs or taste, different universal problems of $U$ can be chosen to be treated as "the" universal problem. In most typical cases the universal problem would be an element of the problem universe or equivalent to such, even though Definition 2.3 does not insist on this. Going back to our earlier examples, as the universal problem of the corresponding universe, one could have chosen the halting problem instead of the acceptance problem for the UTM; or, alternatively, the strictly stronger problem of arithmetical truth could have been chosen, i.e. the problem $\sqcap x\big(True(x) \sqcup \neg True(x)\big)$, where $True(x)$ is the predicate "$x$ is the Gödel number of a true arithmetical sentence". And, in the complexity theory example, the Hamiltonian path problem could have been chosen instead of 3-satisfiability; or, alternatively, there could have been reasons to choose some PSPACE-complete problem as the universal problem instead of an NP-complete problem.

As for "*up to equivalence*", it indicates the somewhat relaxed meaning of "any" in (2) and our subsequent discussion of it. What (2) says, in precise terms, is that, for any universal problem $B$ of $U$, \$ is always allowed to be interpreted as a certain problem which is *equivalent* (in the sense of Section 6 of [10]) to $B$, but not necessarily *equal* to $B$ in the literal sense. Specifically, CL limits the officially admissible interpretations of \$ to what is calls *standard universal problems* of $U$. As will be argued shortly, such a limitation is harmless, yielding no loss of generality or loss of potential applicability of **Int**. Briefly, this is so because, as it turns out, every universal problem of $U$ is equivalent to some standard universal problem of $U$, with standard universal problems thus representing the equivalence classes of all possible universal problems (serving as "standard representatives" of those equivalence classes). But let us see the formal definitions first.

**Definition 2.4** Let $U = \langle A_2, A_3, A_4, \ldots\rangle$ be a problem universe.

1. For a problem $A_1$, the **$A_1$-based standard universal problem of $U$** is the problem $B$ defined as follows:
   - $\mathbf{Lr}^B_e = \{\langle\rangle\} \cup \{\langle\bot i, \Gamma\rangle \mid i \in \{1,2,3,4,\ldots\},\ \Gamma \in \mathbf{Lr}^{A_i}_e\}$.
   - $\mathbf{Wn}^B_e\langle\rangle = \top$; $\quad \mathbf{Wn}^B_e\langle\bot i, \Gamma\rangle = \mathbf{Wn}^{A_i}_e\langle\Gamma\rangle$.
2. A **standard universal problem $B$ of $U$** is the $A_1$-based standard universal problem of $U$ for some problem $A_1$; the latter is said to be the **base** of $B$.

Notice that the $A_1$-based standard universal problem of $U = \langle A_2, A_3, A_4, \ldots\rangle$ is nothing but the infinite choice conjunction

$$B \;=\; A_1 \sqcap A_2 \sqcap A_3 \sqcap A_4 \sqcap \ldots .$$

It is also easy to see that such a $B$ is indeed a universal problem of $U$. A strategy that wins $B \rightarrow A_i$ for any $A_i \in U$ is to select the conjunct $A_i$ of the infinite conjunction $B$ thus bringing the game down to $A_i \rightarrow A_i$, and then continue the rest of the play using the standard copy-cat methods (namely, the strategy $\mathcal{CCS}$ defined in Section 4).

**Definition 2.5** An **interpretation** is a function that sends every atom $Q$ of the language of **Int** to a static game $Q^*$ such that, where $P_1, P_2, \ldots$ is the lexicographic list of all nonlogical atoms, $\$^*$ is a standard universal problem of the universe $\langle P_1^*, P_2^*, \ldots \rangle$. This function extends to compound **Int**-formulas by stipulating that:

$$\begin{aligned} (E \circ\!\!- F)^* &= E^* \circ\!\!- F^*; \\ (E \sqcup F)^* &= E^* \sqcup F^*; \\ (E \sqcap F)^* &= E^* \sqcap F^*. \end{aligned}$$

Next, as in [10], an **Int**-formula $E$ is said to be:

- **valid** iff, for every interpretation $^*$, $\models E^*$;
- **uniformly valid** iff there is an EPM or HPM $\mathcal{M}$, called a **uniform solution** for $E$, such that, for every interpretation $^*$, $\mathcal{M} \models E^*$.

As in [10], $\Vdash E$ is a symbolic way to say that $E$ is valid, $\Vvdash E$ means that $E$ is uniformly valid, and $\mathcal{M} \Vvdash E$ means that $\mathcal{M}$ is a uniform solution for $E$.

It is clear that the promised completeness of **Int** with respect to the semantics of CL (i.e. with respect to validity and/or uniform validity) would automatically remain true under various — whatever — liberalizations of our definitions of the concepts of universal problem and interpretation. But, in fact, this is so not only for completeness. One can also show that **Int** would remain sound with respect to validity even if Definition 2.3 did not have the 'furthermore' clause 2, and Definition 2.5 allowed \$ to be interpreted as any (rather than necessarily standard) universal problem of the universe $\langle P_1^*, P_2^*, \ldots \rangle$ or $\langle F_1^*, F_2^*, \ldots \rangle$. The reason for our choosing the present, seemingly unnecessarily capricious, definitions is to ensure the soundness of **Int** not only with respect to validity, but with respect to uniform validity as well. Our choice does achieve this highly desirable effect while, at the same time, sacrifices no generality or applicability. The point is that when it comes to computability, equivalent problems are "the same" in all relevant aspects and hence can be thought of as being identical. Specifically, when $A$ and $A'$ are equivalent, then $A$ is computable if and only if $A'$ is so, and any compound problem where $A$ is a component (e.g. $A \circ\!\!- C$, $C \sqcup (A \sqcap D)$, etc.) is computable if and only if the result of replacing in it $A$ with $A'$ is so. But, as proven in [6] (Proposition 23.4), we have:

> *Any universal problem $B$ of any given problem universe is equivalent to the $B$-based standard universal problem $B'$ of the same universe.*

So, whether \$ is translated as the above $B$ or $B'$ has no effect on the computability status of the problems represented by formulas. Furthermore, one could show that there is an effective procedure for converting any solution of the problem represented by an arbitrary formula $F$ when \$ is read as $B$ into a solution of the problem represented by the same formula when \$ is read as $B'$, and vice versa. This makes the difference between the two readings of \$ negligible, and allows us to safely restrict our focus only to the official interpretation of \$ as $B'$, even if the universal problem of our original and direct interest was $B$.

### 2.4 Main theorem

By a $\Sigma_1$**-predicate** we mean a predicate that can be written as $\exists zA$ for some decidable finitary predicate (elementary game) $A$. And a **Boolean combination** of $\Sigma_1$-predicates is a $\neg, \wedge, \vee$-combination of such games. We denote the set of all Boolean combinations of $\Sigma_1$-predicates by

$$\Sigma_1^B.$$

Next, we let

$$\sqcup\Sigma_1^B$$

stand for the set of all games of the form $\sqcup xA$, where $A \in \Sigma_1^B$.

Next, let $^*$ be an interpretation and $C$ a set of static games, such as $\Sigma_1^B$ or $\sqcup\Sigma_1^B$. We say that $^*$ is **of complexity** $C$ iff the following two conditions are satisfied:

- for every nonlogical atom $P$, $P^*$ is in $C$;
- the base of $\$^*$ is also in $C$.

**Theorem 2.6** *For any* **Int**-*formula* $K$, $\mathbf{Int} \vdash K$ *iff* $\Vdash K$ *iff* $\Vdash\!\!\vdash K$. *Furthermore:*

*(a) There is an effective procedure that takes an arbitrary* **Int**-*proof of an arbitrary formula* $K$ *and constructs a uniform solution for* $K$.

*(b) If an* **Int**-*formula* $K$ *is not provable in* **Int**, *then* $K^*$ *is not computable for some interpretation* $^*$ *of complexity* $\sqcup\Sigma_1^B$.

The soundness part of this theorem in the strong sense of clause (a) will be proven in Section 4. And a proof of the completeness part in the strong sense of clause (b) will be given in Sections 6 through 10.

## 3 Affine logic

The present paper is only concerned with intuitionistic logic. Yet we need to take a close and fresh look at the more expressive (even though not deductively stronger) affine logic, because it is going to be a heavily exploited tool in our soundness and completeness proof for **Int**.

## 3.1 Syntax and computability-logic semantics

The language of what we in this paper refer to as **affine logic** (**Al**) has the same set of atoms as the language of **Int**, that is, infinitely many nonlogical atoms plus one logical atom \$. Its connectives are the unary $\neg$,⫯,⫰ and the variable-arity — $n$-ary for each $n \geq 2$ — operators $\vee, \wedge, \sqcup, \sqcap$. In the formulas of this language, to which we refer as **Al-formulas**, the operators $\rightarrow$ and $\circ\!\!-$ are understood as abbreviations, defined by

- $E \rightarrow F = \neg E \vee F$;
- $E \circ\!\!- F = ⫰E \rightarrow F$.

Also, when applied to nonatomic formulas, $\neg$ is understood as an abbreviation defined by:

- $\neg\neg E = E$;
- $\neg ⫯E = ⫰\neg E$;
- $\neg ⫰E = ⫯\neg E$;
- $\neg(E_1 \vee \ldots \vee E_n) = \neg E_1 \wedge \ldots \wedge \neg E_n$;
- $\neg(E_1 \wedge \ldots \wedge E_n) = \neg E_1 \vee \ldots \vee \neg E_n$;
- $\neg(E_1 \sqcup \ldots \sqcup E_n) = \neg E_1 \sqcap \ldots \sqcap \neg E_n$;
- $\neg(E_1 \sqcap \ldots \sqcap E_n) = \neg E_1 \sqcup \ldots \sqcup \neg E_n$.

We continue to use $X$, $Y$, $Z$, $T$, $P$, $Q$, $R$, $W$ as metavariables for nonlogical atoms, $E$, $F$, $G$, $H$, $K$ for formulas, and $\underline{G}$, $\underline{H}$ for finite sequences of formulas. We will also be underlining complex expressions, as in $\underline{⫯G}$. The latter should be understood as a sequence $⫯G_1, \ldots, ⫯G_n$ of ⫯-prefixed formulas.

The definition of an **interpretation** for the language of **Al** is literally the same as that for the language of **Int** given in Subsection 2.3, and any interpretation $^*$ extends to compound **Al**-formulas in the same way as in the case of **Int**-formulas, commuting with all connectives seen as the same-name game operations. That is, $(\neg E)^* = \neg E^*$, $(⫰E)^* = ⫰E^*$, $(E_1 \wedge \ldots \wedge E_n)^* = E_1^* \wedge \ldots \wedge E_n^*$, etc. Next, the definitions of **validity** ($\Vdash$), **uniform validity** ($\Vvdash$) and **uniform solution** ($\mathcal{M}\Vvdash$) for **Int**-formulas verbatim extend to all **Al**-formulas.

Based on what we know from [10], when it comes to interpretations, it makes virtually no difference whether $\rightarrow$, $\circ\!\!-$ and $\neg$ (the latter when applied to compound formulas) are understood as primitives or just as abbreviations. This is so because, either by definition or as easily derived facts, for any games $A, B, \ldots$ we have $A \rightarrow B = \neg A \vee B$, $A \circ\!\!- B = ⫰A \rightarrow B$, $\neg\neg A = A$, $\neg(A \sqcup B) = \neg A \sqcap \neg B$, etc. Hence, we can and will safely think of the (officially $\circ\!\!-$-containing) language of **Int** as a sublanguage of the language of **Al**, with all **Int**-formulas thus automatically being also **Al**-formulas.

An **Al-sequent** is any finite sequence of **Al**-formulas. Thus, unlike the case with **Int** where sequents were two-sided, in (the axiomatization of) **Al** we choose one-sided sequents. Below is a Gentzen-style axiomatization of **Al**. As always, a formula $E$ is considered provable iff $E$, as a one-formula sequent, is so. The symbol **Al** $\vdash$ will be used for **Al**-provability, whether it be for sequents or formulas.

The **axioms** of **Al** are all **Al**-sequents of the form

$$\neg E, E,$$

and the **rules of inference** are given by the following schemata, where $n \geq 2$ and $1 \leq i \leq n$:

**Exchange**

$$\frac{\underline{G}, E, F, \underline{H}}{\underline{G}, F, E, \underline{H}}$$

**Weakening**

$$\frac{\underline{G}}{\underline{G}, E}$$

**⫯-Contraction**

$$\frac{\underline{G}, ⫯E, ⫯E}{\underline{G}, ⫯E}$$

**$\sqcup$-Introduction**

$$\frac{\underline{G}, E_i}{\underline{G}, E_1 \sqcup \ldots \sqcup E_n}$$

**$\sqcap$-Introduction**

$$\frac{\underline{G}, E_1 \quad \bullet \; \bullet \; \bullet \quad \underline{G}, E_n}{\underline{G}, E_1 \sqcap \ldots \sqcap E_n}$$

**$\vee$-Introduction**

$$\frac{\underline{G}, E_1, \ldots, E_n}{\underline{G}, E_1 \vee \ldots \vee E_n}$$

**$\wedge$-Introduction**

$$\frac{\underline{G_1}, E_1 \quad \bullet \; \bullet \; \bullet \quad \underline{G_n}, E_n}{\underline{G_1}, \ldots, \underline{G_n}, E_1 \wedge \ldots \wedge E_n}$$

**⫯-Introduction**

$$\frac{\underline{G}, E}{\underline{G}, ⫯E}$$

**⫰-Introduction**

$$\frac{\underline{⫯G}, E}{\underline{⫯G}, ⫰E}$$

The above formulation of **Al** can be called **strict**. We will, however, usually prefer to work with the equivalent **relaxed** version of **Al**, in which sequents are seen as multisets rather than sequences of formulas. This eliminates the annoying need for explicitly referring to one or several obvious steps of applying the exchange rule.

Seeing \$ as an ordinary — nonlogical — atom, syntactically there is no difference between **Al** and the propositional, unit-free fragment of Girard's [4] affine logic. Therefore, in view of known syntactic results for the latter, **Al** is closed under the following two rules:

**Cut**

$$\frac{\underline{G}, E \qquad \neg E, \underline{H}}{\underline{G}, \underline{H}}$$

**Modus ponens**

$$\frac{E \qquad E \to F}{F}$$

**Fact 3.1 (The soundness of Al:)** *If* $\mathbf{Al} \vdash K$*, then* $\Vdash K$ *and hence* $\Vdash K$ *(any* **Al***-formula* $K$*).*

*Furthermore, there is an effective procedure that takes an arbitrary* **Al***-proof of an arbitrary* **Al***-formula* $K$ *and constructs a uniform solution for* $K$*.*

**Proof.** Remember logic **AL** from Section 11 of [10]. If we think of \$ as an ordinary (nonlogical) atom, the language of **Al** is just a fragment of the language of **AL**. Furthermore, as the two systems are analytical, **AL** is a conservative extension of **Al**: the fact that \$ counts as a logical atom in **Al** while it is seen as a nonlogical atom in **AL**, of course, is irrelevant, as there are no special rules for \$ in **Al**. The special treatment of \$ in **Al** only shows itself in semantics. Specifically, ignoring the here irrelevant difference that the language of **Al** is propositional while that of **AL** is not, the set of functions that we now call interpretations is a proper subset of the set of functions called interpretations in [10]. With a moment's thought this can be seen to mean that the concepts of validity, uniform validity or uniform solution in the sense of [10] are stronger than the same concepts in our present sense. For this reason, Fact 3.1 immediately follows from the theorem of Section 11 of [10], which says about **AL** the same as what Fact 3.1 says about **Al**. □

In concordance with the notational conventions of [10], in what follows {EPMs} stands for the set of all EPMs.

**Fact 3.2** *For any* **Al***-formulas* $E$*,*$F$ *and any interpretation* $^*$*, we have:*
*a) if* $\models E^*$ *and* $\models E^* \to F^*$*, then* $\models F^*$*;*
*b) if* $\Vdash E$ *and* $\Vdash E \to F$*, then* $\Vdash F$*.*
*Furthermore, there is an effective function* $f : \{EPMs\} \times \{EPMs\} \longrightarrow \{EPMs\}$ *such that, for any* **Al***-formulas* $E, F$*, interpretation* $^*$ *and EPMs* $\mathcal{E}, \mathcal{C}$*, we have:*
*a) if* $\mathcal{E} \models E^*$ *and* $\mathcal{C} \models E^* \to F^*$*, then* $f(\mathcal{E}, \mathcal{C}) \models F^*$*;*
*b) if* $\mathcal{E} \Vdash E$ *and* $\mathcal{C} \Vdash E \to F$*, then* $f(\mathcal{E}, \mathcal{C}) \Vdash F$*.*

**Proof.** According to the theorem of Section 10 of [10], computability is closed under modus ponens, in the constructive sense that there is an effective function $f$ that converts any solutions of arbitrary games $A$ and $A \to B$ into a solution of $B$, as long as $A$ and $B$ are static. Of course, such a function $f$ (also) works exactly as clause (a) of the present fact states, for $E^*$ and $F^*$ are always static games. And, with a moment's thought, clause (b) can be seen to be an immediate corollary of clause (a). $\Box$

**Fact 3.3** *For any* **Al**-*formula $E$ and any interpretation $^*$, we have:*
*a) if $\models E^*$, then $\models ⫰E^*$;*
*b) if $\Vdash\!\!\vdash E$, then $\Vdash\!\!\vdash ⫰E$.*
*Furthermore, there is an effective function $h : \{EPMs\} \longrightarrow \{EPMs\}$ such that, for any* **Al**-*formula $E$, interpretation $^*$ and EPM $\mathcal{E}$, we have:*
*a) if $\mathcal{E} \models E^*$, then $h(\mathcal{E}) \models ⫰E^*$;*
*b) if $\mathcal{E}\Vdash\!\!\vdash E$, then $h(\mathcal{E})\Vdash\!\!\vdash ⫰E^*$.*

**Proof.** According to one of the lemmas of Section 13 of [10], computability is closed under the rule "*from $A$ to $⫰A$*" in the constructive sense that there is an effective function $h$ that converts any solution of an arbitrary static game $A$ into a solution of $⫰A$. As in the proof of Fact 3.2, such a function $h$ can be seen to (also) work exactly as the above clause (a) states. And, again, clause (b) is an immediate corollary of clause (a). $\Box$

## 3.2 The replacement lemma

An occurrence $O$ of a subformula in a given **Al**-formula is said to be **positive** iff it is not in the scope of $\neg$. An occurrence of a nonatomic formula would thus always be positive, because officially $\neg$ cannot be applied to such a formula.

**Lemma 3.4** *Let $G_1$, $G_2$, $H_1$, $H_2$ be* **Al**-*formulas such that $H_2$ is the result of replacing in $H_1$ a certain positive occurrence of $G_1$ by $G_2$. Then* **Al** $\vdash ⫯(G_1 \wedge \neg G_2), \neg H_1, H_2$.

**Proof.** Assume the conditions of the lemma are satisfied. Let $O$ be the occurrence of $G_1$ in $H_1$ under question. Our proof proceeds by induction on the complexity of $H_1$ — more precisely, the complexity of $H_1$ minus the complexity of $G_1$, for the complexity of the $G_1$ part of $H_1$ is irrelevant. We will be working with the relaxed version of **Al**.

*Case 1*: $H_1 = G_1$. Then $H_2 = G_2$, and thus what we need to show is that **Al** proves $⫯(G_1 \wedge \neg G_2), \neg G_1, G_2$. But indeed, this sequent follows from $G_1 \wedge \neg G_2, \neg G_1, G_2$ by $⫯$-introduction, and the latter, in turn, follows from the axioms $G_1, \neg G_1$ and $\neg G_2, G_2$ by $\wedge$-introduction.

*Case 2*: $H_1 = ⫰E_1$, and the occurrence $O$ of $G_1$ in $H_1$ is the occurrence $O'$ of $G_1$ in $E_1$. Let $E_2$ be the result of replacing in $E_1$ the occurrence $O'$ by $G_2$. We thus have $H_2 = ⫰E_2$. By the induction hypothesis, **Al** $\vdash ⫯(G_1\wedge\neg G_2), \neg E_1, E_2$. From here, by first applying $\vee$-introduction and then $⫰$-introduction, we get **Al** $\vdash ⫯(G_1\wedge\neg G_2), ⫰(\neg E_1\vee E_2)$, i.e.

$$\mathbf{Al} \vdash ⫯(G_1 \wedge \neg G_2),\ \neg ⫯(E_1 \wedge \neg E_2). \tag{3}$$

Next, we have:

1. **Al** $\vdash E_1 \wedge \neg E_2,\ \neg E_1,\ E_2$ (from axioms $E_1, \neg E_1$ and $\neg E_2, E_2$ by $\wedge$-introduction)

2. $\mathbf{Al} \vdash ⫯(E_1 \wedge \neg E_2),\ ⫯\neg E_1,\ E_2$ (from 1 by $⫯$-introduction applied twice)

3. $\mathbf{Al} \vdash ⫯(E_1 \wedge \neg E_2),\ ⫯\neg E_1,\ ⫰E_2$ (from 2 by $⫰$-introduction)

4. $\mathbf{Al} \vdash ⫯(G_1 \wedge \neg G_2),\ ⫯\neg E_1,\ ⫰E_2$ (from (3) and 3 by cut)

Now, the desired $\mathbf{Al} \vdash ⫯(G_1 \wedge \neg G_2), \neg H_1, H_2$ is nothing but an abbreviation of 4.

*Case 3*: $H_1 = ⫯E_1$, and the occurrence $O$ of $G_1$ in $H_1$ is the occurrence $O'$ of $G_1$ in $E_1$. This case is symmetric to the previous one. Let $E_2$ be the result of replacing in $E_1$ the occurrence $O'$ by $G_2$. We thus have $H_2 = ⫯E_2$. For the same reasons as in Case 2, the provability (3) holds. Next, we have:

1. $\mathbf{Al} \vdash E_1 \wedge \neg E_2,\ \neg E_1,\ E_2$ (from axioms $E_1, \neg E_1$ and $\neg E_2, E_2$ by $\wedge$-introduction)

2. $\mathbf{Al} \vdash ⫯(E_1 \wedge \neg E_2),\ \neg E_1,\ ⫯E_2$ (from 1 by $⫯$-introduction applied twice)

3. $\mathbf{Al} \vdash ⫯(E_1 \wedge \neg E_2),\ ⫰\neg E_1,\ ⫯E_2$ (from 2 by $⫰$-introduction)

4. $\mathbf{Al} \vdash ⫯(G_1 \wedge \neg G_2),\ ⫰\neg E_1,\ ⫯E_2$ (from (3) and 3 by cut)

Now, the desired $\mathbf{Al} \vdash ⫯(G_1 \wedge \neg G_2), \neg H_1, H_2$ is nothing but an abbreviation of 4.

*Case 4*: $H_1 = E_1^1 \wedge \ldots \wedge E_1^n$, and the occurrence $O$ of $G_1$ in $H_1$ is the occurrence $O'$ of $G_1$ in $E_1^i$ ($1 \leq i \leq n$). Let $E_2^i$ be the result of replacing in $E_1^i$ the occurrence $O'$ by $G_2$; for any other ($\neq i$) number $j$ with $1 \leq j \leq n$, let $E_2^j = E_1^j$. Thus, $H_2 = E_2^1 \wedge \ldots \wedge E_2^n$. By the induction hypothesis, $\mathbf{Al} \vdash ⫯(G_1 \wedge \neg G_2), \neg E_1^i, E_2^i$. From here, by $\vee$-introduction and $⫰$-introduction, we get

$$\mathbf{Al} \vdash ⫯(G_1 \wedge \neg G_2), ⫰(\neg E_1^i \vee E_2^i),$$

i.e.

$$\mathbf{Al} \vdash ⫯(G_1 \wedge \neg G_2),\ \neg ⫯(E_1^i \wedge \neg E_2^i). \tag{4}$$

Next, we have:

1. $\mathbf{Al} \vdash \neg E_2^i,\ E_2^i$ (axiom)

2. $\mathbf{Al} \vdash \neg E_1^j,\ E_2^j$ for each $j \in \{1, \ldots, i-1, i+1, \ldots, n\}$ (axiom, because $E_1^j = E_2^j$)

3. $\mathbf{Al} \vdash \neg E_1^1,\ \ldots,\ \neg E_1^{i-1},\ \neg E_2^i,\ \neg E_1^{i+1},\ \ldots,\ \neg E_1^n,\ E_2^1 \wedge \ldots \wedge E_2^n$
(from 1 and 2 by $\wedge$-introduction)

4. $\mathbf{Al} \vdash E_1^i,\ \neg E_1^i$ (axiom)

5. $\mathbf{Al} \vdash E_1^i \wedge \neg E_2^i,\ \neg E_1^1,\ \ldots,\ \neg E_1^n,\ E_2^1 \wedge \ldots \wedge E_2^n$ (from 3 and 4 by $\wedge$-introduction)

6. $\mathbf{Al} \vdash ⫯(E_1^i \wedge \neg E_2^i),\ \neg E_1^1,\ \ldots,\ \neg E_1^n,\ E_2^1 \wedge \ldots \wedge E_2^n$ (from 5 by $⫯$-introduction)

7. $\mathbf{Al} \vdash ⫯(E_1^i \wedge \neg E_2^i),\ \neg E_1^1 \vee \ldots \vee \neg E_1^n,\ E_2^1 \wedge \ldots \wedge E_2^n$ (from 6 by $\vee$-introduction)

Now, 7 can be rewritten as $\mathbf{Al} \vdash ⫯(E_1^i \wedge \neg E_2^i), \neg H_1, H_2$. This, together with (4), by cut, implies the desired $\mathbf{Al} \vdash ⫯(G_1 \wedge \neg G_2), \neg H_1, H_2$.

*Case 5*: $H_1 = E_1^1 \vee \ldots \vee E_1^n$. This case symmetric to the previous one, and we leave details to the reader.

*Case 6*: $H_1 = E_1^1 \sqcap \ldots \sqcap E_1^n$, and the occurrence $O$ of $G_1$ in $H_1$ is the occurrence $O'$ of $G_1$ in $E_1^i$ ($1 \leq i \leq n$). Let $E_2^i$ be the result of replacing in $E_1^i$ the occurrence $O'$ by $G_2$; for any other ($\neq i$) number $j$ with $1 \leq j \leq n$, let $E_2^j = E_1^j$. Thus, $H_2 = E_2^1 \sqcap \ldots \sqcap E_2^n$. Reasoning as we did in Case 4, we again find that (4) holds. Next, we have:

1. $\mathbf{Al} \vdash E_1^i \wedge \neg E_2^i,\ \neg E_1^i,\ E_2^i$ (from axioms $E_1^i, \neg E_1^i$ and $\neg E_2^i, E_2^i$ by $\wedge$-introduction)
2. $\mathbf{Al} \vdash E_1^i \wedge \neg E_2^i,\ \neg E_1^1 \sqcup \ldots \sqcup \neg E_1^n,\ E_2^i$ (from 1 by $\sqcup$-introduction)
3. $\mathbf{Al} \vdash \neg E_1^j,\ E_2^j$ for each $j \in \{1, \ldots, i-1, i+1, \ldots, n\}$ (axiom, because $E_1^j = E_2^j$)
4. $\mathbf{Al} \vdash E_1^i \wedge \neg E_2^i,\ \neg E_1^j,\ E_2^j$ for each $j \in \{1, \ldots, i-1, i+1, \ldots, n\}$ (from 3 by weakening)
5. $\mathbf{Al} \vdash E_1^i \wedge \neg E_2^i,\ \neg E_1^1 \sqcup \ldots \sqcup \neg E_1^n,\ E_2^j$ for each $j \in \{1, \ldots, i-1, i+1, \ldots, n\}$ (from 4 by $\sqcup$-introduction)
6. $\mathbf{Al} \vdash E_1^i \wedge \neg E_2^i,\ \neg E_1^1 \sqcup \ldots \sqcup \neg E_1^n,\ E_2^1 \sqcap \ldots \sqcap E_2^n$ (from 2 and 5 by $\sqcap$-introduction)
7. $\mathbf{Al} \vdash ⫯(E_1^i \wedge \neg E_2^i),\ \neg E_1^1 \sqcup \ldots \sqcup \neg E_1^n,\ E_2^1 \sqcap \ldots \sqcap E_2^n$ (from 6 by $⫯$-introduction)

Now, 7 can be rewritten as $\mathbf{Al} \vdash ⫯(E_1^i \wedge \neg E_2^i), \neg H_1, H_2$. From here and (4), by cut, we infer the desired $\mathbf{Al} \vdash ⫯(G_1 \wedge \neg G_2), \neg H_1, H_2$.

*Case 7*: $H_1 = E_1^1 \sqcup \ldots \sqcup E_1^n$. Symmetric to the previous case, with details left to the reader. □

**Lemma 3.5** *Let $G_1$, $G_2$, $H_1$, $H_2$ be* **Al***-formulas, $\mathcal{E}$, $\mathcal{C}$ be EPMs, and $^*$ be an interpretation such that the following three conditions are satisfied:*

1. *$H_2$ is the result of replacing in $H_1$ a certain positive occurrence of $G_1$ by $G_2$;*
2. *$\mathcal{E} \models H_1^*$ (resp. $\mathcal{E} \Vdash\!\vdash H_1$);*
3. *$\mathcal{C} \models G_1^* \rightarrow G_2^*$ (resp. $\mathcal{C} \Vdash\!\vdash G_1 \rightarrow G_2$).*

*Then there is an EPM $\mathcal{D}$ with $\mathcal{D} \models H_2^*$ (resp. $\mathcal{D} \Vdash\!\vdash H_2$).*

*Furthermore, there is a ($^*$-independent) effective procedure that constructs such an EPM $\mathcal{D}$ from arbitrary $G_1$, $G_2$, $H_1$, $H_2$, $\mathcal{E}$, $\mathcal{C}$.*

**Proof.** Assume all of the conditions of the lemma are satisfied. Let $f$ and $h$ be the functions whose existence is stated in Facts 3.2 and 3.3, respectively. By Fact 3.3, the assumption $\mathcal{C} \models G_1^* \rightarrow G_2^*$ (resp. $\mathcal{C} \Vdash\!\vdash G_1 \rightarrow G_2$) implies $h(\mathcal{C}) \models ⫰(G_1^* \rightarrow G_2^*)$ (resp. $h(\mathcal{C}) \Vdash\!\vdash ⫰(G_1 \rightarrow G_2)$). Next, by Lemma 3.4, there is an **Al**-proof of $⫯(G_1 \wedge \neg G_2), \neg H_1, H_2$ and hence — applying $\vee$-introduction twice and abbreviating the result — an **Al**-proof of $⫰(G_1 \rightarrow G_2) \rightarrow (H_1 \rightarrow H_2)$. Of course, such a proof can be found effectively.[5] Therefore, in view of Fact 3.1, we can effectively construct an EPM $\mathcal{B}$ with

[5]After all, **Al** is recursively enumerable, let alone that it is also known to be decidable. Alternatively, a procedure constructing an **Al**-proof of $⫰(G_1 \rightarrow G_2) \rightarrow (H_1 \rightarrow H_2)$ can be directly extracted from our proof of Lemma 3.4.

$\mathcal{B} \Vdash \circ\!\!\downarrow (G_1 \to G_2) \to (H_1 \to H_2)$ and hence also with $\mathcal{B} \models \circ\!\!\downarrow (G_1^* \to G_2^*) \to (H_1^* \to H_2^*)$. Then, by Fact 3.2, $f(h(\mathcal{C}), \mathcal{B}) \models H_1^* \to H_2^*$ (resp. $f(h(\mathcal{C}), \mathcal{B}) \Vdash H_1 \to H_2$). But, by condition 2 of the present lemma, $\mathcal{E} \models H_1^*$ (resp. $\mathcal{E} \Vdash H_1$). Hence, again by Fact 3.2, $f(\mathcal{E}, f(h(\mathcal{C}), \mathcal{B})) \models H_2^*$ (resp. $f(\mathcal{E}, f(h(\mathcal{C}), \mathcal{B})) \Vdash H_2$). Letting the sought $\mathcal{D}$ be $f(\mathcal{E}, f(h(\mathcal{C}), \mathcal{B}))$, we have just demonstrated how to construct $\mathcal{D}$ from $G_1$, $G_2$, $H_1$, $H_2$, $\mathcal{E}$, $\mathcal{C}$. □

# 4 The soundness of intuitionistic logic

In this and some subsequent sections we will be constructing EPMs $\mathcal{E}$ with the claim that $\mathcal{E}$ wins a certain game or a certain class of games. In our description of such an $\mathcal{E}$ and further analysis of its behavior, we will be relying (usually implicitly) on the **clean environment assumption**. According to it, the adversary of $\mathcal{E}$ never makes moves that are illegal for the game(s) under question. As pointed out in [10], such an assumption is perfectly safe and legitimate for, if the adversary makes an illegal move, then $\mathcal{E}$ will be the winner no matter what happens afterwards, and making $\mathcal{E}$ the winner is always the very purpose of our construction.

In some of our proofs we will employ a uniform EPM-solution for $P \to P$ called the **copy-cat strategy** ($\mathcal{CCS}$). This strategy consists in mimicking, in the antecedent, the moves made by the environment in the consequent, and vice versa. More formally, the algorithm that $\mathcal{CCS}$ follows is an infinite loop, in every iteration of which $\mathcal{CCS}$ keeps granting permission until the environment makes a move $1.\alpha$ (resp. $2.\alpha$), to which the machine responds by the move $2.\alpha$ (resp. $1.\alpha$). As shown in the proof of Proposition 22.1 of [6], this strategy is successful for every game of the form $A \to A$. An important detail is that $\mathcal{CCS}$ never looks at the past history of the game, i.e. the movement of its scanning head on the run tape is exclusively left-to-right. This means that, even if the original game was something else and it only evolved to $A \to A$ later as a result of making a series of moves, switching to the $\mathcal{CCS}$ after the game has been brought down to $A \to A$ guarantees success no matter what happened in the past.

**Lemma 4.1** *For any* **Int***-formula $K$, $\Vdash \$ \to K$. Furthermore, there is an effective procedure that takes any* **Int***-formula $K$ and constructs a uniform solution $\mathcal{E}$ for $\$ \to K$.*

**Proof.** By induction on the complexity of $K$, below we show how to construct the above uniform solution $\mathcal{E}$ in the form of an EPM. Certain routine details of a verification of the fact that $\mathcal{E}$ indeed always wins the game represented by $\$ \to K$ will be left to the reader. Some — arbitrary and largely irrelevant — interpretation $^*$ will be assumed to be fixed throughout most of our analysis of the work of $\mathcal{E}$, even though, in all but one case, it will be present only implicitly: by innocent abuse of terminology and notation, we will not be distinguishing between a formula $G$ and the game $G^*$ represented by it, and write $G$ even when, strictly speaking, what is meant is $G^*$ rather than $G$.

*Case 1:* $K = \$$. Just let $\mathcal{E} = \mathcal{CCS}$.

*Case 2:* $K$ is a nonlogical atom. Then, for some number $m$, $K^*$ should be conjunct $\#m$ of the infinite $\sqcap$-conjunction $\$^*$ (note that $m$ does not depend on $^*$). We define

$\mathcal{E}$ to be the EPM that acts as follows. At the beginning, it finds the above number $m$ and makes the move '$1.m$', which can be seen to bring the game $\$ \rightarrow K$ down to $K \rightarrow K$. After this move, $\mathcal{E}$ switches to (continues the rest of the play as) $\mathcal{CCS}$.

*Case 3:* $K$ is $E \circ\!\!-\, F$. Evidently $\mathbf{Al} \vdash F \rightarrow (E \circ\!\!-\, F)$ and hence, by Fact 3.1, we can construct an EPM $\mathcal{A}$ with $\mathcal{A} \Vdash F \rightarrow (E \circ\!\!-\, F)$. By the induction hypothesis, we also know how to construct and EPM $\mathcal{C}$ with $\mathcal{C} \Vdash \$ \rightarrow F$. Now, Lemma 3.5 allows us to combine $\mathcal{A}$ and $\mathcal{C}$ and construct the desired $\mathcal{E}$ with $\mathcal{E} \Vdash \$ \rightarrow (E \circ\!\!-\, F)$.

*Case 4:* $K = E_1 \sqcup E_2$. By the induction hypothesis, we know how to construct an EPM $\mathcal{C}$ with $\mathcal{C} \Vdash \$ \rightarrow E_1$. Now we define $\mathcal{E}$ to be the EPM that first makes the move '2.1', and then plays the rest of the game as $\mathcal{C}$ would play. $\mathcal{E}$ can be seen to be successful because its initial move '2.1' brings $\$ \rightarrow K$ down to $\$ \rightarrow E_1$.

*Case 5:* $K = E_1 \sqcap E_2$. By the induction hypothesis, two EPMs $\mathcal{C}_1$ and $\mathcal{C}_2$ can be constructed with $\mathcal{C}_1 \Vdash \$ \rightarrow E_1$ and $\mathcal{C}_2 \Vdash \$ \rightarrow E_2$. We define $\mathcal{E}$ to be the EPM that acts as follows. At the beginning, $\mathcal{E}$ keeps granting permission until the adversary makes a move $\alpha$. With the clean environment assumption in mind, $\alpha$ is '$2.i$' for $i = 1$ or $i = 2$, and it brings $\$ \rightarrow E_1 \sqcap E_2$ down to $\$ \rightarrow E_i$. If and after such a move $\alpha$ is made, $\mathcal{E}$ continues the rest of the play as $\mathcal{C}_i$. $\square$

**Lemma 4.2** *For any* **Int***-formula $K$, $\Vdash ⫰\$ \rightarrow K$. Furthermore, there is an effective procedure that takes any* **Int***-formula $K$ and constructs a uniform solution for $⫰\$ \rightarrow K$.*

**Proof.** By Lemma 4.1, we can construct a uniform solution for $\neg\$ \vee K$. At the same time, evidently $\mathbf{Al} \vdash \neg\$ \rightarrow ⫯\neg\$$ and hence, by Fact 3.1, we can also construct a uniform solution for $\neg\$ \rightarrow ⫯\neg\$$. Now, Lemma 3.5 allows us to combine those two uniform solutions and construct a uniform solution for $⫯\neg\$ \vee K$, i.e. for $⫰\$ \rightarrow K$. $\square$

**Lemma 4.3** *There is an EPM $\mathcal{E}$ such that, for any* **Int***-formulas $E_1$ and $E_2$, $\mathcal{E}$ is a uniform solution for $⫯E_1 \sqcap ⫯E_2 \rightarrow ⫯(E_1 \sqcap E_2)$.*

**Proof.** As in the proof of Lemma 4.1, we will write a formula $G$ to mean the game $G^*$ for whatever (irrelevant) interpretation $^*$. Here is a strategy that the sought $\mathcal{E}$ can follow with a guaranteed success. Keep granting permission until the adversary makes a (legal) move '$2.\epsilon.i$' ($1 \leq i \leq 2$). Such a move can be seen to bring the game $⫯E_1 \sqcap ⫯E_2 \rightarrow ⫯(E_1 \sqcap E_2)$ down to $⫯E_1 \sqcap ⫯E_2 \rightarrow ⫯E_i$. Now make the move '$1.i$', which further brings the game down to $⫯E_i \rightarrow ⫯E_i$. Finally, switch to $\mathcal{CCS}$. $\square$

Remember that, in this paper, an underlined expression such as $\underline{G}$, *when occurring in a sequent*, stands for a sequence $G_1, \ldots, G_n$ of formulas. We now further agree that, *when occurring as a subexpression within a formula*, such an expression will be an abbreviation of $G_1 \wedge \ldots \wedge G_n$. This subexpression will always occur within a "larger" context such as $\underline{G} \wedge F$ or $\underline{G} \rightarrow F$. Our convention is that, when $n = 0$, both $\underline{G} \wedge F$ and $\underline{G} \rightarrow F$ simply mean $F$. And, of course, when $n = 1$, $G_1 \wedge \ldots \wedge G_n$ is just $G_1$.

Next, we agree to understand each **Int**-sequent $\underline{G} \Rightarrow K$ as — and identify with — the **Al**-formula $\underline{⫰G} \rightarrow K$. Remember that where $\underline{G} = G_1, \ldots, G_n$, $\underline{⫰G}$ means $⫰G_1, \ldots, ⫰G_n$, and hence $\underline{⫰G} \rightarrow K$ means $⫰G_1 \wedge \ldots \wedge ⫰G_n \rightarrow K$.

Now we are ready to prove the soundness part of Theorem 2.6. Consider an arbitrary **Int**-sequent $\mathcal{S}$ with **Int** $\vdash \mathcal{S}$. By induction on the length of an **Int**-proof of $\mathcal{S}$, we are going to show that $\mathcal{S}$, understood as a formula according to the above convention, has a uniform solution $\mathcal{E}$. This is sufficient to conclude that **Int** is 'uniformly sound'. Clause (a) of Theorem 2.6 also claims 'constructive soundness', i.e. that such an $\mathcal{E}$ can be effectively built from a given **Int**-proof of $\mathcal{S}$ (at least, when $\mathcal{S}$ is an empty-antecedent sequent). This claim of the theorem will be automatically taken care of by the fact that our proof of the existence of $\mathcal{E}$ is constructive: all of the uniform-validity-related lemmas on which we rely provide a way for actually constructing corresponding uniform solutions. With this remark in mind and for the considerations of readability, in what follows we only talk about uniform validity without explicitly mentioning uniform solutions and without explicitly showing how to construct such solutions. Also, we no longer use $\Rightarrow$ or $⟜$, seeing each sequent $\underline{G} \Rightarrow K$ as the formula $\underline{⫰G} \to K$, and each subformula $E_1 ⟜ E_2$ of such a formula as $⫰E_1 \to E_2$.

There are 11 cases to consider, corresponding to the 11 possible rules (including axioms as "rules" without premises) that might have been used at the last step of an **Int**-derivation of $\mathcal{S}$, with $\mathcal{S}$ being the conclusion of the rule. In each non-axiom case below, "induction hypothesis" means the assumption that the premise(s) of the corresponding rule is (are) uniformly valid. The goal in each case is to show that the conclusion of the rule is also uniformly valid. In what follows, "soundness" should be understood as Fact 3.1, and "modus ponens" as Fact 3.2. We will work with the relaxed version of **Al**, treating **Al**-sequents as multisets rather than sets of formulas.

**Axiom $⫰K \to K$:** Immediately from the obvious fact **Al** $\vdash ⫰K \to K$ by soundness.

**Axiom $⫰\$ \to K$:** Immediately from Lemma 4.2.

**Exchange:** It is easy to verify that **Al** proves

$$(\underline{⫰G} \wedge ⫰E \wedge ⫰F \wedge \underline{⫰H} \to K) \ \to \ (\underline{⫰G} \wedge ⫰F \wedge ⫰E \wedge \underline{⫰H} \to K)$$

and hence, by soundness, this formula is uniformly valid. By the induction hypothesis, so is its antecedent. Hence, by modus ponens, so is the consequent.

**Weakening:** Similar to the previous case, relying on the fact that, due to the presence of weakening in **Al**, the latter proves $(\underline{⫰G} \to K) \to (\underline{⫰G} \wedge ⫰E \to K)$.

**Contraction**: Similar to the previous two cases, relying on the fact that, due to the presence of $⫯$-contraction, **Al** proves $(\underline{⫰G} \wedge ⫰E \wedge ⫰E \to K) \to (\underline{⫰G} \wedge ⫰E \to K)$.

**Left $⟜$:** By the induction hypothesis, we have:

$$\Vdash \underline{⫰G} \wedge ⫰E \to K_1; \tag{5}$$

$$\Vdash \underline{⫰H} \to K_2. \tag{6}$$

Our goal is to show that

$$\Vdash \underline{⫰G} \wedge \underline{⫰H} \wedge ⫰(⫰K_2 \to E) \to K_1. \tag{7}$$

We claim that the following formula is provable in **Al** and hence, by soundness, is uniformly valid:

$$(⫰\underline{G} \wedge ⫰E \to K_1) \to \Bigl(⫰(⫰\underline{H} \to K_2) \to \bigl(⫰\underline{G} \wedge ⫰\underline{H} \wedge ⫰(⫰K_2 \to E) \to K_1\bigr)\Bigr). \quad (8)$$

Below is an **Al**-proof of the above formula for the case when $\underline{G} = G$ and $\underline{H} = H$ (both sequences are just one-formula sequences). Proofs for the cases when $\underline{G}$ and/or $\underline{H}$ are longer or shorter, would take different numbers of steps, but otherwise would be similar.

1. $\neg K_2,\ K_2$ (axiom)
2. $⫰H,\ ⫯\neg H$ (axiom)
3. $⫰H \wedge \neg K_2,\ ⫯\neg H,\ K_2$ (from 2 and 1 by $\wedge$-introduction)
4. $⫯(⫰H \wedge \neg K_2),\ ⫯\neg H,\ K_2$ (from 3 by $⫯$-introduction)
5. $⫯(⫰H \wedge \neg K_2),\ ⫯\neg H,\ ⫰K_2$ (from 4 by $⫰$-introduction)
6. $E, \neg E$ (axiom)
7. $E,\ ⫯(⫰H \wedge \neg K_2),\ ⫯\neg H,\ ⫰K_2 \wedge \neg E$ (from 5 and 6 by $\wedge$-introduction)
8. $E,\ ⫯(⫰H \wedge \neg K_2),\ ⫯\neg H,\ ⫯(⫰K_2 \wedge \neg E)$ (from 7 by $⫯$-introduction)
9. $⫰E,\ ⫯(⫰H \wedge \neg K_2),\ ⫯\neg H,\ ⫯(⫰K_2 \wedge \neg E)$ (from 8 by $⫰$-introduction)
10. $⫰G,\ ⫯\neg G$ (axiom)
11. $⫰G \wedge ⫰E,\ ⫯(⫰H \wedge \neg K_2),\ ⫯\neg G,\ ⫯\neg H,\ ⫯(⫰K_2 \wedge \neg E)$ (from 10 and 9 by $\wedge$-introduction)
12. $\neg K_1,\ K_1$ (axiom)
13. $(⫰G \wedge ⫰E) \wedge \neg K_1,\ ⫯(⫰H \wedge \neg K_2),\ ⫯\neg G,\ ⫯\neg H,\ ⫯(⫰K_2 \wedge \neg E),\ K_1$
    (from 11 and 12 by $\wedge$-introduction)
14. $\bigl((⫰G \wedge ⫰E) \wedge \neg K_1\bigr) \vee \Bigl(⫯(⫰H \wedge \neg K_2) \vee \bigl((⫯\neg G \vee ⫯\neg H \vee ⫯(⫰K_2 \wedge \neg E)) \vee K_1\bigr)\Bigr)$
    (from 13 by $\vee$-introduction applied several times)

Now, (8) is just an abbreviation of 14. By Fact 3.3, (6) implies $\Vdash ⫰(⫰\underline{H} \to K_2)$. From this, together with (5) and $\Vdash$(8), by modus ponens applied twice, we infer the desired (7).

**Right $\circ\!-$:** It is easy to see that the following formula is provable in **Al** and hence, by soundness, is uniformly valid:

$$\bigl(⫰\underline{G} \wedge ⫰E \to K\bigr) \to \bigl(⫰\underline{G} \to (⫰E \to K)\bigr).$$

And, by the induction hypothesis, $\Vdash ⫰\underline{G} \wedge ⫰E \to K$. Applying modus ponens, we get $\Vdash ⫰\underline{G} \to (⫰E \to K)$.

**Left $\sqcup$:** By the induction hypothesis, we have:

$$\Vdash ⫰\underline{G} \wedge ⫰E_1 \to K; \quad (9)$$

$$\Vdash ⫰\underline{G} \wedge ⫰E_2 \rightarrow K. \tag{10}$$

Our goal is to show that

$$\Vdash ⫰\underline{G} \wedge ⫰(E_1 \sqcup E_2) \rightarrow K. \tag{11}$$

We claim that the following formula is provable in **Al** and hence, by soundness, is uniformly valid:

$$\bigl(⫰\underline{G} \wedge ⫰E_1 \rightarrow K\bigr) \rightarrow \Bigl(\bigl(⫰\underline{G} \wedge ⫰E_2 \rightarrow K\bigr) \rightarrow \bigl((\neg ⫰G \vee (⫯\neg E_1 \sqcap ⫯\neg E_2)) \vee K\bigr)\Bigr). \tag{12}$$

As was done when discussing Left $\circ\!\!-$, below we only give an **Al**-proof of the above formula for the case when $\underline{G} = G$. The case when $\underline{G}$ is empty is simpler, and the cases when $\underline{G}$ contains more than one formula would take more steps but otherwise would be similar.

1. $⫰G \wedge ⫰E_1,\ \neg ⫰G,\ ⫯\neg E_1$ (from axioms $⫰G, \neg ⫰G$ and $⫰E_1, ⫯\neg E_1$ by $\wedge$-introduction)
2. $(⫰G \wedge ⫰E_1) \wedge \neg K,\ \neg ⫰G,\ ⫯\neg E_1,\ K$ (from 1 and axiom $\neg K, K$ by $\wedge$-introduction)
3. $(⫰G \wedge ⫰E_1) \wedge \neg K,\ (⫰G \wedge ⫰E_2) \wedge \neg K,\ \neg ⫰G,\ ⫯\neg E_1,\ K$ (from 2 by weakening)
4. $(⫰G \wedge ⫰E_1) \wedge \neg K,\ (⫰G \wedge ⫰E_2) \wedge \neg K,\ \neg ⫰G,\ ⫯\neg E_2,\ K$ (similar to 3)
5. $(⫰G \wedge ⫰E_1) \wedge \neg K,\ (⫰G \wedge ⫰E_2) \wedge \neg K,\ \neg ⫰G,\ ⫯\neg E_1 \sqcap ⫯\neg E_2,\ K$
   (from 3 and 4 by $\sqcap$-introduction)
6. $\bigl((⫰G \wedge ⫰E_1) \wedge \neg K\bigr) \vee \Bigl(\bigl((⫰G \wedge ⫰E_2) \wedge \neg K\bigr) \vee \bigl((\neg ⫰G \vee (⫯\neg E_1 \sqcap ⫯\neg E_2)) \vee K\bigr)\Bigr)$
   (from 5 by $\vee$-introduction applied several times)

Now, (12) is just an abbreviation of 6. From (9), (10) and $\Vdash$(12), by modus ponens applied twice, we infer

$$\Vdash (\neg ⫰G \vee (⫯\neg E_1 \sqcap ⫯\neg E_2)) \vee K.$$

But, by Lemma 4.3, $\Vdash ⫯\neg E_1 \sqcap ⫯\neg E_2 \rightarrow ⫯(\neg E_1 \sqcap \neg E_2)$. Applying Lemma 3.5, we find $\Vdash (\neg ⫰G \vee ⫯(\neg E_1 \sqcap \neg E_2)) \vee K$, which is nothing but a disabbreviation of the desired (11).

**Right $\sqcup$:** By the induction hypothesis, $\Vdash ⫰\underline{G} \rightarrow K_i$. Of course, we also have $\mathbf{Al} \vdash K_i \rightarrow K_1 \sqcup K_2$ and thus $\Vdash K_i \rightarrow K_1 \sqcup K_2$. Hence, by Lemma 3.5, $\Vdash ⫰\underline{G} \rightarrow K_1 \sqcup K_2$.

**Left $\sqcap$:** By the induction hypothesis, $\Vdash ⫰\underline{G} \wedge ⫰E_i \rightarrow K$, i.e. $\Vdash (\underline{⫯\neg G} \vee ⫯\neg E_i) \vee K$. Obviously we also have $\mathbf{Al} \vdash ⫯\neg E_i \rightarrow ⫯(\neg E_1 \sqcup \neg E_2)$ and thus $\Vdash ⫯\neg E_i \rightarrow ⫯(\neg E_1 \sqcup \neg E_2)$. Hence, by Lemma 3.5, $\Vdash (\underline{⫯\neg G} \vee ⫯(\neg E_1 \sqcup \neg E_2)) \vee K$, i.e. $\Vdash ⫰\underline{G} \wedge ⫰(E_1 \sqcap \neg E_2) \rightarrow K$.

**Right $\sqcap$:** By the induction hypothesis, $\Vdash ⫰\underline{G} \rightarrow K_1$ and $\Vdash ⫰\underline{G} \rightarrow K_2$. Next, one could easily verify that **Al** proves

$$(⫰\underline{G} \rightarrow K_1) \rightarrow \bigl((⫰\underline{G} \rightarrow K_2) \rightarrow (⫰\underline{G} \rightarrow K_1 \sqcap K_2)\bigr),$$

so this formula is uniformly valid. Applying modus ponens twice yields the desired $\Vdash ⫰\underline{G} \rightarrow K_1 \sqcap K_2$.

# 5 Machines vs. machines

As noted earlier, the rest of this paper is devoted to a proof of the completeness part of Theorem 2.6. The present section borrows a discussion from [7], providing certain background information necessary for our completeness proof but missing in [10], the only external source on computability logic on which this paper was promised to rely.

Remember that $\neg\Gamma$, when $\Gamma$ is a run, means the result of reversing all labels in $\Gamma$. For a run $\Gamma$ and a computation branch $B$ of an HPM or EPM, we say that $B$ **cospells** $\Gamma$ iff $B$ spells $\neg\Gamma$ in the sense of Section 6 of [10]. Intuitively, when a machine $\mathcal{M}$ plays as $\bot$ (rather than $\top$), then the run that is generated by a given computation branch $B$ of $\mathcal{M}$ is the run cospelled (rather than spelled) by $B$, for the moves that $\mathcal{M}$ makes get the label $\bot$, and the moves that its adversary makes get the label $\top$.

We say that an EPM $\mathcal{E}$ is **fair** iff, for every valuation $e$, every $e$-computation branch of $\mathcal{E}$ is fair in the sense of Section 6 of [10].

**Lemma 5.1** *Assume $\mathcal{E}$ is a fair EPM, $\mathcal{H}$ is any HPM, and $e$ is any valuation. There are a uniquely defined $e$-computation branch $B_{\mathcal{E}}$ of $\mathcal{E}$ and a uniquely defined $e$-computation branch $B_{\mathcal{H}}$ of $\mathcal{H}$ — which we respectively call* **the $(\mathcal{E}, e, \mathcal{H})$-branch** *and* **the $(\mathcal{H}, e, \mathcal{E})$-branch** *— such that the run spelled by $B_{\mathcal{H}}$, called* **the $\mathcal{H}$ vs. $\mathcal{E}$ run on** *$e$, is the run cospelled by $B_{\mathcal{E}}$.*

When $\mathcal{H}, \mathcal{E}, e$ are as above, $\Gamma$ is the $\mathcal{H}$ vs. $\mathcal{E}$ run on $e$ and $A$ is a game with $\mathbf{Wn}^A_e\langle\Gamma\rangle = \top$ (resp. $\mathbf{Wn}^A_e\langle\Gamma\rangle = \bot$), we say that $\mathcal{H}$ **wins** (resp. **loses**) $A$ **against** $\mathcal{E}$ **on** $e$.

A strict proof of the above lemma can be found in [6] (Lemma 20.4), and we will not reproduce the formal proof here. Instead, the following intuitive explanation should suffice:

**Proof idea.** Assume $\mathcal{H}$, $\mathcal{E}$, $e$ are as in Lemma 5.1. The play that we are going to describe is the unique play generated when the two machines play against each other, with $\mathcal{H}$ in the role of $\top$, $\mathcal{E}$ in the role of $\bot$, and $e$ spelled on the valuation tapes of both machines. We can visualize this play as follows. Most of the time during the process $\mathcal{H}$ remains inactive (sleeping); it is woken up only when $\mathcal{E}$ enters a permission state, on which event $\mathcal{H}$ makes a (one single) transition to its next computation step — that may or may not result in making a move — and goes back to sleep that will continue until $\mathcal{E}$ enters a permission state again, and so on. From $\mathcal{E}$'s perspective, $\mathcal{H}$ acts as a patient adversary who makes one or zero move only when granted permission, just as the EPM-model assumes. And from $\mathcal{H}$'s perspective, who, like a person in a comma, has no sense of time during its sleep and hence can think that the wake-up events that it calls the beginning of a clock cycle happen at a constant rate, $\mathcal{E}$ acts as an adversary who can make any finite number of moves during a clock cycle (i.e. while $\mathcal{H}$ was sleeping), just as the HPM-model assumes. This scenario uniquely determines an $e$-computation branch $B_{\mathcal{E}}$ of $\mathcal{E}$ that we call the $(\mathcal{E}, e, \mathcal{H})$-branch, and an $e$-computation branch $B_{\mathcal{H}}$ of $\mathcal{H}$ that we call the $(\mathcal{H}, e, \mathcal{E})$-branch. What we call the $\mathcal{H}$ vs. $\mathcal{E}$ run on $e$ is the run generated in this play. In particular — since we let $\mathcal{H}$ play in the role of $\top$ — this is the run spelled by $B_{\mathcal{H}}$. $\mathcal{E}$, who plays in the role of $\bot$, sees the same run,

only it sees the labels of the moves of that run in negative colors. That is, $B_{\mathcal{E}}$ cospells rather than spells that run. This is exactly what Lemma 5.1 asserts. $\diamond$

# 6 Dedollarization

As we remember, the language of **Int**, unlike that of **Al**, treats $\sqcap$ as a strictly binary operator. Yet in this section we will be writing expressions $E_1 \sqcap \ldots \sqcap E_n$ for unspecified $n \geq 1$ and treating them as legitimate (subformulas of) **Int**-formulas. Such an expression should be understood as an abbreviation of $E_1 \sqcap (E_2 \sqcap (\ldots \sqcap E_n) \ldots))$; when $n = 1$, $E_1 \sqcap \ldots \sqcap E_n$ simply means $E_1$.

An **Int**-formula is said to be **dollarless** iff it does not contain \$; and an **Int**-sequent is **dollarless** iff all of its formulas are so.

**Definition 6.1** Let $K$ be an **Int**-formula, $P_1, \ldots, P_n$ all of its nonlogical atoms listed according to the lexicographic order, and $P_0$ lexicographically the smallest nonlogical atom not occurring in $K$. Then what we call the **dedollarization** of $K$ is the result of replacing in $K$ every occurrence of \$ by the formula $P_0 \sqcap P_1 \sqcap \ldots \sqcap P_n$.

Note that the dedollarization of a given formula is (indeed) a dollarless formula.

**Lemma 6.2** *Whenever the dedollarization of an* **Int***-formula $K$ is provable in* **Int***, so is $K$ itself.*

**Proof.** Consider an arbitrary **Int**-formula $K$, and let $F$ be its dedollarization. We assume that $K$, $P_0$, $P_1, \ldots, P_n$ are as in Definition 6.1. Suppose **Int** $\vdash F$. Let $G$ be the result of replacing in $F$ every occurrence of $P_0$ by \$. **Int**-provability is known to be closed under replacing nonlogical atoms by whatever formulas, so we have **Int** $\vdash G$. Next, it is obvious that $\Vert\!\models \$ \circ\!\!-\!\!\circ \$ \sqcap P_1 \sqcap \ldots \sqcap P_n$. This, by Lemma 2.1, implies that $\Vert\!\models G \circ\!\!-\!\!\circ K$, because $K$ is nothing but the result of replacing in $G$ (every occurrence of) $\$ \sqcap P_1 \sqcap \ldots \sqcap P_n$ by \$. Now, in view of the soundness and completeness of **Int** with respect to Kripke semantics, the facts **Int** $\vdash G$ and $\Vert\!\models G \circ\!\!-\!\!\circ K$ can be seen to immediately imply **Int** $\vdash K$. $\square$

**Lemma 6.3** *Assume $K$ is an* **Int***-formula, $F$ its dedollarization, and $^\circ$ an interpretation of a complexity $C$ such that $\not\models F^\circ$. Then there is an interpretation $^*$ of the same complexity $C$ such that $\not\models K^*$.*

**Proof.** Assume $K$, $F$, $^\circ$ and $C$ are as in the conditions of Lemma 6.3. We further assume that $K$, $P_0$, $P_1, \ldots, P_n$ are as in Definition 6.1. We select $^*$ to be the unique interpretation such that:

- $P_0^* = P_0^\circ$, $P_1^* = P_1^\circ$, $\ldots$, $P_n^* = P_n^\circ$;
- for any nonlogical atom $Q$ with $Q \notin \{P_0, P_1, \ldots, P_n\}$, $Q^* = P_0^*$ (i.e. $Q^* = P_0^\circ$);
- the base of $\$^*$ is $P_0^*$ (i.e. $P_0^\circ$).

Obviously $^*$ remains of complexity $C$. We claim that

$$\begin{array}{l} \models \$^* \rightarrow P_0^* \sqcap P_1^* \sqcap \ldots \sqcap P_n^*; \\ \models P_0^* \sqcap P_1^* \sqcap \ldots \sqcap P_n^* \rightarrow \$^*. \end{array} \qquad (13)$$

To prove the above, let us fix $Q_1, Q_2, \ldots$ as the lexicographic list of all nonlogical atoms of the language of **Int**. $\$^*$ thus can and will be seen as the infinite conjunction $P_0^* \sqcap Q_1^* \sqcap Q_2^* \sqcap \ldots$. In what follows, as always, we rely on the clean environment assumption. Also, we pretend that $P_0^* \sqcap P_1^* \sqcap \ldots \sqcap P_n^*$ is literally $P_0^* \sqcap P_1^* \sqcap \ldots \sqcap P_n^*$ rather than an abbreviation of $P_0^* \sqcap (P_1^* \sqcap (\ldots \sqcap P_n^*) \ldots))$ which it, strictly speaking, is. This innocent and easy-to-correct inaccuracy is just to simplify our descriptions of winning strategies.

Here is a relaxed description and analysis of an EPM-strategy for

$$\$^* \rightarrow P_0^* \sqcap P_1^* \sqcap \ldots \sqcap P_n^*,$$

i.e. for

$$(P_0^* \sqcap Q_1^* \sqcap Q_2^* \sqcap \ldots) \ \rightarrow \ (P_0^* \sqcap P_1^* \sqcap \ldots \sqcap P_n^*).$$

Wait until the environment selects a conjunct $P_i^*$ $(0 \leq i \leq n)$ in the consequent. $P_i$ is $Q_j$ for a certain $j$. So, select the conjunct $Q_j^*$ in the antecedent, and then switch to the copy-cat strategy $\mathcal{CCS}$. A win is guaranteed because the first two moves have brought the game down to $Q_j^* \rightarrow P_i^*$, i.e. $P_i^* \rightarrow P_i^*$.

Next, here is a strategy for

$$P_0^* \sqcap P_1^* \sqcap \ldots \sqcap P_n^* \rightarrow \$^*,$$

i.e. for

$$(P_0^* \sqcap P_1^* \sqcap \ldots \sqcap P_n^*) \ \rightarrow \ (P_0^* \sqcap Q_1^* \sqcap Q_2^* \sqcap \ldots).$$

Wait till the environment selects a conjunct of the consequent. If the selected conjunct is $Q_i^*$ such that $Q_i = P_j$ for some $0 \leq j \leq n$, then select the conjunct $P_j^*$ in the antecedent. This brings the game down to $P_j^* \rightarrow Q_i^*$, i.e. $P_j^* \rightarrow P_j^*$, so switching to $\mathcal{CCS}$ at this point guarantees success. And otherwise, if the selected conjunct of the consequent was $P_0^*$, or $Q_i^*$ such that $Q_i \notin \{P_0, \ldots, P_n\}$ so that (by our choice of $^*$) $Q_i^* = P_0^*$, then select the conjunct $P_0^*$ in the antecedent. This brings the game down to $P_0^* \rightarrow P_0^*$, so, again, switching to $\mathcal{CCS}$ at this point guarantees success. (13) is thus proven.

Now, it is our assumption that $\not\models F^\circ$. But $^*$ agrees with $^\circ$ on all atoms of $F$, and hence $F^\circ = F^*$. So, $\not\models F^*$. This, in view of (13) and Lemma 3.5, can be seen to imply that $\not\models K^*$, because $F$ is the result of replacing in $K$ (all occurrences of) \$ by $P_0 \sqcap P_1 \sqcap \ldots \sqcap P_n$. $\square$

# 7 Standardization

We say that an **Int**-sequent is **standard** iff it is

$$\left. \begin{array}{llll} X_1 \circ\!\!- (Y_1 \circ\!\!- Z_1 \sqcup T_1), & \ldots, & X_k \circ\!\!- (Y_k \circ\!\!- Z_k \sqcup T_k), \\ (P_1 \circ\!\!- Q_1) \circ\!\!- R_1, & \ldots, & (P_k \circ\!\!- Q_k) \circ\!\!- R_k \end{array} \right\} \ \Rightarrow \ W$$

where $k \geq 0$, and the $X_i, Y_i, Z_i, T_i, P_i, Q_i, R_i$ $(1 \leq i \leq k)$ and $W$ are nonlogical atoms.

Note that, by definition, a standard **Int**-sequent is always dollarless.

Where $K$ is a dollarless **Int**-formula and $H$ is a subformula of it, throughout this section we will be using the notation

$$H^K$$

for a certain atom intended to be the "standard atomic name" assigned by us to $H$ as a subformula of $K$. Specifically, let $G_1, \ldots, G_m$ be all of the non-atomic subformulas of $K$ — including $K$ itself — listed according to the lexicographic order, and let $W_1, \ldots, W_m$ be the first (in the lexicographic order) $m$ nonlogical atoms of the language of **Int** not occurring in $K$. Then we define the atom $H^K$ by stipulating that:

- $H^K = H$ if $H$ is atomic;
- $H^K = W_i$ if $H = G_i$ $(1 \leq i \leq m)$.

**Definition 7.1** Let $K$ be a dollarless **Int**-formula, and let

$$\begin{array}{lllllll} A_1 & = & B_1 \circ\!- C_1, & \ldots, & A_a & = & B_a \circ\!- C_a, \\ D_1 & = & E_1 \sqcup F_1, & \ldots, & D_d & = & E_d \sqcup F_d, \\ G_1 & = & H_1 \sqcap I_1, & \ldots, & G_g & = & H_g \sqcap I_g \end{array}$$

be all of its non-atomic subformulas, with each of the three groups listed according to the lexicographic order. Then we define the **standardization** of $K$ as the **Int**-sequent obtained by appending $3d+3g$ extra copies of the last formula $(B_a^K \circ\!- C_a^K) \circ\!- A_a^K$ to the antecedent of the following sequent:

$$\left.\begin{array}{lll} A_1^K \circ\!- (B_1^K \circ\!- C_1^K \sqcup C_1^K), & \ldots, & A_a^K \circ\!- (B_a^K \circ\!- C_a^K \sqcup C_a^K), \\ E_1^K \circ\!- (E_1^K \circ\!- D_1^K \sqcup D_1^K), & \ldots, & E_d^K \circ\!- (E_d^K \circ\!- D_d^K \sqcup D_d^K), \\ F_1^K \circ\!- (F_1^K \circ\!- D_1^K \sqcup D_1^K), & \ldots, & F_d^K \circ\!- (F_d^K \circ\!- D_d^K \sqcup D_d^K), \\ D_1^K \circ\!- (D_1^K \circ\!- E_1^K \sqcup F_1^K), & \ldots, & D_d^K \circ\!- (D_d^K \circ\!- E_d^K \sqcup F_d^K), \\ G_1^K \circ\!- (G_1^K \circ\!- H_1^K \sqcup H_1^K), & \ldots, & G_g^K \circ\!- (G_g^K \circ\!- H_g^K \sqcup H_g^K), \\ G_1^K \circ\!- (G_1^K \circ\!- I_1^K \sqcup I_1^K), & \ldots, & G_g^K \circ\!- (G_g^K \circ\!- I_g^K \sqcup I_g^K), \\ H_1^K \circ\!- (I_1^K \circ\!- G_1^K \sqcup G_1^K), & \ldots, & H_g^K \circ\!- (I_g^K \circ\!- G_g^K \sqcup G_g^K), \\ (B_1^K \circ\!- C_1^K) \circ\!- A_1^K, & \ldots, & (B_a^K \circ\!- C_a^K) \circ\!- A_a^K \end{array}\right\} \Rightarrow \quad K^K.$$

Observe that the standardization of a dollarless **Int**-formula is always (indeed) a standard sequent.

**Lemma 7.2** *Let $K$ be a dollarless* **Int***-formula, $\underline{G} \Rightarrow W$ its standardization, and $\mathcal{K}$ a Kripke model with $\mathcal{K} \models \underline{G}$. Then every subformula $H$ of $K$ is $\mathcal{K}$-equivalent to $H^K$. Specifically, as $K^K = W$, we have $\mathcal{K} \models K \circ\!\!-\!\!\circ W$.*

**Proof.** Let $K$, $\underline{G} \Rightarrow W$, $\mathcal{K} = (\mathcal{W}, \mathcal{R}, \models)$ be as the lemma assumes, and $H$ be any subformula of $K$. We need to show that $\mathcal{K} \| \models H \circ\!\!-\!\!\circ H^K$, which will be done by induction on the complexity of $H$.

*Case 1*: $H$ is atomic. Then $H^K = H$, so this case is trivial.

*Case 2*: $H = E \circ\!\!- F$. Then $H^K \circ\!\!- (E^K \circ\!\!- F^K \sqcup F^K)$ and $(E^K \circ\!\!- F^K) \circ\!\!- H^K$ are among the formulas of $\underline{G}$. Therefore, as $\mathcal{K} \| \models \underline{G}$, we have

$$\mathcal{K} \| \models H^K \circ\!\!- (E^K \circ\!\!- F^K \sqcup F^K),\ (E^K \circ\!\!- F^K) \circ\!\!- H^K. \tag{14}$$

But, by the induction hypothesis, $\mathcal{K} \| \models E^K \circ\!\!-\!\!\circ E$ and $\mathcal{K} \| \models F^K \circ\!\!-\!\!\circ F$. Of course, we also have $\mathcal{K} \| \models F^K \sqcup F^K \circ\!\!-\!\!\circ F^K$. Therefore, in view of Fact 2.1, (14) can be rewritten as

$$\mathcal{K} \| \models H^K \circ\!\!- (E \circ\!\!- F),\ (E \circ\!\!- F) \circ\!\!- H^K.$$

As $(E \circ\!\!- F) = H$, the above means nothing but that $\mathcal{K} \| \models H \circ\!\!-\!\!\circ H^K$.

*Case 3*: $H = E \sqcup F$. Then

$$\mathcal{K} \| \models E^K \circ\!\!- (E^K \circ\!\!- H^K \sqcup H^K),\ F^K \circ\!\!- (F^K \circ\!\!- H^K \sqcup H^K),\ H^K \circ\!\!- (H^K \circ\!\!- E^K \sqcup F^K),$$

because $\underline{G}$ contains the above three formulas and $\mathcal{K} \| \models \underline{G}$. But, by the induction hypothesis, $E^K$ and $F^K$ are $\mathcal{K}$-equivalent to $E$ and $F$, respectively. Hence, by Fact 2.1,

$$\mathcal{K} \| \models E \circ\!\!- (E \circ\!\!- H^K \sqcup H^K),\ F \circ\!\!- (F \circ\!\!- H^K \sqcup H^K) \tag{15}$$

and

$$\mathcal{K} \| \models H^K \circ\!\!- (H^K \circ\!\!- E \sqcup F). \tag{16}$$

Consider any world $p$ with $p \models H^K$. By (16), we also have $p \models H^K \circ\!\!- (H^K \circ\!\!- E \sqcup F)$. Hence, with a moment's thought, it can be seen that $p \models E \sqcup F$, i.e. $p \models H$. As $p$ was arbitrary, we conclude that

$$\mathcal{K} \| \models H^K \circ\!\!- H. \tag{17}$$

Next, consider any world $p$ with $p \models H$, i.e. $p \models E \sqcup F$. We should have $p \models E$ or $p \models F$. Suppose the former is the case (the latter is, of course, similar). From (15) we also know that $p \models E \circ\!\!- (E \circ\!\!- H^K \sqcup H^K)$. Hence it can be seen that $p \models H^K$. Again, as $p$ was arbitrary, we conclude that

$$\mathcal{K} \| \models H \circ\!\!- H^K. \tag{18}$$

Now, (17) and (18) together mean nothing but that $\mathcal{K} \| \models H \circ\!\!-\!\!\circ H^K$.

*Case 4*: $H = E \sqcap F$. Then, for reasons already familiar to us, we have

$$\mathcal{K} \| \models H^K \circ\!\!- (H^K \circ\!\!- E^K \sqcup E^K),\ H^K \circ\!\!- (H^K \circ\!\!- F^K \sqcup F^K),\ E^K \circ\!\!- (F^K \circ\!\!- H^K \sqcup H^K).$$

By the induction hypothesis, $\mathcal{K} \| \models E^K \circ\!\!-\!\!\circ E$ and $\mathcal{K} \| \models F^K \circ\!\!-\!\!\circ F$, and hence

$$\mathcal{K} \| \models H^K \circ\!\!- (H^K \circ\!\!- E \sqcup E),\ H^K \circ\!\!- (H^K \circ\!\!- F \sqcup F) \tag{19}$$

and

$$\mathcal{K} \models\!\!\models E \circ\!\!- (F \circ\!\!- H^K \sqcup H^K). \qquad (20)$$

Consider any world $p$. Suppose $p \models\!\!\models H$, i.e. $p \models\!\!\models E \sqcap F$, meaning that $p \models\!\!\models E$ and $p \models\!\!\models F$. Then, from (20), we can see that $p \models\!\!\models H^K$. Now, for vice versa, suppose $p \models\!\!\models H^K$. Then, from (19), we can see that $p \models\!\!\models E$ and $p \models\!\!\models F$, which means that $p \models\!\!\models E \sqcap F$, i.e. $p \models\!\!\models H$. Thus, $p \models\!\!\models H$ iff $p \models\!\!\models H^K$. This, as $p$ was arbitrary, means that $\mathcal{K} \models\!\!\models H \circ\!\!-\!\!\circ H^K$. $\square$

**Lemma 7.3** *Assume $K$ is a dollarless* **Int***-formula, and $\underline{G} \Rightarrow W$ is its standardization. Then* **Int** $\vdash K, \underline{G} \Rightarrow W$.

**Proof.** Assume, for a contradiction, that $\underline{G} \Rightarrow W$ is the standardization of a dollarless **Int**-formula $K$, and **Int** $\not\vdash K, \underline{G} \Rightarrow W$. Then, by the completeness of **Int** with respect to Kripke semantics, there is a model $\mathcal{K} = (\mathcal{W}, \mathcal{R}, \models\!\!\models)$ with $\mathcal{K} \not\models\!\!\models K, \underline{G} \Rightarrow W$, meaning that, for some world $p \in \mathcal{W}$, we have $p \models\!\!\models K$, $p \models\!\!\models \underline{G}$ and $p \not\models\!\!\models W$. Obviously we may assume here that $p = 1$, so that $\mathcal{K} \models\!\!\models \underline{G}$. Then, by Lemma 7.2, $\mathcal{K} \models\!\!\models K \circ\!\!-\!\!\circ W$. But this is a contradiction, because we have $1 \models\!\!\models K$ and $1 \not\models\!\!\models W$. $\square$

**Lemma 7.4** *Assume $K$ is a dollarless* **Int***-formula, $\underline{G} \Rightarrow W$ is its standardization, and* **Int** $\not\vdash K$. *Then* **Int** $\not\vdash \underline{G} \Rightarrow W$.

**Proof.** Consider an arbitrary dollarless **Int**-formula $K$ and its standardization $\underline{G} \Rightarrow W$. We assume that all of the conditions of Definition 7.1 regarding the subformulas of $K$ are satisfied, so that each formula of $\underline{G}$ is one of the formulas of the antecedent of the sequent displayed in that definition.

Suppose **Int** $\not\vdash K$. Then, by the completeness of **Int** with respect to Kripke semantics, there is a model $\mathcal{K}$ with $\mathcal{K} \not\models\!\!\models K$. We may assume here that, for every subformula $H$ of $K$, $\mathcal{K} \models\!\!\models H^K \circ\!\!-\!\!\circ H$. Indeed, when $H$ is atomic, then it is automatically $\mathcal{K}$-equivalent to $H^K$ because $H = H^K$. And if $H$ is not atomic, then $H^K$ is not among the atoms of $K$, and we may make arbitrary assumptions regarding what worlds force $H^K$ without affecting the fact that $\mathcal{K} \not\models\!\!\models K$; so, our assumption is that $H^K$ is forced exactly by the worlds that force $H$.

Now we claim that $\mathcal{K} \models\!\!\models \underline{G}$. To prove this claim, pick an arbitrary formula $G$ of $\underline{G}$. We need to show that $\mathcal{K} \models\!\!\models G$. Looking back at the sequent displayed in Definition 7.1, there are 8 cases to consider, depending on the form of $G$. Here we will only consider the case

$$G = A_i^K \circ\!\!- (B_i^K \circ\!\!- C_i^K \sqcup C_i^K),$$

leaving the other 7 cases as rather similar and equally easy exercises for the reader. Remember that, by our assumptions, for every relevant formula $H$ (here $H \in \{A_i, B_i, C_i\}$) we have $\mathcal{K} \models\!\!\models H^K \circ\!\!-\!\!\circ H$. So, in view of Fact 2.1, in order to verify that $\mathcal{K} \models\!\!\models G$, it would suffice to show that $\mathcal{K} \models\!\!\models A_i \circ\!\!- (B_i \circ\!\!- C_i \sqcup C_i)$. But, by the conditions of Definition 7.1, $A_i = B_i \circ\!\!- C_i$. So, our task reduces to showing that $\mathcal{K} \models\!\!\models (B_i \circ\!\!- C_i) \circ\!\!- (B_i \circ\!\!- C_i \sqcup C_i)$. Since $\mathcal{K} \models\!\!\models C_i \sqcup C_i \circ\!\!-\!\!\circ C_i$, this task further reduces to "showing" the trivial fact that $\mathcal{K} \models\!\!\models (B_i \circ\!\!- C_i) \circ\!\!- (B_i \circ\!\!- C_i)$.

Thus, we consider $\mathcal{K} \Vdash \underline{G}$ proven. Then, by Lemma 7.2, $\mathcal{K} \Vdash W \circ\!\!-\!\!\circ K$. But $\mathcal{K} \nVdash K$, and thus $\mathcal{K} \nVdash W$. This means that $\mathcal{K} \nVdash \underline{G} \Rightarrow W$. Consequently, by the soundness of **Int** with respect to Kripke semantics, **Int** $\nvdash \underline{G} \Rightarrow W$. $\square$

# 8 Desequentization

Consider an arbitrary standard **Int**-sequent

$$\left.\begin{array}{l} X_1 \circ\!- (Y_1 \circ\!- Z_1 \sqcup T_1),\ \ldots,\ X_k \circ\!- (Y_k \circ\!- Z_k \sqcup T_k),\\ (P_1 \circ\!- Q_1) \circ\!- R_1,\ \ldots,\ (P_k \circ\!- Q_k) \circ\!- R_k \end{array}\right\} \Rightarrow W. \tag{21}$$

For any positive integer $n$, we define the $n$**-desequentization** of the above sequent (21) as the following **Al**-formula (22) where, for each $j \in \{1, \ldots, k\}$ and $p \in \{1, \ldots, n\}$,

$$P_j^p = P_j \ \text{ and } \ Q_j^p = Q_j,$$

and where $E_1 \wedge \ldots \wedge E_n$ is understood as just $E_1$ when $n = 1$:

$$\begin{array}{l} \overset{|}{\circ}\big(X_1 \wedge Y_1 \to Z_1 \sqcup T_1\big)\ \wedge\ \ldots\ \wedge\ \overset{|}{\circ}\big(X_k \wedge Y_k \to Z_k \sqcup T_k\big)\ \wedge \\ \overset{|}{\circ}\big((\overset{|}{\circ}P_1^1 \to Q_1^1) \wedge \ldots \wedge (\overset{|}{\circ}P_1^n \to Q_1^n) \to R_1\big)\ \wedge\ \ldots\ \wedge \\ \overset{|}{\circ}\big((\overset{|}{\circ}P_k^1 \to Q_k^1) \wedge \ldots \wedge (\overset{|}{\circ}P_k^n \to Q_k^n) \to R_k\big) \end{array} \to W. \tag{22}$$

**Lemma 8.1** *Assume $n$ is an arbitrary positive integer, $K$ is a dollarless* **Int***-formula, and $D$ is the $n$-desequentization of the standardization of $K$. Then, for any interpretation $^\circ$ with $\models K^\circ$, we have $\models D^\circ$.*

**Proof.** Consider an arbitrary $n \geq 1$, an arbitrary dollarless **Int**-formula $K$ and an arbitrary interpretation $^\circ$ with $\models K^\circ$. We assume that (21) is the standardization of $K$, so that the $n$-desequentization $D$ of the standardization of $K$ is (22).

Using $G_1, \ldots, G_{2k}$ as abbreviations of the $2k$ formulas of the antecedent of (21), we rewrite (21) as

$$G_1, \ldots, G_{2k} \Rightarrow W.$$

According to Lemma 7.3,

$$\textbf{Int} \vdash K, G_1, \ldots, G_{2k} \Rightarrow W.$$

From here, applying the Right $\circ\!-$ rule $2k + 1$ times, we get

$$\textbf{Int} \vdash K \circ\!- (G_1 \circ\!- (G_2 \circ\!- \ldots (G_{2k} \circ\!- W) \ldots))$$

and hence, by the already established soundness of **Int**,

$$\Vvdash K \circ\!- (G_1 \circ\!- (G_2 \circ\!- \ldots (G_{2k} \circ\!- W) \ldots)). \tag{23}$$

Next, in a routine syntactic exercise, one can show that the following formula is provable in **Al** and hence, by the soundness of **Al**, it is uniformly valid:

$$\begin{array}{l}\Big(K \circ\!\!- (G_1 \circ\!\!- (G_2 \circ\!\!- \ldots (G_{2k} \circ\!\!- W)\ldots))\Big)\\ \rightarrow\ \Big(\overset{|}{\circ}K \rightarrow (\overset{|}{\circ}G_1 \wedge \ldots \wedge \overset{|}{\circ}G_{2k} \rightarrow W)\Big).\end{array} \tag{24}$$

From (23) and $\Vdash$(24), in view of Fact 3.2, we find

$$\models \overset{|}{\circ}K^{\circ} \rightarrow (\overset{|}{\circ}G_1 \wedge \ldots \wedge \overset{|}{\circ}G_{2k} \rightarrow W)^{\circ}. \tag{25}$$

By Fact 3.3, our assumption $\models K^{\circ}$ implies $\models \overset{|}{\circ}K^{\circ}$. This, in turn, together with (25), again by Fact 3.2, implies

$$\models (\overset{|}{\circ}G_1 \wedge \ldots \wedge \overset{|}{\circ}G_{2k} \rightarrow W)^{\circ}.$$

If we now disabbreviate the $G_i$s, the formula $\overset{|}{\circ}G_1 \wedge \ldots \wedge \overset{|}{\circ}G_{2s} \rightarrow W$ rewrites as

$$\left.\begin{array}{l}\overset{|}{\circ}\big(\overset{|}{\circ}X_1 \rightarrow (\overset{|}{\circ}Y_1 \rightarrow Z_1 \sqcup T_1)\big) \quad \wedge \ldots \wedge \quad \overset{|}{\circ}\big(\overset{|}{\circ}X_k \rightarrow (\overset{|}{\circ}Y_k \rightarrow Z_k \sqcup T_k)\big)\\ \wedge \quad \overset{|}{\circ}\big(\overset{|}{\circ}(\overset{|}{\circ}P_1 \rightarrow Q_1) \rightarrow R_1\big) \quad \wedge \ldots \wedge \quad \overset{|}{\circ}\big(\overset{|}{\circ}(\overset{|}{\circ}P_k \rightarrow Q_k) \rightarrow R_k\big)\end{array}\right. \rightarrow W,$$

which, in turn, is an abbreviation of

$$\left(\begin{array}{l}\underset{|}{\circ}\Big(\overset{|}{\circ}X_1 \wedge \big(\overset{|}{\circ}Y_1 \wedge \neg(Z_1 \sqcup T_1)\big)\Big) \quad \vee \ldots \vee\\ \underset{|}{\circ}\Big(\overset{|}{\circ}X_k \wedge \big(\overset{|}{\circ}Y_k \wedge \neg(Z_k \sqcup T_k)\big)\Big) \quad \vee\\ \underset{|}{\circ}\Big(\overset{|}{\circ}(\overset{|}{\circ}P_1 \rightarrow Q_1) \wedge \neg R_1\Big) \quad \vee \ldots \vee\\ \underset{|}{\circ}\Big(\overset{|}{\circ}(\overset{|}{\circ}P_k \rightarrow Q_k) \wedge \neg R_k\Big)\end{array}\right) \vee \quad W. \tag{26}$$

So, $\models (26)^{\circ}$. Now, of course, **Al** proves

$$\overset{|}{\circ}X_i \wedge \big(\overset{|}{\circ}Y_i \wedge \neg(Z_i \sqcup T_i)\big) \ \rightarrow\ (\overset{|}{\circ}X_i \wedge \overset{|}{\circ}Y_i) \wedge \neg(Z_i \sqcup T_i)$$

and hence, by the soundness of **Al**, the above formula is uniformly valid for each $1 \leq i \leq k$. Therefore, starting from the fact $\models (26)^{\circ}$ and applying Lemma 3.5 $k$ times, we find that the following formula is uniformly valid:

$$\left(\begin{array}{l}\underset{|}{\circ}\Big((\overset{|}{\circ}X_1 \wedge \overset{|}{\circ}Y_1) \wedge \neg(Z_1 \sqcup T_1)\Big) \quad \vee \ldots \vee\\ \underset{|}{\circ}\Big((\overset{|}{\circ}X_k \wedge \overset{|}{\circ}Y_k) \wedge \neg(Z_k \sqcup T_k)\Big) \quad \vee\\ \underset{|}{\circ}\Big(\overset{|}{\circ}(\overset{|}{\circ}P_1 \rightarrow Q_1) \wedge \neg R_1\Big) \quad \vee \ldots \vee\\ \underset{|}{\circ}\Big(\overset{|}{\circ}(\overset{|}{\circ}P_k \rightarrow Q_k) \wedge \neg R_k\Big)\end{array}\right) \vee \quad W. \tag{27}$$

Next, due to the presence of ⫯-contraction, **Al** clearly proves

$$\text{⫰}(\text{⫰}P_i \rightarrow Q_i) \;\rightarrow\; \text{⫰}(\text{⫰}P_i^1 \rightarrow Q_i^1) \wedge \ldots \wedge \text{⫰}(\text{⫰}P_i^n \rightarrow Q_i^n)$$

(any $1 \leq i \leq k$) and, by soundness, the above formula is uniformly valid. Therefore, starting from $\models (27)^\circ$ and applying Lemma 3.5 $k$ times, we find that the following formula is uniformly valid:

$$\left(\begin{array}{l} \text{⫯}\Big((\text{⫰}X_1 \wedge \text{⫰}Y_1) \wedge \neg(Z_1 \sqcup T_1)\Big) \quad \vee \ldots \vee \\ \text{⫯}\Big((\text{⫰}X_k \wedge \text{⫰}Y_k) \wedge \neg(Z_k \sqcup T_k)\Big) \quad \vee \\ \text{⫯}\Big(\big(\text{⫰}(\text{⫰}P_1^1 \rightarrow Q_1^1) \wedge \ldots \wedge \text{⫰}(\text{⫰}P_1^n \rightarrow Q_1^n)\big) \wedge \neg R_1\Big) \quad \vee \ldots \vee \\ \text{⫯}\Big(\big(\text{⫰}(\text{⫰}P_k^1 \rightarrow Q_k^1) \wedge \ldots \wedge \text{⫰}(\text{⫰}P_k^n \rightarrow Q_k^n)\big) \wedge \neg R_k\Big) \end{array}\right) \vee \;\; W. \tag{28}$$

Next, consider the formula

$$\left(\begin{array}{l} \text{⫯}\Big((X_1 \wedge Y_1) \wedge \neg(Z_1 \sqcup T_1)\Big) \quad \vee \ldots \vee \\ \text{⫯}\Big((X_k \wedge Y_k) \wedge \neg(Z_k \sqcup T_k)\Big) \quad \vee \\ \text{⫯}\Big(\big((\text{⫰}P_1^1 \rightarrow Q_1^1) \wedge \ldots \wedge (\text{⫰}P_1^n \rightarrow Q_1^n)\big) \wedge \neg R_1\Big) \quad \vee \ldots \vee \\ \text{⫯}\Big(\big((\text{⫰}P_k^1 \rightarrow Q_k^1) \wedge \ldots \wedge (\text{⫰}P_k^n \rightarrow Q_k^n)\big) \wedge \neg R_k\Big) \end{array}\right) \vee \;\; W. \tag{29}$$

Observe that (29) is nothing but the result of deleting ⫰ before every subformula $X_i$, $Y_i$ and $(\text{⫰}P_i^p \rightarrow Q_i^p)$ in (28). And, of course, for any (such sub)formula $E$, $\mathbf{Al} \vdash \text{⫰}E \rightarrow E$ and hence $\Vdash \text{⫰}E \rightarrow E$. Therefore, starting from $\models (28)^\circ$ and applying Lemma 3.5 $k+k+nk$ times, we infer that $\models (29)^\circ$. But (29) is a disabbreviation of (22), and thus the desired $\models (22)^\circ$ holds. □

# 9 Main lemma

**Lemma 9.1** *Assume $\underline{G} \Rightarrow K$ is a standard* **Int***-sequent, $\mathcal{K}$ is a Kripke model of size $n$ with $\mathcal{K} \models \underline{G}$ and $\mathcal{K} \not\models W$, and $D$ is the $n$-desequentization of $\underline{G} \Rightarrow W$. Then $D$ is not valid; specifically, there is an interpretation $^\circ$ of complexity $\bigsqcup\Sigma_1^B$ such that $\not\models D^\circ$.*

The present long section is entirely devoted to a proof of this lemma. It will be seen in the last, very short section of this paper that Lemma 9.1, in combination with some earlier-proven lemmas, almost immediately implies the sought completeness of **Int** with respect to the semantics of computability logic.

## 9.1 Main claim

Let us get started with our proof of Lemma 9.1. We pick and fix an arbitrary standard **Int**-sequent

$$\left.\begin{array}{l} X_1 \circ\!\!-\, (Y_1 \circ\!\!-\, Z_1 \sqcup T_1),\ \ldots,\ X_k \circ\!\!-\, (Y_k \circ\!\!-\, Z_k \sqcup T_k), \\ (P_1 \circ\!\!-\, Q_1) \circ\!\!-\, R_1,\ \ldots,\ (P_k \circ\!\!-\, Q_k) \circ\!\!-\, R_k \end{array}\right\} \ \Rightarrow\ \ W \qquad (30)$$

abbreviated as $\underline{G} \Rightarrow W$, a Kripke model $\mathcal{K} = (\mathcal{W}, \mathcal{R}, \models)$ of size $n$ and assume that $\mathcal{K} \models \underline{G}$ and $\mathcal{K} \not\models W$. Let us agree for the rest of this section that:

- $j$ exclusively ranges over $1, \ldots, k$;
- $i$ exclusively ranges over $1, \ldots, 2k$;
- $p, q$ exclusively range over $1, \ldots, n$;
- $w, u$ exclusively range over bit strings.

Next, for each $j$ and $p$, we let

$$P_j^p = P_j \ \text{ and } \ Q_j^p = Q_j.$$

Then the $n$-desequentization of (30) is

$$\begin{array}{l} \overset{\mid}{\circ}\big(X_1 \wedge Y_1 \rightarrow Z_1 \sqcup T_1\big)\ \wedge\ \ldots\ \wedge\ \overset{\mid}{\circ}\big(X_k \wedge Y_k \rightarrow Z_k \sqcup T_k\big)\ \wedge \\ \overset{\mid}{\circ}\big((\overset{\mid}{\circ}P_1^1 \rightarrow Q_1^1) \wedge \ldots \wedge (\overset{\mid}{\circ}P_1^n \rightarrow Q_1^n) \rightarrow R_1\big)\ \wedge\ \ldots\ \wedge \qquad\qquad \rightarrow\ \ W. \qquad (31) \\ \overset{\mid}{\circ}\big((\overset{\mid}{\circ}P_k^1 \rightarrow Q_k^1) \wedge \ldots \wedge (\overset{\mid}{\circ}P_k^n \rightarrow Q_k^n) \rightarrow R_k\big) \end{array}$$

Thus, our goal is to find a counterinterpretation (of complexity $\bigsqcup\Sigma_1^B$) for the formula (31), with a *counterinterpretation* here meaning an interpretation $^\circ$ such that $\not\models (31)^\circ$.

Remember, from Section 7 of [10], the distinction between *general* and *elementary* letters (=atoms when 0-ary). Elementary letters are to be interpreted as predicates, while general letters can be interpreted as arbitrary static games. The atoms of the languages of **Int** and **Al**, as we know, are general rather than elementary. However, when it comes to interpretations, formulas with only elementary letters — called *elementary-base* formulas — are both technically and intuitively easier to deal with than those with general letters. For this reason, we are going to replace (31) with the elementary-base — though no longer propositional — formula (32) of the same form as (31), and then construct a counterinterpretation for (32) rather than (31).

In particular, for each (general, 0-ary) nonlogical atom $A$ of the language of **Int**, we fix a unique 1-ary elementary letter $\dot{A}$ — unique in the sense that whenever $A \neq B$, we also have $\dot{A} \neq \dot{B}$. We also fix a variable $x$ and, for each atom $A$ of the language of **Int**, agree on the abbreviation $\breve{A}$ defined by

$$\breve{A} \ = \ \bigsqcup x \dot{A}(x).$$

Now, the above-mentioned elementary-base formula (32) is simply obtained from (31) through replacing every atom $A$ by $\breve{A}$:

$$\begin{array}{l} ⫰\big(\breve{X}_1 \wedge \breve{Y}_1 \rightarrow \breve{Z}_1 \sqcup \breve{T}_1\big) \ \wedge \ \ldots \ \wedge \ ⫰\big(\breve{X}_k \wedge \breve{Y}_k \rightarrow \breve{Z}_k \sqcup \breve{T}_k\big) \ \wedge \\ ⫰\big((⫰\breve{P}^1_1 \rightarrow \breve{Q}^1_1) \wedge \ldots \wedge (⫰\breve{P}^n_1 \rightarrow \breve{Q}^n_1) \rightarrow \breve{R}_1\big) \ \ \wedge \ \ldots \ \wedge \\ ⫰\big((⫰\breve{P}^1_k \rightarrow \breve{Q}^1_k) \wedge \ldots \wedge (⫰\breve{P}^n_k \rightarrow \breve{Q}^n_k) \rightarrow \breve{R}_k\big) \end{array} \quad \rightarrow \quad \breve{W}. \qquad (32)$$

The language in which (32) is written is a sublanguage of the otherwise more expressive first-order language of Section 7 of [10]. Specifically, every atom of this sublanguage is nonlogical, elementary and 1-ary, and the only variable that may occur in formulas is $x$. Hence, the meanings of the terms "(admissible) interpretation", "valid" and "uniformly valid", when used in this new context, should be understood as defined in [10] which, of course, are fully consistent with the meanings of these terms as (re)defined in the previous sections of this paper for the languages of **Int** and **Al**. For simplicity, we agree that, in contexts dealing with (32), by just "interpretation" we always mean (32)-admissible interpretation (all but some pathological interpretations would be automatically (32)-admissible, anyway). And let us agree to say that such an interpretation $^\star$ is **of complexity** $\Sigma^B_1$ iff, for every atom $A$ of the language of **Int**, the game $\dot{A}^\star(x)$ is in $\Sigma^B_1$ (see Subsection 2.4).

We label the following statement and subsequent similar statements "claims" rather than "lemmas" because they are true only in our particular context, set by the assumptions that we have already made within the present proof of Lemma 9.1.

**Claim 9.2** *There is an interpretation* $^\star$ *of complexity* $\Sigma^B_1$ *such that* $\not\models (32)^\star$.

Before we attempt to prove Claim 9.2, let us see that it implies the main Lemma 9.1. According to this claim, we have $\not\models (32)^\star$, where $^\star$ is a certain interpretation sending every atom of (32) to a Boolean combination of $\Sigma_1$-predicates. Let now $^\circ$ be the interpretation that sends every atom $A$ of the language of **Int** to $\breve{A}^\star$. It is easy to see that then $(31)^\circ = (32)^\star$, so that $\not\models (31)^\circ$. And clearly $^\circ$ is of the desired complexity $\bigsqcup\Sigma^B_1$, because $A^\circ = \breve{A}^\star = \bigsqcup x \dot{A}^\star(x)$. As (31) is the $n$-desequentization of (30), we find Lemma 9.1 proven.

So, the "only" remaining duty within our proof of the main Lemma 9.1 is to prove Claim 9.2. The rest of this section is solely devoted to that task.

## 9.2 Terminology and notation

Note that, since each atom of (32) is to be interpreted as an elementary game, the structure (**Lr** component) of the game $(32)^\star$ does not depend on the selection of an interpretation $^\star$. This nice property of elementary-base formulas was one of our reasons for choosing to deal with (32) instead of (31). In many contexts, it allows us to terminologically treat (32) as if it was a game, even though, strictly speaking, it is just a formula, and becomes a game only after an interpretation is applied to it. Namely, we can and will unambiguously say "legal run of (32)", meaning "legal run of $(32)^\star$ for some (= every) interpretation $^\star$".

We will often need to differentiate between subformulas of (31) or (32) and particular occurrences of such. It should be remembered that the expressions "$P^p_j$", "$\breve{P}^p_j$", "$\breve{X}_j \wedge \breve{Y}_j$", etc. are metaexpressions, denoting subformulas of (31) or (32). As it happens, for different occurrences of subformulas of (31) or (32) we have chosen different metaexpressions, so those occurrences can be safely identified with the corresponding metaexpressions. To avoid possible notational confusions, we will write "$\lfloor P^p_j \rfloor$", "$\lfloor \breve{P}^p_j \rfloor$", "$\lfloor \breve{X}_j \wedge \breve{Y}_j \rfloor$", etc. to indicate that we mean **metaexpressions** (= particular occurrences of subformulas) rather than the formulas for which those expressions stand. So, say, when we write $\breve{X}_i = \breve{X}_j$ or $\breve{X}_i = \breve{Z}_j$, we mean that $\breve{X}_i$ and $\breve{X}_j$, or $\breve{X}_i$ and $\breve{Z}_j$, are identical as formulas; on the other hand, by writing $\lfloor \breve{X}_i \rfloor = \lfloor \breve{X}_j \rfloor$ we will mean that the two metaexpressions "$\breve{X}_i$" and "$\breve{X}_j$" are graphically identical, i.e., that $\breve{X}_i$ and $\breve{X}_j$ stand for the same occurrence of the same subformula of (32), which implies that $i = j$. And, as the expressions "$\breve{X}$" and "$\breve{Z}$" are graphically different from each other, we would never have $\lfloor \breve{X}_i \rfloor = \lfloor \breve{Z}_j \rfloor$, no matter what $i$ and $j$ are.

Consider any particular legal position or run $\Gamma$ of (32). Since (32) is a $\rightarrow$-combination of games, every move of $\Gamma$ has the form $1.\alpha$ or $2.\alpha$. Intuitively, $1.\alpha$ means the move $\alpha$ made in the antecedent of (32), and $2.\alpha$ the move $\alpha$ made in the consequent. Correspondingly, we think of $\Gamma$ as consisting of two subruns which, using the notational conventions of Subsection 4.3 of [10], are denoted by $\Gamma^{1.}$ and $\Gamma^{2.}$. We will be referring to $\Gamma^{2.}$ as the **$\Gamma$-residual position of** $\lfloor \breve{W} \rfloor$, because, intuitively, $\Gamma^{2.}$ is what remains of $\Gamma$ after discarding in it everything but the part that constitutes a run in the $\lfloor \breve{W} \rfloor$ component of (32). Note that we wrote $\lfloor \breve{W} \rfloor$ here. Using just $\breve{W}$ instead could have been ambiguous, for $\breve{W}$, as a formula, may (and probably does) have many occurrences in (32), while $\lfloor \breve{W} \rfloor$ refers to the occurrence of that formula in the consequent and only there. Also, we said "position" rather than "run". It is safe to do so because $\Gamma^{2.}$, which has to be a legal run of the game $\breve{W} = \bigsqcup x \dot{W}(x)$ (for otherwise $\Gamma$ would not be a legal run of (32)), contains at most one labmove — namely, it is $\langle\rangle$ or $\langle \top a \rangle$ for some constant $a$.

As for $\Gamma^{1.}$, it is a legal run of the negation of the antecedent of (32) rather than the antecedent itself. That is so because, as we remember, a game $A \rightarrow B$ is defined as $\neg A \vee B$. And this means nothing but that $\neg\Gamma^{1.}$ is a legal run of the antecedent of (32). Accordingly, we will be interested in $\neg\Gamma^{1.}$ rather than $\Gamma^{1.}$, because we find it more convenient to see the antecedent of (32) as it is, without a negation. The antecedent of (32), in turn, is a $\wedge$-conjunction, and we think of $\neg\Gamma^{1.}$ as consisting of as many subruns as the number of conjuncts. Namely, each such subrun is $\neg\Gamma^{1.i.}$ for some $i$. If here $i \in \{1, \ldots, k\}$, we call $\neg\Gamma^{1.i.}$ the **$\Gamma$-residual run of** $\lfloor \circ\!\!\mid (\breve{X}_i \wedge \breve{Y}_i \rightarrow \breve{Z}_i \sqcup \breve{T}_i) \rfloor$, and if $i = k + j$, we call $\neg\Gamma^{1.i.}$ the **$\Gamma$-residual run of**

$$\lfloor \circ\!\!\mid \big( (\circ\!\!\mid \breve{P}^1_j \rightarrow \breve{Q}^1_j) \wedge \ldots \wedge (\circ\!\!\mid \breve{P}^n_j \rightarrow \breve{Q}^n_j) \;\rightarrow\; \breve{R}_j \big) \rfloor.$$

As in the case of $\lfloor \breve{W} \rfloor$, such names correspond to the intuitive meanings of $\neg\Gamma^{1.i.}$. For example, where $1 \leq j \leq k$, $\neg\Gamma^{1.j.}$ can be characterized as the part of $\Gamma$ that constitutes a run in the $\lfloor \circ\!\!\mid (\breve{X}_j \wedge \breve{Y}_j \rightarrow \breve{Z}_j \sqcup \breve{T}_j) \rfloor$ component. Such a run should be a legal run of the game (represented by) $\circ\!\!\mid (\breve{X}_j \wedge \breve{Y}_j \rightarrow \breve{Z}_j \sqcup \breve{T}_j)$, for otherwise $\Gamma$ would not be a legal run of (32).

Assume $\Psi$ is the $\Gamma$-residual run of $\lfloor \circ\!\!\mid(\breve{X}_j \wedge \breve{Y}_j \rightarrow \breve{Z}_j \sqcup \breve{T}_j)\rfloor$. We will be referring to the bitstring tree $\mathit{Tree}^{\circ\!\mid(\breve{X}_j\wedge\breve{Y}_j\rightarrow\breve{Z}_j\sqcup\breve{T}_j)}\langle\Psi\rangle$ (see Subsection 4.6 of [10]) as the **$\Gamma$-residual $\lfloor \circ\!\!\mid(\breve{X}_j \wedge \breve{Y}_j \rightarrow \breve{Z}_j \sqcup \breve{T}_j)\rfloor$-tree**. Intuitively, this is the underlying BT structure of the subrun of $\Gamma$ that is taking place in the $\lfloor \circ\!\!\mid(\breve{X}_j \wedge \breve{Y}_j \rightarrow \breve{Z}_j \sqcup \breve{T}_j)\rfloor$ component of (32). Then the run $\Psi$ is further thought of as consisting of multiple legal runs of $\breve{X}_j \wedge \breve{Y}_j \rightarrow \breve{Z}_j \sqcup \breve{T}_j$, specifically, the run $\Psi^{\preceq w}$ (again, see Subsection 4.6 of [10]) for each complete branch $w$ of the $\Gamma$-residual $\lfloor \circ\!\!\mid(\breve{X}_j \wedge \breve{Y}_j \rightarrow \breve{Z}_j \sqcup \breve{T}_j)\rfloor$-tree. Notice that such a $\Psi^{\preceq w}$, as a legal run of $\breve{X}_j \wedge \breve{Y}_j \rightarrow \breve{Z}_j \sqcup \breve{T}_j$, would be finite, containing at most 4 labmoves. Hence we can refer to it as "position" rather than "run". We call such a position $\Psi^{\preceq w}$ the **$\Gamma$-residual position of $\lfloor \breve{X}_j \wedge \breve{Y}_j \rightarrow \breve{Z}_j \sqcup \breve{T}_j\rfloor^w$**. The **$\Gamma$-residual**

$$\lfloor \circ\!\!\mid\big((\circ\!\!\mid\breve{P}_j^1 \rightarrow \breve{Q}_j^1) \wedge \ldots \wedge (\circ\!\!\mid\breve{P}_j^n \rightarrow \breve{Q}_j^n) \;\rightarrow\; \breve{R}_j\big)\rfloor\textbf{-tree}$$

and the **$\Gamma$-residual run of**

$$\lfloor((\circ\!\!\mid\breve{P}_j^1 \rightarrow \breve{Q}_j^1) \wedge \ldots \wedge (\circ\!\!\mid\breve{P}_j^n \rightarrow \breve{Q}_j^n) \;\rightarrow\; \breve{R}_j\rfloor^w$$

(where $w$ is a complete branch of that tree) are defined similarly. In this case, for safety, we say *run* rather than *position*, for the game (represented by) $(\circ\!\!\mid\breve{P}_j^1 \rightarrow \breve{Q}_j^1) \wedge \ldots \wedge (\circ\!\!\mid\breve{P}_j^n \rightarrow \breve{Q}_j^n) \;\rightarrow\; \breve{R}_j$ is not finite-depth because of its $\circ\!\!\mid\breve{P}_j^p$-components, and hence the corresponding $\Psi^{\preceq w}$ may be infinite.

Assume $w$ is a complete branch of the $\Gamma$-residual $\lfloor \circ\!\!\mid(\breve{X}_j \wedge \breve{Y}_j \rightarrow \breve{Z}_j \sqcup \breve{T}_j)\rfloor$-tree, and $\Theta$ is the $\Gamma$-residual position of $\lfloor \breve{X}_j \wedge \breve{Y}_j \rightarrow \breve{Z}_j \sqcup \breve{T}_j\rfloor^w$. Such a $\Theta$ is thought of as consisting of two subpositions: $\neg\Theta^{1.}$ and $\Theta^{2.}$. We respectively refer to these as the **$\Gamma$-residual position of $\lfloor \breve{X}_j \wedge \breve{Y}_j\rfloor^w$** and the **$\Gamma$-residual position of $\lfloor \breve{Z}_j \sqcup \breve{T}_j\rfloor^w$**. In turn, $\neg\Theta^{1.}$ is further seen as consisting of two subpositions $\neg\Theta^{1.1.}$ and $\neg\Theta^{1.2.}$, to which we respectively refer as the **$\Gamma$-residual position of $\lfloor \breve{X}_j\rfloor^w$** and the **$\Gamma$-residual position of $\lfloor \breve{Y}_j\rfloor^w$**. Similarly, if $w$ is a complete branch of the $\Gamma$-residual $\lfloor \circ\!\!\mid\big((\circ\!\!\mid\breve{P}_j^1 \rightarrow \breve{Q}_j^1) \wedge \ldots \wedge (\circ\!\!\mid\breve{P}_j^n \rightarrow \breve{Q}_j^n) \rightarrow \breve{R}_j\big)\rfloor$-tree and $\Theta$ is the $\Gamma$-residual run of

$$\lfloor(\circ\!\!\mid\breve{P}_j^1 \rightarrow \breve{Q}_j^1) \wedge \ldots \wedge (\circ\!\!\mid\breve{P}_j^n \rightarrow \breve{Q}_j^n) \rightarrow \breve{R}_j\rfloor^w,$$

we respectively refer to

1. $\neg\Theta^{1.}$,
2. $\Theta^{2.}$,
3. $\neg\Theta^{1.p.}$,
4. $\Theta^{1.p.1.}$ and
5. $\neg\Theta^{1.p.2.}$

as the **$\Gamma$-residual runs of**

1. $\lfloor(\circ\!\!\mid\breve{P}_j^1 \rightarrow \breve{Q}_j^1) \wedge \ldots \wedge (\circ\!\!\mid\breve{P}_j^n \rightarrow \breve{Q}_j^n)\rfloor^w$,

2. $\lfloor \breve{R}_j \rfloor^w$,

3. $\lfloor ⫰\breve{P}^p_j \rightarrow \breve{Q}^p_j \rfloor^w$,

4. $\lfloor ⫰\breve{P}^p_j \rfloor^w$ and

5. $\lfloor \breve{Q}^p_j \rfloor^w$,

respectively (in the cases of $\lfloor \breve{R}_j \rfloor^w$ and $\lfloor \breve{Q}^p_j \rfloor^w$ we can always say "position" instead of "run", of course).

Assume $w$ is a complete branch of the $\Gamma$-residual $\lfloor ⫰\big((⫰\breve{P}^1_j \rightarrow \breve{Q}^1_j) \wedge \ldots \wedge (⫰\breve{P}^n_j \rightarrow \breve{Q}^n_j) \rightarrow \breve{R}_j\big)\rfloor$-tree, and $\Upsilon$ is the $\Gamma$-residual run of $\lfloor ⫰\breve{P}^p_j \rfloor^w$. We call $\mathit{Tree}^{⫰\breve{P}^p_j}\langle\Upsilon\rangle$ the **$\Gamma$-residual $\lfloor ⫰\breve{P}^p_j \rfloor^w$-tree**. $\Upsilon$ is then further seen as consisting of multiple legal positions of $\breve{P}^p_j$, specifically, the position $\Upsilon^{\preceq u}$ for each complete branch $u$ of the $\Gamma$-residual $\lfloor ⫰\breve{P}^p_j \rfloor^w$-tree. We refer to such a position $\Upsilon^{\preceq u}$ as the **$\Gamma$-residual position of** $\lfloor \breve{P}^p_j \rfloor^w_u$.

In the above terminological conventions we have started writing "$\lfloor \breve{X}_j \wedge \breve{Y}_j \rfloor^w$", "$\lfloor \breve{P}^p_j \rfloor^w_u$", etc. Formally these, just like simply "$\lfloor \breve{X}_i \wedge \breve{Y}_i \rfloor$" or "$\lfloor \breve{P}_i \rfloor$", are *metaexpressions*. If, say, we write $\lfloor \breve{X}_i \rfloor^w = \lfloor \breve{X}_j \rfloor^u$, we imply that the two components are graphically the same, here meaning that $i = j$ and $w = u$. Note that such metaexpressions would not always be finite. For instance, $\lfloor \breve{X}_i \rfloor^w$ would be infinite if the bitstring $w$ is so; this, however, can only be the case when $\Gamma$ is an infinite run.

What we call the **residual molecules of** $\Gamma$, or simply **$\Gamma$-molecules**, are the following metaexpressions:

- $\lfloor \breve{W} \rfloor$;
- $\lfloor \breve{X}_j \rfloor^w$, $\lfloor \breve{Y}_j \rfloor^w$ and $\lfloor \breve{Z}_j \sqcup \breve{T}_j \rfloor^w$ for each $j$ and each complete branch $w$ of the $\Gamma$-residual $\lfloor ⫰(\breve{X}_j \wedge \breve{Y}_j \rightarrow \breve{Z}_j \sqcup \breve{T}_j)\rfloor$-tree.
- $\lfloor \breve{Q}^p_j \rfloor^w$ and $\lfloor \breve{R}_j \rfloor^w$ for each $j$, $p$ and each complete branch $w$ of the $\Gamma$-residual $\lfloor ⫰\big((⫰\breve{P}^1_j \rightarrow \breve{Q}^1_j) \wedge \ldots \wedge (⫰\breve{P}^n_j \rightarrow \breve{Q}^n_j) \rightarrow \breve{R}_j\big)\rfloor$-tree.
- $\lfloor \breve{P}^p_j \rfloor^w_u$ for each $j, p$, each complete branch $w$ of the $\Gamma$-residual $\lfloor ⫰\big((⫰\breve{P}^1_j \rightarrow \breve{Q}^1_j) \wedge \ldots \wedge (⫰\breve{P}^n_j \rightarrow \breve{Q}^n_j) \rightarrow \breve{R}_j\big)\rfloor$-tree and each complete branch $u$ of the $\Gamma$-residual $\lfloor ⫰\breve{P}^p_j \rfloor^w$-tree.

We may say just "**molecule**" instead of "$\Gamma$-molecule" when $\Gamma$ is fixed in a given context or is irrelevant.

We say that the **types** of the above 7 sorts of molecules are $W$, $X_j$, $Y_j$, $Z_j \sqcup T_j$, $Q_j$, $R_j$ and $P_j$, respectively. When irrelevant, the index $j$ can be omitted here. We differentiate between types and what we call **metatypes**. The metatype of $\lfloor \breve{W} \rfloor$ is the metaexpression $\lfloor W \rfloor$; the metatype of $\lfloor \breve{P}^p_j \rfloor^w_u$ is the metaexpression $\lfloor P^p_j \rfloor$; the metatype of $\lfloor \breve{Z}_j \sqcup \breve{T}_j \rfloor^w$ is the metaexpression $\lfloor Z_j \sqcup T_j \rfloor$; the metatype of $\lfloor \breve{X}_j \rfloor^w$ is the metaexpression $\lfloor X_j \rfloor$, and similarly for $\lfloor \breve{Y}_j \rfloor^w$, $\lfloor \breve{Q}^p_j \rfloor^w$, $\lfloor \breve{R}_j \rfloor^w$. As in the case of types, the

indices $j$ and/or $p$ can be omitted here when irrelevant. Notice that if two molecules have different types, then they also have different metatypes, but not vice versa: for instance, $\lfloor \check{X}_j \rfloor^w$ and $\lfloor \check{Y}_j \rfloor^v$ would always have different metatypes (because "$X$"$\neq$"$Y$"), but their types may be identical, meaning that so are the atoms of (31) for the occurrences of which $X_j$ and $Y_j$ stand. And, where $q \neq p$, the two molecules $\lfloor Q_j^p \rfloor^w$ and $\lfloor Q_j^q \rfloor^w$ would always have identical types but different metatypes.

$\lfloor W \rfloor$-, $\lfloor X \rfloor$-, $\lfloor Y \rfloor$- and $\lfloor Q \rfloor$-metatype molecules are said to be **positive**; and $\lfloor Z \sqcup T \rfloor$-, $\lfloor R \rfloor$- and $\lfloor P \rfloor$-metatype molecules are said to be **negative**. When one of two molecules $M_1, M_2$ is positive and the other is negative, we say that $M_1$ and $M_2$ have **opposite genders**. Mark the fact that, for a $\Gamma$-residual molecule $M$, the $\Gamma$-residual position of $M$ is always $\langle\rangle$ or $\langle \top a \rangle$ for some constant $a$ unless $M$ is $\lfloor Z \sqcup T \rfloor$-metatype, in which case the $\Gamma$-residual position of $M$ is always $\langle\rangle$, $\langle \top d \rangle$ or $\langle \top d, \top a \rangle$ for some constant $a$ and some $d \in \{1, 2\}$. Note that here, when $M$ is negative, the moves (if any) are made by player $\bot$ even though they are $\top$-labeled.

Let $M$ be a $\Gamma$-molecule, and $\Phi$ the $\Gamma$-residual position of $M$. To what we will be referring as the **content of** $M$ **in** $\Gamma$, or the **$\Gamma$-content of** $M$, or **the content of** $M$ **as a $\Gamma$-residual molecule**, or — when $\Gamma$ is clear from the context or is irrelevant — simply the **content of** $M$, is the formula $C$ such that:

- If the type of $M$ is $A$ with $A \in \{W, X_j, Y_j, Q_j, R_j, P_j \mid 1 \leq j \leq k\}$, then:
  - if $\Phi = \langle\rangle$, then $C = \bigsqcup x \dot{A}(x)$;
  - if $\Phi = \langle \top a \rangle$, then $C = \dot{A}(a)$.
- If the type of $M$ is $Z_j \sqcup T_j$, then:
  - if $\Phi = \langle\rangle$, then $C = \bigsqcup x \dot{Z}_j(x) \sqcup \bigsqcup x \dot{T}_j(x)$;
  - if $\Phi = \langle \top 1 \rangle$, then $C = \bigsqcup x \dot{Z}_j(x)$;
  - if $\Phi = \langle \top 1, \top a \rangle$, then $C = \dot{Z}_j(a)$;
  - if $\Phi = \langle \top 2 \rangle$, then $C = \bigsqcup x \dot{T}_j(x)$;
  - if $\Phi = \langle \top 2, \top a \rangle$, then $C = \dot{T}_j(a)$.

Note that two molecules may have identical contents even if their metatypes (but not types!) are different.

Let $M$ be a $\Gamma$-residual molecule, and $C$ its content in $\Gamma$. We say that $M$, as a $\Gamma$-residual molecule (or in $\Gamma$) is:

- **grounded** iff $C$ is of the form $\dot{A}(a)$ (some constant $a$ and atom $A$ of (30));
- $\bigsqcup x$**-contentual** iff $C$ is of the form $\bigsqcup x \dot{A}(x)$;
- $\sqcup$**-contentual** iff $C$ is of the form $\bigsqcup x \dot{A}(x) \sqcup \bigsqcup x \dot{B}(x)$.

Intuitively, the content of a given $\Gamma$-molecule $M$ is the game to which the corresponding subgame of (32) has evolved as a result of the moves of $\Gamma$ made within the $M$ component. For example, for the molecule $\lfloor \breve{W} \rfloor$, being $\bigsqcup x$-contentual means that the moves of $\Gamma$ have not affected this component, so that, as a (sub)game, it remains $\bigsqcup x\dot{W}(x)$; and being grounded means that the moves of $\Gamma$ — in fact, one of such moves — has brought the game $\bigsqcup x\dot{W}(x)$ down to $\dot{W}(a)$ for some constant $a$. Similarly for molecules whose metatypes are $\lfloor X \rfloor$, $\lfloor Y \rfloor$, $\lfloor P \rfloor$, $\lfloor Q \rfloor$, $\lfloor R \rfloor$. On the other hand, for $\lfloor Z \sqcup T \rfloor$-metatype molecules, it is being $\sqcup$-contentual rather than $\bigsqcup x$-contentual that means not having been affected by the moves of $\Gamma$.

We say that a $\Gamma$-molecule $M_1$ is **matchingly grounded** (in $\Gamma$) iff $M_1$ is grounded and there is another grounded $\Gamma$-molecule $M_2$ such that $M_1$ and $M_2$ have opposite genders but identical contents. If $M_1$ is grounded but not matchingly so, then we say that it is **non-matchingly grounded**.

Note that, for a $\Gamma$-residual molecule $M$, the content of $M$, as well as whether $M$ is (matchingly) grounded, $\bigsqcup x$-contentual or $\sqcup$-contentual, depends on $\Gamma$. That is why, unless $\Gamma$ is fixed or clear from the context, for safety we should say "the content of $M$ in $\Gamma$" instead of just "the content of $M$", say "$M$ is grounded in $\Gamma$" or "$M$, as a $\Gamma$-molecule, is grounded" instead of just "$M$ is grounded", etc. The point is that the same metaexpression $M$ can be a residual molecule of two different runs. Relevant to our interests are only the cases when one run, say $\Gamma_1$, is an initial segment of the other run, say $\Gamma_2$. Assume this is so for the rest of the present paragraph, and assume $M$ is a $\Gamma_1$-molecule. If $M = \lfloor \breve{W} \rfloor$, $M$ will also be a $\Gamma_2$-molecule. It is possible, however, that $M$ is $\bigsqcup x$-contentual in $\Gamma_1$ while grounded in $\Gamma_2$; and if so, the content of $M$ in $\Gamma_1$ will be different from that in $\Gamma_2$. And it is also generally possible that $M$ is non-matchingly grounded in $\Gamma_1$ while matchingly grounded in $\Gamma_2$. Similarly when the metatype of $M$ is anything other than $\lfloor W \rfloor$. However, when $M$ is a non-$\lfloor W \rfloor$-metatype $\Gamma_1$-molecule such as, say, $\lfloor \breve{X}_j \rfloor^w$, then there is no guarantee that $M$ is also a $\Gamma_2$-molecule. For instance, if ($\Gamma_2 \neq \Gamma_1$ and) in position $\Gamma_1$ the player $\top$ made a replicative move in the $\lfloor ⫰(\breve{X}_j \wedge \breve{Y}_j \rightarrow \breve{Z}_j \sqcup \breve{T}_j) \rfloor$ component of the game which split (i.e. extended to $w0$ and $w1$) the leaf $w$ of the $\Gamma_1$-residual $\lfloor ⫰(\breve{X}_j \wedge \breve{Y}_j \rightarrow \breve{Z}_j \sqcup \breve{T}_j) \rfloor$-tree, then $w$ would be just an internal node rather than a complete branch of the $\Gamma_2$-residual $\lfloor ⫰(\breve{X}_j \wedge \breve{Y}_j \rightarrow \breve{Z}_j \sqcup \breve{T}_j) \rfloor$-tree, meaning that $M = \lfloor \breve{X}_j \rfloor^w$ is not a residual molecule of $\Gamma_2$. Instead of $\lfloor \breve{X}_j \rfloor^w$, in the general case, $\Gamma_2$ could have many residual molecules of the form $\lfloor \breve{X}_j \rfloor^{w'}$, where $w \preceq w'$. Let us agree to say about each such $\Gamma_2$-molecule $\lfloor \breve{X}_j \rfloor^{w'}$ that it **descends** from $\lfloor \breve{X}_j \rfloor^w$, or that $\lfloor \breve{X}_j \rfloor^w$ is **the $\Gamma_1$-predecessor** — or simply **a predecessor** if we do not care about details — of $\lfloor \breve{X}_j \rfloor^{w'}$. Similarly for the cases when $M$ is $\lfloor \breve{Y}_j \rfloor^w$, $\lfloor \breve{Z}_j \sqcup \breve{T}_j \rfloor^w$, $\lfloor \breve{Q}_j^p \rfloor^w$ or $\lfloor \breve{R}_j \rfloor^w$. And rather similarly for the case $M = \lfloor \breve{P}_j^p \rfloor_u^w$: in this case $M$ will be said to be the **$\Gamma_1$-predecessor** (or just **a predecessor**) of every $\Gamma_2$-molecule $\lfloor \breve{P}_j^p \rfloor_{u'}^{w'}$ such that $w \preceq w'$ and $u \preceq u'$; and, correspondingly, every such $\lfloor \breve{P}_j^p \rfloor_{u'}^{w'}$ will be said to **descend** from $\lfloor \breve{P}_j^p \rfloor_u^w$. Extending this terminology to the remaining case of $M = \lfloor \breve{W} \rfloor$, the latter is always its own (single) **descendant** and **predecessor**.

Here comes some more terminology. In the context of a given legal position $\Gamma$ of (32), where $M$ is a $\bigsqcup x$-contentual $\Gamma$-molecule, to **ground** $M$ intuitively means to make

a (legal) move that makes $M$ grounded; we say that such a move is **patient** if $M$ is the only molecule whose content it modifies. Informally speaking, a patient move for a non-$\lfloor W \rfloor$-metatype molecule $M$ means that the move is made in the corresponding leaf (or two nested leaves if $M$ is $\lfloor P \rfloor$-metatype) rather than internal node(s) of the corresponding underlying BT(s), for a move in an internal node $v$ would simultaneously affect several molecules — all those that are associated with leaves $r$ such that $v \preceq r$. In precise terms, we have:

- To ground $\lfloor \breve{W} \rfloor$ means to make the move $2.a$ for some constant $a$. This sort of a move is automatically patient.
- To ground $\lfloor \breve{X}_j \rfloor^w$ (resp. $\lfloor \breve{Y}_j \rfloor^w$, resp. $\lfloor \breve{Z}_j \sqcup \breve{T}_j \rfloor^w$) means to make the move $1.j.w'.1.1.a$ (resp. $1.j.w'.1.2.a$, resp. $1.j.w'.2.a$) for some bitstring $w' \preceq w$ and some constant $a$. Such a move is patient iff $w' = w$.
- To ground $\lfloor \breve{R}_j \rfloor^w$ means to make the move $1.i.w'.2.a$ for $i = k+j$, some bitstring $w' \preceq w$ and some constant $a$. Again, such a move is patient iff $w' = w$.
- To ground $\lfloor \breve{Q}^p_j \rfloor^w$ means to make the move $1.i.w'.1.p.2.a$ for $i = k + j$, some bitstring $w' \preceq w$ and some constant $a$. Again, such a move is patient iff $w' = w$.
- To ground $\lfloor \breve{P}^p_j \rfloor^w_u$ means to make the move $1.i.w'.1.p.1.u'.a$ for $i = k + j$, some bitstrings $w' \preceq w$ and $u' \preceq u$, and some constant $a$. Such a move patient iff both $w' = w$ and $u' = u$.

Notice that player $\top$ can only ground positive molecules, while player $\bot$ can only ground negative molecules. Every grounding move thus has the form $\alpha.a$ for some constant $a$. Let us call such a constant $a$ *the choice constant* of the grounding move. Then we say that a given act (move) of grounding is done **diversifyingly** iff the move is patient, and its choice constant is the smallest constant that has never been used before in the play (run) as the choice constant of some grounding move.

Only $\bigsqcup x$-contentual molecules can be grounded. As for $\sqcup$-contentual molecules, before they can be grounded, they should be **dedisjunctionized**. That is, a move should be made that changes their status from $\sqcup$-contentual to $\bigsqcup x$-contentual. Precisely, for a $\sqcup$-contentual molecule $\lfloor \breve{Z}_j \sqcup \breve{T}_j \rfloor^w$, such a move is $1.j.w.2.1$ or $1.j.w.2.2$.[6] The effect of the move $1.j.w.2.1$ is turning the molecule's content $\bigsqcup x\dot{Z}_j(x) \sqcup \bigsqcup x\dot{T}_j(x)$ into $\bigsqcup x\dot{Z}_j(x)$, and therefore we say "to $Z$**-dedisjunctionize**" for making such a dedisjunctionizing move. And the effect of the move $1.j.w.2.2$ is turning the molecule's content into $\bigsqcup x\dot{T}_j(x)$, so we say "to $T$**-dedisjunctionize**" is this case. Note that, since $\lfloor \breve{Z}_j \sqcup \breve{T}_j \rfloor^w$ is negative, it can only be dedisjunctionized by player $\bot$, even though (or therefore), in the resulting position, the move $1.j.w.2.1$ or $1.j.w.2.2$ will be $\top$-labeled.

As before, assume $\Gamma$ is a (context-setting) legal run of (32). By a **$\Gamma$-supermolecule** we mean a grounded residual molecule $M$ of some subposition (i.e. a finite initial segment) $\Phi$ of $\Gamma$ such that for no proper initial segment $\Psi$ of $\Phi$ is the $\Psi$-predecessor of

[6]Well, just as in the case of grounding, the moves $1.j.w'.2.1$ or $1.j.w'.2.2$, where $w' \preceq w$ (rather than necessarily $w' = w$) would achieve the same "dedisjunctionizing" effect; however, we do not consider such cases as they are never going to really emerge in our construction.

$M$ grounded. We refer to such a $\Phi$ as the **position of grounding** of $M$, and refer to the length of $\Phi$ as the **time of grounding** of $M$.

Obviously any grounded residual molecule $M$ (of $\Gamma$ or any of its initial segments) descends from some unique supermolecule. We call the supermolecule from which $M$ descends the **essence** of $M$. Every supermolecule is thus its own essence. Only grounded molecules have essences. Therefore, if we say "the essence of $M$", the claim that $M$ is grounded is automatically implied. Whether a grounded molecule $M$ is a supermolecule or not, by the **position of grounding** and **time of grounding** of $M$ we mean those of the essence of $M$. Intuitively, the position of grounding of such an $M$ is the position in which $M$ — more precisely, a predecessor of $M$ — first became grounded, and the time of grounding tells us how soon after the start of the play this happened. Mark the obvious fact that the content of any grounded molecule is the same as that of the essence of that molecule.

We define a **$\Gamma$-chain** as any nonempty finite sequence $\langle M_1, \ldots, M_m \rangle$ of $\Gamma$-supermolecules such that:

1. $M_1$ and only $M_1$ is $\lfloor P \rfloor$-metatype.

2. For each odd $k$ with $1 \leq k < m$, $M_k$ is negative, $M_{k+1}$ is positive, and the two supermolecules have identical contents.

3. For each odd $k$ with $3 \leq k \leq m$, we have:

   - if $M_k = \lfloor \breve{Z}_j \sqcup \breve{T} \rfloor^w$ (some $j, w$), then $M_{k-1}$ is the essence of $\lfloor \breve{X}_j \rfloor^w$ or the essence of $\lfloor \breve{Y}_j \rfloor^w$;
   - if $M_k = \lfloor \breve{R}_j \rfloor^w$ (some $j, w$), then $M_{k-1}$ is the essence of $\lfloor \breve{Q}^p_j \rfloor^w$ for some $p \in \{1, \ldots, n\}$.

We will simply say "chain" instead of "$\Gamma$-chain" when $\Gamma$ is fixed or irrelevant. As an aside, observe that, according to the above definition, if $\lfloor \breve{W} \rfloor$ is in a chain, then it can only be the last element of the chain, and such a chain is of an even length. All internal (neither the first nor the last) odd-numbered elements of a chain have the metatype $\lfloor Z \sqcup T \rfloor$ or $\lfloor R \rfloor$, and all internal even-numbered elements have the metatype $\lfloor X \rfloor$, $\lfloor Y \rfloor$ or $\lfloor Q \rfloor$.

When $C = \langle M_1, \ldots, M_m \rangle$ is a chain, we say that $C$ **hits** $M_m$, and that $C$ **originates** from $M_1$, or that $M_1$ is the **origin** of $C$.

We say that a chain $C$ is **open** iff, where $\lfloor P^p_j \rfloor$ is the metatype of the origin of $C$, no element of $C$ has the metatype $\lfloor Q^p_j \rfloor$ (for the same $j, p$).

Where $\Gamma$ (as before) is a legal run of (32) and $M$ is a $\Gamma$-supermolecule, by the **$\Gamma$-base** of $M$ we mean the set of all worlds $p \in \mathcal{W}$ such that there is an open $\Gamma$-chain hitting $M$ whose origin is of the metatype $\lfloor P^p_j \rfloor$ for some $j$ (and that very $p$). If $M$ is a grounded $\Gamma$-molecule but not necessarily a supermolecule, then the **$\Gamma$-base** of $M$ is defined as the $\Gamma$-base of the essence of $M$. In contexts where $\Gamma$ is fixed, we denote the $\Gamma$-base of $M$ by **Base**$(M)$.

### 9.3 The counterstrategy

In this subsection we set up a counterstrategy for (32) in the form of an EPM $\mathcal{E}$, which will act in the role of $\bot$ in a play over (32). $\mathcal{E}$ is a *universal counterstrategy* for (32), in the sense that, as will be shown later, no HPM wins $(32)^\star$ against this particular, one-for-all EPM for the yet-to-be-constructed interpretation $^\star$. Since $\mathcal{E}$ plays as $\bot$ rather than $\top$, we will be interested in the run *cospelled* rather than spelled by any given computation branch of $\mathcal{E}$. That is, in a play, the moves made by $\mathcal{E}$ get the label $\bot$, and the moves made by its adversary get the label $\top$. In our description of $\mathcal{E}$ and the further analysis of its behavior, as we did in earlier sections, we will be relying on the clean environment assumption, here meaning that the adversary of a $\mathcal{E}$ never makes illegal moves. From the definition of $\mathcal{E}$ it will be also immediately clear that $\mathcal{E}$ itself does not make any illegal moves either. Since all runs that $\mathcal{E}$ generates are thus legal, we usually omit the word "legal", and by a run or position we will always mean a legal run or position of (32).

The work of $\mathcal{E}$ consists in sequentially performing the first two or all three (depending on how things evolve) of the following procedures FIRST, SECOND and THIRD. In the descriptions of these procedures, "current" should be understood as $\Phi$-residual, where $\Phi$ is the position of the play at the time when a given step is performed. This word may be omitted, and by just saying "molecule" we mean current molecule. Similarly, "$\sqcup$-contentual", "$\bigsqcup x$-contentual", "(matchingly) grounded", "to ground", etc. should be understood in the context of the then-current position. Similarly, for a grounded molecule $M$, $\mathbf{Base}(M)$ means the $\Phi$-base of $M$, where $\Phi$ is the then-current position. Also, since the current position is always a position (finite run), it is safe to say "leaf" instead of "complete branch" when talking about bitstring trees in our description of the work of $\mathcal{E}$.

**PROCEDURE** FIRST: Diversifyingly ground all $\lfloor P \rfloor$-metatype molecules, and go to SECOND.

**PROCEDURE** SECOND: If $\lfloor \breve{W} \rfloor$ is matchingly grounded, go to THIRD. Else perform each of the following routines:

**Routine 1.** For each $j$ and each leaf $w$ of the current $\lfloor \circ\!\!\mid(\breve{X}_j \wedge \breve{Y}_j \to \breve{Z}_j \sqcup \breve{T}_j)\rfloor$-tree, whenever both $\lfloor \breve{X}_j \rfloor^w$ and $\lfloor \breve{Y}_j \rfloor^w$ are matchingly grounded and $\lfloor \breve{Z}_j \sqcup \breve{T}_j \rfloor^w$ is $\sqcup$-contentual, do the following:

(i): if $p \models Z$ for every world $p$ such that $p$ is accessible from each element of $\mathbf{Base}(\lfloor \breve{X}_j \rfloor^w) \cup \mathbf{Base}(\lfloor \breve{Y}_j \rfloor^w)$, then $Z$-dedisjunctionize $\lfloor \breve{Z}_j \sqcup \breve{T}_j \rfloor^w$, after which diversifyingly ground (the now $\bigsqcup x$-contentual) $\lfloor \breve{Z}_j \sqcup \breve{T}_j \rfloor^w$;

(ii): else $T$-dedisjunctionize $\lfloor \breve{Z}_j \sqcup \breve{T}_j \rfloor^w$, after which diversifyingly ground (the now $\bigsqcup x$-contentual) $\lfloor \breve{Z}_j \sqcup \breve{T}_j \rfloor^w$.

**Routine 2.** For each $j$ and each leaf $w$ of the current $\lfloor \circ\!\!\mid\big((\circ\!\!\mid \breve{P}^1_j \to \breve{Q}^1_j) \wedge \ldots \wedge (\circ\!\!\mid \breve{P}^n_j \to \breve{Q}^n_j) \to \breve{R}_j\big)\rfloor$-tree, whenever each $\lfloor \breve{Q}^p_j \rfloor^w$ $(1 \leq p \leq n)$ is matchingly grounded and $\lfloor \breve{R}_j \rfloor^w$ is $\bigsqcup x$-contentual, diversifyingly ground $\lfloor \breve{R}_j \rfloor^w$.

**Routine 3.** Grant permission, and repeat SECOND.

**PROCEDURE** THIRD:
*Step 1*: $Z$-dedisjunctionize all (remaining) $\sqcup$-contentual $\lfloor Z \sqcup T \rfloor$-metatype molecules.
*Step 2*: Diversifyingly ground all $\bigsqcup x$-contentual $\lfloor Z \sqcup T \rfloor$- and $\lfloor R \rfloor$-metatype molecules.
*Step 3*: Go into an infinite loop within a permission state.

Remember that a fair EPM is one whose every $e$-computation branch (any valuation $e$) is fair, i.e. permission is granted infinitely many times in each branch. Before we go any further, let us make the straightforward observation that

$$\mathcal{E} \textit{ is a fair EPM.} \tag{33}$$

This is so because THIRD grants permission infinitely many times within Step 3, and if THIRD is never reached by $\mathcal{E}$, then SECOND will be iterated infinitely many times, with each iteration granting permission within Routine 3.

Our ultimate goal is to show that (32) is *not valid*, which, as mentioned, will be achieved by finding an interpretation $^\star$ such that no HPM wins $(32)^\star$ against $\mathcal{E}$. We approach this goal by first proving the weaker fact that (32) is *not uniformly valid*. In particular, below we are going to show that, for any valuation $e$ and any $e$-computation branch $B$ of $\mathcal{E}$, there is an interpretation $^\dagger$ such that the run cospelled by $B$ is a $\bot$-won run of $(32)^\dagger$. As will be observed at the beginning of Subsection 9.6, this fact immediately implies the non-uniform-validity of (32). Such a $B$-depending counterinterpretation is going to be what in [8] is called **perfect**, in our particular case meaning that for any predicate letter $\dot{A}$ of (32), $\dot{A}^\dagger(x)$ is a finitary predicate that does not depend on any variables except $x$. This can be easily seen to make $\breve{A}^\dagger$ and hence $(32)^\dagger$ a constant game, allowing us to safely ignore the valuation parameter $e$ in most contexts, which is irrelevant because neither the game $e[\breve{A}^\dagger]$ nor (notice) the work of $\mathcal{E}$ depends on $e$. To define such an interpretation, it is sufficient to indicate what constant atomic formulas are made by it true and what constant atomic formulas are made false. Here and later, for simplicity, by "constant atomic formulas" we mean formulas of the form $\dot{A}(a)$, where $\dot{A}$ is a predicate letter occurring in (32) and $a$ is a constant. For obvious reasons, how $^\dagger$ interprets any other atoms is irrelevant, and we may safely pretend that such atoms simply do not exist in the language.

So, fix any valuation $e$ spelled on the valuation tape of $\mathcal{E}$, and any $e$-computation branch $B$ of $\mathcal{E}$. Let $\Gamma$ be the run cospelled by $B$. Let us agree to say "**ultimate**" (run, position, tree) for "$\Gamma$-residual" (run, position, tree). To $\Gamma$ itself we refer as the **ultimate run of** $\lfloor(32)\rfloor$, or simply the **ultimate run**. By just saying "**molecule**" we mean a residual molecule of $\Gamma$ or of any initial segment of it. And $\Gamma$-molecules we call **ultimate molecules**. Any molecule would thus be an ultimate molecule or a predecessor of such. When $M$ is an ultimate molecule, by just saying that $M$ is $\sqcup$- or $\bigsqcup x$-contentual, grounded or matchingly grounded we mean that $M$ is so in $\Gamma$. Also, "chain" now always means $\Gamma$-chain, and, for any grounded molecule $M$, $\mathbf{Base}(M)$ means the $\Gamma$-base of $M$.

We will say that the branch $B$ is **short** iff, in the process of playing it up, $\mathcal{E}$ never entered the THIRD stage, thus forever remaining in SECOND. Otherwise $B$ is **long**.

The scope of all this $B$- and $\Gamma$-dependent terminology and conventions extends to the following two subsections, throughout which $B$ and $\Gamma$ are fixed.

Our construction of a counterinterpretation for (32) depends on whether $B$ is short or long. We consider these two cases separately.

## 9.4 Constructing a counterinterpretation when $B$ is short

Assume $B$ is short. This, looking at the first line of the description of SECOND, means that, in the ultimate run of $\lfloor(32)\rfloor$ (in $\Gamma$, that is), $\breve{W}$ is not matchingly grounded. We select our

$$\textit{counterinterpretation}\ ^{\dagger}$$

to be the perfect interpretation that makes the contents of all positive non-matchingly grounded ultimate molecules false, and all other constant atomic formulas true.

**Convention 9.3**

- Where $\dot{A}$ is a predicate letter of (32) and $a$ is any constant, we say that $\dot{A}(a)$ is **true** iff $\dot{A}^{\dagger}(a)$ is true.
- We say that $\lfloor(32)\rfloor$ is **true** iff the ultimate run of $\lfloor(32)\rfloor$ is a $\top$-won run of $(32)^{\dagger}$.
- We say that $\lfloor\breve{W}\rfloor$ is **true** iff the ultimate position of $\lfloor\breve{W}\rfloor$ is a $\top$-won position of $\breve{W}^{\dagger}$.
- We say that $\lfloor\circ\!\!\!\!|\,(\breve{X}_j\wedge\breve{Y}_j\rightarrow\breve{Z}_j\sqcup\breve{T}_j)\rfloor$ is **true** iff the ultimate run of $\lfloor\circ\!\!\!\!|\,(\breve{X}_j\wedge\breve{Y}_j\rightarrow\breve{Z}_j\sqcup\breve{T}_j)\rfloor$ is a $\top$-won run of $\big(\circ\!\!\!\!|\,(\breve{X}_j\wedge\breve{Y}_j\rightarrow\breve{Z}_j\sqcup\breve{T}_j)\big)^{\dagger}$. Similarly for $\lfloor\circ\!\!\!\!|\,\big((\circ\!\!\!\!|\,\breve{P}^1_j\rightarrow\breve{Q}^1_j)\wedge\ldots\wedge(\circ\!\!\!\!|\,\breve{P}^n_j\rightarrow\breve{Q}^n_j)\rightarrow\breve{R}_j\big)\rfloor$.
- Where $w$ is a complete branch of the ultimate $\lfloor\circ\!\!\!\!|\,(\breve{X}_j\wedge\breve{Y}_j\rightarrow\breve{Z}_j\sqcup\breve{T}_j)\rfloor$-tree, we say that $\lfloor\breve{Z}_j\sqcup\breve{T}_j\rfloor^w$ is **true** iff the ultimate run (position) of $\lfloor\breve{Z}_j\sqcup\breve{T}_j\rfloor^w$ is a $\top$-won run of $(\breve{Z}_j\sqcup\breve{T}_j)^{\dagger}$. Similarly for $\lfloor\breve{X}_j\rfloor^w$, $\lfloor\breve{Y}_j\rfloor^w$, $\lfloor\breve{X}_j\wedge\breve{Y}_j\rfloor^w$, $\lfloor\breve{Q}^p_j\rfloor^w$, $\lfloor\breve{R}_j\rfloor^w$, $\lfloor\circ\!\!\!\!|\,\breve{P}^p_j\rightarrow\breve{Q}_j\rfloor^w$.
- Where $w$ is a complete branch of the ultimate $\lfloor\circ\!\!\!\!|\,\big((\circ\!\!\!\!|\,\breve{P}^1_j\rightarrow\breve{Q}^1_j)\wedge\ldots\wedge(\circ\!\!\!\!|\,\breve{P}^n_j\rightarrow\breve{Q}^n_j)\rightarrow\breve{Z}_j\big)\rfloor$-tree and $u$ is a complete branch of the ultimate $\lfloor\circ\!\!\!\!|\,\breve{P}^p_j\rfloor^w$-tree, we say that $\lfloor\breve{P}^p_j\rfloor^w_u$ is **true** iff the ultimate position of $\lfloor\breve{P}^p_j\rfloor^w_u$ is a $\top$-won position of $(\breve{P}_j)^{\dagger}$.

"**False**" will mean "not true".

Based on the definitions of $\sqcup$ and $\bigsqcup$, with a moment's thought we can see that the following claim is valid:

**Claim 9.4** *Let $M$ be an arbitrary ultimate molecule.*

**(i)** *If $M$ is $\sqcup$- or $\bigsqcup x$-contentual, then $M$ is false.*

**(ii)** *If $M$ is grounded, then $M$ is true iff its content is true.*

Our goal is to show that $\lfloor(32)\rfloor$ is false. Throughout the rest of this subsection, we agree to understand the terms "grounded", "$\bigsqcup x$-contentual", etc. in the context of the ultimate run.

Note that $\lfloor\breve{W}\rfloor$ is guaranteed to be false. Indeed, if $\lfloor\breve{W}\rfloor$ is $\bigsqcup x$-contentual, it is false by clause (i) of Claim 9.4. And, if $\lfloor\breve{W}\rfloor$ is grounded, then, as no switch to THIRD has occured, $\lfloor\breve{W}\rfloor$ must be non-matchingly grounded. Then, with clause (ii) of Claim 9.4 in mind, $\lfloor\breve{W}\rfloor$ can be seen to be false by our choice of $^\dagger$.

As $\lfloor\breve{W}\rfloor$ is false, in order to show that $\lfloor(32)\rfloor$ is false, it would suffice to verify that, for each $j$, both $\lfloor\circ\!\!\!\mid(\breve{X}_j\wedge\breve{Y}_j\rightarrow\breve{Z}_j\sqcup\breve{T}_j)\rfloor$ and $\lfloor\circ\!\!\!\mid\big((\circ\!\!\!\mid\breve{P}^1_j\rightarrow\breve{Q}^1_j)\wedge\ldots\wedge(\circ\!\!\!\mid\breve{P}^n_j\rightarrow\breve{Q}^n_j)\rightarrow\breve{R}_j\big)\rfloor$ are true. Why this would suffice can be seen directly from the definitions of $\wedge$ and $\rightarrow$. In turn, based on the definition of $\circ\!\!\!\mid$, the truth of $\lfloor\circ\!\!\!\mid(\breve{X}_j\wedge\breve{Y}_j\rightarrow\breve{Z}_j\sqcup\breve{T}_j)\rfloor$ and $\lfloor\circ\!\!\!\mid\big((\circ\!\!\!\mid\breve{P}^1_j\rightarrow\breve{Q}^1_j)\wedge\ldots\wedge(\circ\!\!\!\mid\breve{P}^n_j\rightarrow\breve{Q}^n_j)\rightarrow\breve{R}_j\big)\rfloor$ means nothing but that:

(a) for every complete branch $w$ of the ultimate $\lfloor\circ\!\!\!\mid(\breve{X}_j\wedge\breve{Y}_j\rightarrow\breve{Z}_j\sqcup\breve{T}_j)\rfloor$-tree, $\lfloor\breve{X}_j\wedge\breve{Y}_j\rightarrow\breve{Z}_j\sqcup\breve{T}_j\rfloor^w$ is true, and

(b) for every complete branch $w$ of the ultimate $\lfloor\circ\!\!\!\mid\big((\circ\!\!\!\mid\breve{P}^1_j\rightarrow\breve{Q}^1_j)\wedge\ldots\wedge(\circ\!\!\!\mid\breve{P}^n_j\rightarrow\breve{Q}^n_j)\rightarrow\breve{R}_j\big)\rfloor$-tree, $\lfloor(\circ\!\!\!\mid\breve{P}^1_j\rightarrow\breve{Q}^1_j)\wedge\ldots\wedge(\circ\!\!\!\mid\breve{P}^n_j\rightarrow\breve{Q}^n_j)\rightarrow\breve{R}_j\rfloor^w$ is true.

Pick any $j$, and assume $w$ is as in (a). If $\lfloor\breve{Z}_j\sqcup\breve{T}_j\rfloor^w$ is grounded, then, with Claim 9.4(ii) in mind, it is true. This is so because, by our choice of $^\dagger$, only the contents of positive non-matchingly grounded (ultimate) molecules are made false by this interpretation; $\lfloor\breve{Z}_j\sqcup\breve{T}_j\rfloor^w$ is not positive, nor is its content the same as that of some positive non-matchingly grounded ultimate molecule, for then that molecule would not be non-matchingly grounded. The truth of $\lfloor\breve{Z}_j\sqcup\breve{T}_j\rfloor^w$, in turn, by the definition of $\rightarrow$, can be seen to imply the truth of $\lfloor\breve{X}_j\wedge\breve{Y}_j\rightarrow\breve{Z}_j\sqcup\breve{T}_j\rfloor^w$. Suppose now $\lfloor\breve{Z}_j\sqcup\breve{T}_j\rfloor^w$ is not grounded. Analyzing how $\mathcal{E}$ acts in Routine 1 of SECOND, this fact can be seen to indicate that at least one of the two ultimate molecules $\lfloor\breve{X}_j\rfloor^w$ and $\lfloor\breve{Y}_j\rfloor^w$ — let us assume it is $\lfloor\breve{X}_j\rfloor^w$ — is not matchingly grounded. If $\lfloor\breve{X}_j\rfloor^w$ is $\bigsqcup x$-contentual, it is false as are all $\bigsqcup x$-contentual molecules. And if $\lfloor\breve{X}_j\rfloor^w$ is non-matchingly grounded, it is again false by our choice of $^\dagger$. In either case $\lfloor\breve{X}_j\rfloor^w$ is thus false and hence, as can be seen from the definition of $\wedge$, so is $\lfloor\breve{X}_j\wedge\breve{Y}_j\rfloor^w$. This, in turn, by the definition of $\rightarrow$, makes $\lfloor\breve{X}_j\wedge\breve{Y}_j\rightarrow\breve{Z}_j\sqcup\breve{T}_j\rfloor^w$ true. Statement (a) is taken care of.

Assume now $w$ is as in (b). If $\lfloor\breve{R}_j\rfloor^w$ is grounded, then we are done for the same reasons as in the previous paragraph when discussing the similar case for $\lfloor\breve{Z}_j\sqcup\breve{T}_j\rfloor^w$. Suppose now $\lfloor\breve{R}_j\rfloor^w$ is $\bigsqcup x$-contentual. Reasoning as we did in the corresponding case of the previous paragraph for "$\lfloor\breve{X}_j\rfloor^w$ or $\lfloor\breve{Y}_j\rfloor^w$" (only appealing to Routine 2 of SECOND instead of Routine 1), we find that, for some $p$, $\lfloor\breve{Q}^p_j\rfloor^w$ is false. Therefore, in order to conclude that $\lfloor(\circ\!\!\!\mid\breve{P}^1_j\rightarrow\breve{Q}^1_j)\wedge\ldots\wedge(\circ\!\!\!\mid\breve{P}^n_j\rightarrow\breve{Q}^n_j)\rightarrow\breve{R}_j\rfloor^w$ is true, it would be sufficient to show that $\lfloor\circ\!\!\!\mid\breve{P}^p_j\rfloor^w$ is true. The truth of $\lfloor\circ\!\!\!\mid\breve{P}^p_j\rfloor^w$ means nothing but that, for every complete branch $u$ of the ultimate $\lfloor\circ\!\!\!\mid\breve{P}^p_j\rfloor^w$-tree, $\lfloor\breve{P}^p_j\rfloor^w_u$ is true. But this is indeed so. Consider any complete branch $u$ of the ultimate $\lfloor\circ\!\!\!\mid\breve{P}^p_j\rfloor^w$-tree. $\lfloor\breve{P}^p_j\rfloor^w_u$ is grounded due

to the actions of $\mathcal{E}$ during FIRST. And, as this molecule is negative, it can be seen to be true by our choice of $^\dagger$. Statement (b) is thus also taken care of.

So, $\lfloor(32)\rfloor$ is false, meaning that $\Gamma$ is a $\bot$-won run of $(32)^\dagger$, i.e. $\mathbf{Wn}^{(32)^\dagger}\langle\Gamma\rangle = \bot$.

## 9.5 Constructing a counterinterpretation when $B$ is long

Throughout this subsection, we assume $B$ is long, i.e., at some point of playing $B$ up, $\mathcal{E}$ switched from SECOND to THIRD. This means that $\lfloor\breve{W}\rfloor$ is matchingly grounded in the ultimate run $\Gamma$. We let $\Delta$ denote the (finite) initial segment of $\Gamma$ consisting of all of the (lab)moves made before the switch to THIRD occurred.

Throughout this subsection, $\Gamma$ will remain the default, context-setting run for run-relative terminology and notation such as "supermolecule", "grounded", "$\mathbf{Base}(M)$", etc.

**Claim 9.5** *No two different negative supermolecules have identical contents.*

**Proof.** This is so because negative molecules are grounded by $\mathcal{E}$, which always does grounding in a diversifying way — in a way that ensures that the content of the resulting molecule is different from that of any other grounded molecule. $\square$

The supermolecules whose position of termination is not longer than $\Delta$ we call **old-generation supermolecules**. In other words, old-generation supermolecules are those that became grounded while $\mathcal{E}$ was still working within FIRST of SECOND.

We say that a supermolecule $M$ is **well-behaved** iff, whenever there is an open chain hitting $M$ and originating from a $\lfloor P^p_j\rfloor$-metatype supermolecule, we have $p \models P_j$.

**Claim 9.6** *Suppose $M$ is a well-behaved negative old-generation supermolecule, $\dot{A}(a)$ is its content, and $p$ is a world accessible from every element of* $\mathbf{Base}(M)$. *Then $p \models A$.*

**Proof.** Let $M$, $\dot{A}(a)$, $p$ be as above, and let $\Phi$ be the position of grounding of $M$. Obviously it is sufficient to assume here — and we will assume — that $p$ is the *$\mathcal{R}$-smallest* world accessible from every world of $\mathbf{Base}(M)$, in the sense that, for any $q \in \mathcal{W}$, if $q$ is accessible from every element of $\mathbf{Base}(M)$, then $q$ is also accessible from $p$. This means that $\mathcal{R}$ is a linear order on $\mathbf{Base}(M)$ with $p$ being its greatest element, or otherwise $\mathbf{Base}(M)$ is empty and $p$ is the root world 1. We proceed by induction on the time of grounding of $M$, i.e. the length of $\Phi$. There are three cases to consider, depending on the metatype of $M$.

*Case 1:* $M$ is $\lfloor P\rfloor$-metatype, specifically, $M = \lfloor\breve{P}^q_j\rfloor^w$ for some $q, j, w$. Then $\mathbf{Base}(M) = \{q\}$ and thus $p = q$ because, according to our assumptions, $p$ is the $\mathcal{R}$-smallest world accessible from every element of $\mathbf{Base}(M)$. And, as $M$ is well-behaved, we have $p \models P_j$.

*Case 2:* $M$ is $\lfloor Z \sqcup T\rfloor$-metatype, specifically, $M = \lfloor\breve{Z}_j \sqcup \breve{T}_j\rfloor^w$. Since $M$ is second-generation, $\lfloor\breve{X}_j\rfloor^w$ should be matchingly grounded in $\Phi$ (in fact in a proper initial segment of $\Phi$), for otherwise $\mathcal{E}$ would not have grounded $\lfloor\breve{Z}_j \sqcup \breve{T}_j\rfloor^w$. Let $N$ be the essence of $\lfloor\breve{X}_j\rfloor^w$, and let $L$ be the essence of a negative grounded $\Phi$-molecule whose

content is the same as that of ($\lfloor \breve{X}_j \rfloor^w$ and hence of) $N$. The time of grounding of ($N$ and hence of) $L$ is smaller than that of $M$ and thus, just like $M$, $L$ is an old-generation supermolecule. Observe that every $L$-hitting open chain turns into an $M$-hitting open chain after appending $N$ and $M$ to it. For this reason, $\mathbf{Base}(L) \subseteq \mathbf{Base}(M)$ and, since $p$ is accessible from all elements of $\mathbf{Base}(M)$, so is it from all elements of $\mathbf{Base}(L)$. Also, as $M$ is well-behaved, so is $L$. Hence, by the induction hypothesis,

$$p \models X_j. \tag{34}$$

A similar argument convinces us that

$$p \models Y_j. \tag{35}$$

According to our original assumptions regarding $\mathcal{K}$, every world forces every formula of the antecedent of (30). So, $p \models X_j \circ\!\!- (Y_j \circ\!\!- Z_j \sqcup T_j)$. This, together with (34) and (35), implies that $p \models Z_j \sqcup T_j$. Therefore, either $p \models Z_j$ or $p \models T_j$. It is rather obvious that $\mathbf{Base}(M) = \mathbf{Base}(\lfloor X_j \rfloor^w) \cup \mathbf{Base}(\lfloor Y_j \rfloor^w)$. Keeping this fact in mind together with our assumption that $p$ is the $\mathcal{R}$-smallest world accessible from all worlds of $\mathbf{Base}(M)$, and analyzing the work of $\mathcal{E}$ within Routine 1 of SECOND, we find that, if $p \models Z_j$, then $A = Z_j$, so that $p \models A$; and if $p \not\models Z_j$ and hence $p \models T_j$, then $A = T_j$, so that, again, $p \models A$. In either case we thus have $p \models A$, as desired.

*Case 3:* $M$ is $\lfloor R \rfloor$-metatype, specifically, $M = \lfloor \breve{R}_j \rfloor^w$. Then, of course, $A = R_j$. We want to show that $p \models P_j \circ\!\!- Q_j$, from which the desired $p \models R_j$ (i.e. $p \models A$) follows because, as noted in the previous case, every formula of the antecedent of (30) is forced by $p$, including $(P_j \circ\!\!- Q_j) \circ\!\!- R_j$.

For a contradiction, assume $p \not\models P_j \circ\!\!- Q_j$. Then there is a world $q$ accessible from $p$ such that $q \models P_j$ and $q \not\models Q_j$. From Routine 2 of SECOND we see that $\lfloor \breve{Q}^q_j \rfloor^w$ should be matchingly grounded in the position preceeding $\Phi$ (otherwise $\mathcal{E}$ would not have grounded $M$ in that position). Arguing for $\lfloor Q^q_j \rfloor^w$ as we did in Case 2 for $\lfloor \breve{X}_j \rfloor^w$, we find that there is a negative old-generation supermolecule $L$ whose time of grounding is smaller than that of $M$ and whose content is the same as that of $\lfloor Q^q_j \rfloor^w$, so that the type of $L$ is $Q_j$. Note that every open chain $C$ hitting $L$ can be extended to a chain hitting $M$ by adding to $C$ the essence of $\lfloor Q^q_j \rfloor^w$ and then $M$; such a chain will remain open unless the origin of $C$ was a $\lfloor P^q_j \rfloor$-metatype molecule. Therefore, remembering that $q \models P_j$, we find that $\mathbf{Base}(L)$ is well-behaved, because so is $\mathbf{Base}(M)$. With the above observation in mind, we also find that $q$ can be the only world that is in $\mathbf{Base}(L)$ but not in $\mathbf{Base}(M)$. Hence, remembering that $p$ is accessible from every element of $\mathbf{Base}(M)$ and $p\mathcal{R}q$, we find that $q$ is accessible from every element of $\mathbf{Base}(L)$. All this allows us to apply the induction hypothesis to $q$ and $L$ and infer that $q \models Q_j$, which is a contradiction. □

**Claim 9.7** *There is an open chain hitting* $\lfloor \breve{W} \rfloor$.

**Proof.** Since $B$ is long, $\lfloor \breve{W} \rfloor$ is matchingly grounded (already) in $\Delta$, so $\lfloor \breve{W} \rfloor$ is an old-generation supermolecule. Let $N$ be a negative grounded $\Delta$-molecule whose content is identical with that of $\lfloor \breve{W} \rfloor$, and let $M$ be the essence of $N$. The type of

($N$ and hence) $M$, of course, should be $W$. Remember our assumption that $\mathcal{K}\|\not\models W$, i.e. $1 \not\models W$. It implies that $\mathbf{Base}(M) \neq \emptyset$, for otherwise — vacuously — $M$ would be well-behaved and 1 be accessible from every element of $\mathbf{Base}(M)$, whence, by Claim 9.6, we would have $1 \models W$. The fact that $\mathbf{Base}(M)$ is nonempty means that there is an open chain hitting $M$. The result of adding $\lfloor \breve{W} \rfloor$ to that chain obviously remains an open chain. $\Box$

Let us select and fix some — say, lexicographically the smallest — open chain hitting $\lfloor \breve{W} \rfloor$, and call it the **master chain**. According to Claim 9.7, such a chain exists. Let us call the supermolecules that are in the master chain **master supermolecules**.

Now we are ready to define the

*counterinterpretation* $^\dagger$.

We define it as the perfect interpretation that makes the content of each master supermolecule false, and makes all other constant atomic formulas true.

In what follows, we fully adopt the terminology of Convention 9.3, with the only difference that now the underlying interpretation $^\dagger$ on which that terminology is based is $^\dagger$ as defined in the present subsection rather than as defined in Subsection 9.4. As in Subsection 9.4, our goal is to show that $\lfloor(32)\rfloor$ is false.

That $\lfloor \breve{W} \rfloor$ is false is immediate from our choice of $^\dagger$. Hence, as in Subsection 9.4, in order to show that $\lfloor(32)\rfloor$ is false, it would suffice to verify that, for each $j$, we have:

(a) for every complete branch $w$ of the ultimate $\lfloor \circ\!\!\!\mid(\breve{X}_j \wedge \breve{Y}_j \rightarrow \breve{Z}_j \sqcup \breve{T}_j)\rfloor$-tree, $\lfloor \breve{X}_j \wedge \breve{Y}_j \rightarrow \breve{Z}_j \sqcup \breve{T}_j \rfloor^w$ is true, and

(b) for every complete branch $w$ of the ultimate $\lfloor \circ\!\!\!\mid\big((\circ\!\!\!\mid\breve{P}^1_j \rightarrow \breve{Q}^1_j) \wedge \ldots \wedge (\circ\!\!\!\mid\breve{P}^n_j \rightarrow \breve{Q}^n_j) \rightarrow \breve{R}_j\big)\rfloor$-tree, $\lfloor(\circ\!\!\!\mid\breve{P}^1_j \rightarrow \breve{Q}^1_j) \wedge \ldots \wedge (\circ\!\!\!\mid\breve{P}^n_j \rightarrow \breve{Q}^n_j) \rightarrow \breve{R}_j\rfloor^w$ is true.

Pick any $j$, and assume $w$ is as in (a). $\lfloor \breve{Z}_j \sqcup \breve{T}_j \rfloor^w$ is grounded at least due to the actions of $\mathcal{E}$ during THIRD (in case it otherwise managed to stay non-grounded throughout SECOND). If the essence of $\lfloor \breve{Z}_j \sqcup \breve{T}_j \rfloor^w$ is not in the master chain, then its content is true, because the latter, in view of Claim 9.5 and with a little thought, can be seen to be different from the content of any master supermolecule, and, by our choice of $^\dagger$, this interpretation only falsifies the contents of master supermolecules. In turn, from the truth of the content of the essence of $\lfloor \breve{Z}_j \sqcup \breve{T}_j \rfloor^w$ and hence the truth of $\lfloor \breve{Z}_j \sqcup \breve{T}_j \rfloor^w$ we infer that $\lfloor \breve{X}_j \wedge \breve{Y}_j \rightarrow \breve{Z}_j \sqcup \breve{T}_j \rfloor^w$ is true. Suppose now the essence of $\lfloor \breve{Z}_j \sqcup \breve{T}_j \rfloor^w$ is in the master chain. Then, from the definition of chain and with some thought, we can see that either the essence of $\lfloor \breve{X}_j \rfloor^w$ or the essence of $\lfloor \breve{Y}_j \rfloor^w$ should be there, too. Specifically, such a supermolecule would be the one immediately preceding the essence of $\lfloor \breve{Z}_j \sqcup \breve{T}_j \rfloor^w$ in the master chain. Hence, by our choice of $^\dagger$, either $\lfloor \breve{X}_j \rfloor^w$ or $\lfloor \breve{Y}_j \rfloor^w$ is false, which makes $\lfloor \breve{Z}_j \wedge \breve{Y}_j \rightarrow \breve{Z}_j \sqcup \breve{T}_j \rfloor^w$ true.

Assume now $w$ is as in (b). As was the case with $\lfloor \breve{Z}_j \sqcup \breve{T}_j \rfloor^w$ in the previous paragraph, $\lfloor R_j \rfloor^w$ is grounded, and, if its essence is not in the master chain, then its content and hence $\lfloor(\circ\!\!\!\mid\breve{P}^1_j \rightarrow \breve{Q}^1_j) \wedge \ldots \wedge (\circ\!\!\!\mid\breve{P}^n_j \rightarrow \breve{Q}^n_j) \rightarrow \breve{R}_j\rfloor^w$ is true. Suppose now the essence of $\lfloor \breve{R}_j \rfloor^w$ is in the master chain. Arguing as we did in the previous

case for $\lfloor \breve{X}_j \rfloor^w$ or $\lfloor \breve{Y}_j \rfloor^w$, we find that, for some $p$, the essence of $\lfloor \breve{Q}_j^p \rfloor^w$ is a master supermolecule and hence $\lfloor \breve{Q}_j^p \rfloor^w$ is false. Therefore, in order to conclude that $\lfloor (\circ\!\!\mid \breve{P}_j^1 \to \breve{Q}_j^1) \wedge \ldots \wedge (\circ\!\!\mid \breve{P}_j^n \to \breve{Q}_j^n) \to \breve{R}_j \rfloor^w$ true, it would suffice to show that $\lfloor \circ\!\!\mid \breve{P}_j^p \rfloor^w$ is true. The truth of $\lfloor \circ\!\!\mid \breve{P}_j^p \rfloor^w$ means nothing but that, for every complete branch $u$ of the ultimate $\lfloor \circ\!\!\mid \breve{P}_j^p \rfloor^w$-tree, $\lfloor \breve{P}_j^p \rfloor_u^w$ is true. But this is indeed so. Consider any complete branch $u$ of the ultimate $\lfloor \circ\!\!\mid \breve{P}_j^p \rfloor^w$-tree. $\lfloor \breve{P}_j^p \rfloor_u^w$ is grounded due to the actions during FIRST. Its essence cannot be in the master chain because the essence of $\lfloor \breve{Q}_j^p \rfloor^w$ is there and the master chain is open. But then, in view of Claim 9.5, the content of $\lfloor \breve{P}_j^p \rfloor_u^w$ is different from that of any master supermolecule, which, by our choice of $^\dagger$, makes $\lfloor \breve{P}_j^p \rfloor_u^w$ true. Statement (b) is thus also taken care of.

So, $\lfloor (32) \rfloor$ is false, meaning that $\Gamma$ is a $\bot$-won run of $(32)^\dagger$, i.e. $\mathbf{Wn}^{(32)^\dagger}\langle \Gamma \rangle = \bot$.

## 9.6 From non-uniform-validity to non-validity

In the previous two subsections we in fact showed that

$$\textit{(32) is not uniformly valid.} \qquad (36)$$

Indeed, suppose, for a contradiction, that (32) is uniformly valid. Let then $\mathcal{H}$ be a uniform HPM-solution for (32) — an HPM that wins $(32)^*$ for every interpretation $^*$. Pick an arbitrary valuation $e$ (which is irrelevant here anyway), and let $B$ be the $(\mathcal{E}, e, \mathcal{H})$-branch. As observed in (33), $\mathcal{E}$ is fair, so the conditions of Lemma 5.1 are met and such a branch $B$ is defined. Let then $^\dagger$ be the interpretation constructed from $B$ as in Subsection 9.4 or 9.5, depending on whether $B$ is short or long. Then, as we showed in those two subsections, the run $\Gamma$ cospelled by $B$ is a $\top$-lost run of $(32)^\dagger$ ($= e[(32)^\dagger]$, because the interpretation is perfect). But, by Lemma 5.1, the same $\Gamma$ is a run spelled by some $e$-computation branch of $\mathcal{H}$ — specifically, it is the $\mathcal{H}$ vs. $\mathcal{E}$ run on $e$. This means that $\mathcal{H}$ does not win $(32)^\dagger$, contrary to our assumption that $\mathcal{H}$ is a uniform solution for (32).

In turn, (36) can be eventually rather easily translated into the fact of completeness of **Int** with respect to uniform validity.

Our goal, however, is to show the completeness of **Int** with respect to validity rather than just uniform validity — the goal which, as we remember, has been reduced to showing the non-validity of (32). The fact of non-validity of (32), of course, is stronger than the fact of its non-uniform-validity. The point is that the counterinterpretation $^\dagger$ constructed in Subsections 9.4 and 9.5 depends on the computation branch $B$ and hence on the HPM $\mathcal{H}$ that plays against $\mathcal{E}$. That is, different $\mathcal{H}$s could require different $^\dagger$s. In order to show that (32) is not valid, we need to construct a *one*, common-for-all-$\mathcal{H}$ counterinterpretation $^\star$, with the property that every HPM $\mathcal{H}$ loses $(32)^\star$ against $\mathcal{E}$ for that very interpretation $^\star$ on some valuation $e$. Not to worry, we can handle that.

Let us fix the sequence

$$\mathcal{H}_1,\ \mathcal{H}_2,\ \mathcal{H}_3,\ \ldots$$

of all HPMs, enumerated according to the lexicographic order of their standardized descriptions. Next, we select a variable $z$ different from $x$ and, for each constant $c$, define

$$e_c$$

to be the valuation that sends $z$ to $c$, and (arbitrarily) sends all other variables to 1.

For each constant $c$, let

$$B_c$$

be the $(\mathcal{E}, e_c, \mathcal{H}_c)$-branch, and let

$$\Gamma_c$$

be the run cospelled by $B_c$, i.e. (by Lemma 5.1) the $\mathcal{H}_c$ vs. $\mathcal{E}$ run on $e_c$.

Next, for any constant $c$, let

$$^{\ddagger c}$$

be the interpretation constructed from $B_c$ and $\Gamma_c$ in the way we constructed the interpretation $^{\dagger}$ from $B$ and $\Gamma$ in:

Subsection 9.4 if $B_c$ is short;
Subsection 9.5 if $B_c$ is long.

From the results of Subsections 9.4 and 9.5 we thus have:

$$\textit{For any constant } c,\ \mathbf{Wn}^{(32)^{\ddagger c}}\langle \Gamma_c \rangle = \bot. \tag{37}$$

Consider any atom $\dot{A}(x)$ of (32). For each particular constant $c$, the predicate $\dot{A}^{\ddagger c}(x)$ is unary, depending only on $x$. But if here $c$ is seen as a variable and, as such, renamed into $z$, then the predicate becomes binary, depending on $x$ and $z$. Let us denote such a predicate by $\ddot{A}(x,z)$. That is, we define $\ddot{A}(x,z)$ as the predicate such that, for any constants $a$ and $c$,

$$\ddot{A}(a,c) = \dot{A}^{\ddagger c}(a).$$

We now define our ultimate

$$\textit{counterinterpretation} \quad ^{\star}$$

by stipulating that, for any atom $\dot{A}(x)$ of (32),[7] $\dot{A}^{\star}(x)$ is nothing but the above predicate $\ddot{A}(x,z)$. Note that, unlike $^{\ddagger c}$ (any particular constant $c$), $^{\star}$ is not a perfect interpretation, for $\dot{A}^{\star}(x)$ depends on the "hidden" variable $z$.

**Claim 9.8** *For any constant $c$,* $(32)^{\ddagger c} = e_c[(32)^{\star}]$.

[7] As noted for $^{\dagger}$ in Subsection 9.3, how $^{\star}$ interprets the atoms that are not in (32) is irrelevant and not worth bothering at this point.

**Proof.** Consider any predicate letter $\dot{A}$ of (32), and any constant $c$. We claim that

$$\big(\sqcup x\dot{A}(x)\big)^{\ddagger c} = e_c[\big(\sqcup x\dot{A}(x)\big)^{\star}]. \tag{38}$$

The games on the two sides of the above equation can be rewritten as $\sqcup x\dot{A}^{\ddagger c}(x)$ and $e_c[\sqcup x\dot{A}^{\star}(x)]$, respectively. $e_c[\sqcup x\dot{A}^{\star}(x)]$ can be further rewritten as $e_c[\sqcup x\ddot{A}(x,z)]$. Thus, in order to verify (38), we need to show that

$$\sqcup x\dot{A}^{\ddagger c}(x) = e_c[\sqcup x\ddot{A}(x,z)]. \tag{39}$$

As noted earlier, $\sqcup x\dot{A}^{\ddagger c}(x)$ is a constant game due to the fact that the interpretation $^{\ddagger c}$ is perfect. So, it is its own instance, and $\sqcup x\dot{A}^{\ddagger c}(x)$ can be safely written instead of $e[\sqcup x\dot{A}^{\ddagger c}(x)]$ for whatever (irrelevant) valuation $e$. Of course, the two games of (39) have the same legal runs, with every such run being $\langle\rangle$ or $\langle\top a\rangle$ for some constant $a$. So, we only need to verify that the **Wn** components of those two games are also identical. And, of course, considering only legal runs when comparing the two **Wn** components is sufficient. Furthermore, the empty run is a $\bot$-won run of both games. So, we only need to focus on legal runs of length 1. Consider any such run $\langle\top a\rangle$. By the definition of $\sqcup$, $\mathbf{Wn}^{\sqcup x\dot{A}^{\ddagger c}(x)}\langle\top a\rangle = \top$ iff $\mathbf{Wn}^{\dot{A}^{\ddagger c}(a)}\langle\rangle = \top$; in turn, $\mathbf{Wn}^{\dot{A}^{\ddagger c}(a)}\langle\rangle = \top$ means nothing but that $\dot{A}^{\ddagger c}(a)$ is true; and, by the definition of $\ddot{A}$, $\dot{A}^{\ddagger c}(a) = \ddot{A}(a,c)$. Thus,

$$\mathbf{Wn}^{\sqcup x\dot{A}^{\ddagger c}(x)}\langle\top a\rangle = \top \textit{ iff } \ddot{A}(a,c) \textit{ is true.} \tag{40}$$

Next, again by the definition of $\sqcup$, we have

$$\mathbf{Wn}_{e_c}^{\sqcup x\ddot{A}(x,z)}\langle\top a\rangle = \mathbf{Wn}_{e_c}^{\ddot{A}(a,z)}\langle\rangle.$$

As $\ddot{A}(a,z)$ only depends on $z$ and $e_c$ sends this variable to $c$, we also have $e_c[\ddot{A}(a,z)] = \ddot{A}(a,c)$. And, of course, $\mathbf{Wn}^{\ddot{A}(a,c)}\langle\rangle = \top$ means nothing but that $\ddot{A}(a,c)$ is true. So,

$$\mathbf{Wn}_{e_c}^{\sqcup x\ddot{A}(x,z)}\langle\top a\rangle = \top \textit{ iff } \ddot{A}(a,c) \textit{ is true.}$$

The above, together with (40), implies

$$\mathbf{Wn}^{\sqcup x\dot{A}^{\ddagger c}(x)}\langle\top a\rangle = \mathbf{Wn}_{e_c}^{\sqcup x\ddot{A}(x,z)}\langle\top a\rangle,$$

thus completing our proof of (39) and hence of (38).

Now, by induction, (38) extends from formulas of the form $\sqcup x\dot{A}(x)$ to all more complex subformulas of (32) including (32) itself, meaning that $(32)^{\ddagger c} = e_c[(32)^{\star}]$. The steps of this induction are straightforward, because each of the operations $^{\star}$, $^{\ddagger c}$ and $e_c[\ldots]$ commutes with each of the connectives $\wedge, ⫰$ and $\rightarrow$. $\Box$

Putting (37) and Claim 9.8 together, we find that, for any constant $c$, $\Gamma_c$ is a $\bot$-won run of $e_c[(32)^{\star}]$. Now, remember that $\Gamma_c$ is the $\mathcal{H}_c$ vs. $\mathcal{E}$ run on $e_c$, according to Lemma 5.1 meaning that $\Gamma_c$ is the run spelled by an $e_c$-computation branch of $\mathcal{H}_c$. So, every $\mathcal{H}_c$ loses $(32)^{\star}$ on valuation $e_c$, i.e. no $\mathcal{H}_c$ wins $(32)^{\star}$. But every HPM is $\mathcal{H}_c$ for some $c$. Thus, no HPM wins $(32)^{\star}$. In other words, (32) is not valid.

This almost completes our proof of the main Claim 9.2 and hence our proof of the main Lemma 9.1 and hence our proof of the completeness of **Int**. What remains to verify for the official completion of our proof of Theorem 2.6 is that $^\star$ is of complexity $\Sigma_1^B$. Such a verification is given in the following subsection.

## 9.7 The complexity of the counterinterpretation

The counterinterpretation $^\star$ constructed in the previous subsection interprets each atom $\dot{A}(x)$ of (32) as the binary predicate $\ddot{A}(x,z)$. With "true in the sense of Subsection 9.4 (resp. 9.5)" below understood as truth in the sense of the corresponding subsection where $B_c$ is taken in the role of $B$ (and, accordingly, $\Gamma_c$ in the role of $\Gamma$) when constructing the counterinterpretation $^\dagger$ from it, the meaning of the proposition $\ddot{A}(a,c)$ for any given constants $a,c$ is in fact the disjunction of the following two statements:

1. "$B_c$ is short and $\dot{A}(a)$ is true in the sense of Subsection 9.4";
2. "$B_c$ is long and $\dot{A}(a)$ is true in the sense of Subsection 9.5".

Note that arbitrarily long initial segments of $B_c$ can be effectively constructed from $c$. This can be done by first constructing the machine $\mathcal{H}_c$ from number $c$, and then tracing, step by step, how the play between $\mathcal{H}_c$ and $\mathcal{E}$ evolves on valuation $e_c$ according to the scenario described in the proof idea for Lemma 5.1. This makes it clear that the predicate "$B_c$ is long" (with $c$ here treated as a variable) is of complexity $\Sigma_1$, because it says nothing but that there is a computation step in $B_c$ at which $\lfloor \breve{W} \rfloor$ gets matchingly grounded.

Next, "$B_c$ is short" is just the negation of "$B_c$ is long".

Next, some thought can show "$\dot{A}(a)$ is true in the sense of Subsection 9.4" (with $a$, together with the hidden $c$, here treated as a variable) to mean nothing but that there is no computation step in $B_c$ at which $\dot{A}(a)$ becomes the content of some positive non-matchingly grounded residual molecule of the then-current position. So this is the negation of a $\Sigma_1$-predicate.

Finally, "$\dot{A}(a)$ is true in the sense of Subsection 9.5" means that there is a — lexicographically smallest — open chain $C$ hitting $\lfloor \breve{W} \rfloor$, and $\dot{A}(a)$ is not the content of anything in $C$. With a little thought we can see that, to find such a chain $C$, it would be sufficient to trace $B_c$ only up to the computation step at which $\lfloor \breve{W} \rfloor$ gets grounded. So, the complexity of the predicate "$\dot{A}(a)$ is true in the sense of Subsection 9.5" is $\Sigma_1$.

To summarize, where $\dot{A}(x)$ is an atom of (32), $\ddot{A}(x,z)$, i.e $\dot{A}^\star(x)$, is indeed a Boolean combination of $\Sigma_1$-predicates. As for all other elementary atoms, we may make arbitrary assumptions — without affecting the incomputability of $(32)^\star$ — about how they are interpreted by $^\star$. So, we assume that they, too, are interpreted as Boolean combinations of $\Sigma_1$-predicates.

## 10 Finishing the completeness proof for intuitionistic logic

Suppose $K$ is an **Int**-formula with **Int** $\not\vdash K$. Let $F$ be the dedollarization of $K$. According to Lemma 6.2, **Int** $\not\vdash F$. Let $\underline{G} \Rightarrow W$ be the standardization of $F$. By Lemma 7.4, **Int** $\not\vdash \underline{G} \Rightarrow W$. Hence, by the completeness of **Int** with respect to Kripke semantics, there is a model $\mathcal{K}$ with $\mathcal{K} \not\Vdash \underline{G} \Rightarrow W$. Obviously here we may assume that $\mathcal{K} \Vdash \underline{G}$ and $\mathcal{K} \not\Vdash W$. Let $n$ be the size of $\mathcal{K}$. Then, by Lemma 9.1, there is an interpretation $^\circ$ of complexity $\sqcup\Sigma_1^B$ such that $\not\models D^\circ$, where $D$ is the $n$-desequentization of $\underline{G} \Rightarrow W$. But then, by Lemma 8.1, $\not\models F^\circ$. But then, by Lemma 6.3, there is an interpretation $^*$ of the same complexity $\sqcup\Sigma_1^B$ such that $\not\models K^*$. This proves the completeness of **Int** in the strong sense of clause (b) of Theorem 2.6, and here our story ends.

# Index